\documentclass[preprint,authoryear,3p]{elsarticle}

\usepackage{lineno, hyperref}
\usepackage{lipsum}
\journal{}

\input{macros.sty}
\graphicspath{{Figures/}}

\setcitestyle{square,numbers}

\usepackage{floatrow}
\begin{document}

\begin{frontmatter}

\title{Bayesian Learning of Coupled Biogeochemical-Physical Models}

\author[a1,a2]{Abhinav Gupta}
\ead{guptaa@mit.edu}

\author[a1,a2]{Pierre F. J. Lermusiaux\corref{correspondingauthor}}
\ead{pierrel@mit.edu}

\address[a1]{Department of Mechanical Engineering, Massachusetts Institute of Technology, 77 Mass. Ave., Cambridge, MA - 02139}

\address[a2]{Center for Computational Science and Engineering, Massachusetts Institute of Technology, 77 Mass. Ave., Cambridge, MA - 02139}

\cortext[correspondingauthor]{Corresponding author}

\begin{abstract}

Predictive dynamical models for marine ecosystems are used for a variety of needs.
% research and societal needs.
%, involving ocean, weather, climate, energy, food, sustainability, and security applications. 
%For biogeochemical ocean models, 
Due to the sparse measurements and limited understanding of the myriad of ocean processes, 
there is however significant uncertainty. There is model uncertainty in the parameter values, functional forms with diverse parameterizations, and level of complexity needed, and thus in the state variable fields.
% The prediction of the biogeochemical-physical state and parameter values is thus challenging and uncertain.
% Thus, the present work
% simultaneously estimates the probabilities of state variable fields, parameter values, and model equations themselves using novel dynamics-based Bayesian inference from sparse observations. 
%
%
We develop a Bayesian model learning methodology that allows interpolation in the space of candidate dynamical models and discovery of new models from noisy, sparse, and indirect observations, all while estimating state variable fields and parameter values, as well as the joint probability distributions of all learned quantities. %\AG{Should we also say, "in a computationally efficient way"?}
We address the challenges of
high-dimensional and multidisciplinary dynamics governed by partial differential equations (PDEs) by 
using state augmentation and the computationally efficient Gaussian Mixture Model - Dynamically Orthogonal
%(GMM-DO) 
filter.
%
%Thus, we develop a novel PDE-based Bayesian learning framework for high-dimensional and multidisciplinary dynamical system, combining the Dynamically Orthogonal (DO) differential equations for adaptive reduced-order stochastic evolution, and the Gaussian Mixture Model-DO (GMM-DO) filter for simultaneous nonlinear inference in the augmented space of state variables, parameters, and model equations. 
%
Our innovations include
stochastic formulation parameters and stochastic complexity parameters to unify candidate models into a single general model as well as stochastic expansion parameters within piecewise function approximations to generate dense candidate model spaces. 
These innovations allow handling many compatible and embedded candidate models, possibly none of which are accurate, and learning elusive unknown functional forms that augment these models.
Our new Bayesian methodology is generalizable and
interpretable. It seamlessly and
rigorously discriminates among existing models, but also extrapolates out of the space of models to discover new ones.
%It provides associated joint probability distributions for all the learned quantities.
%
%All of this was achieved just at the cost of a single stochastic model simulation with parameter estimation, enabling both discrimination and discovery of models.
 % This novel approach helps to avoid the need for computation of marginal likelihood for each individual candidate model, and not only gives us the ability to discriminate between existing models, but to also extrapolate in the space of models to discover newer ones. 
% The novel theory and method not only seamlessly and rigorously discriminates between existing models, but also extrapolates in the space of models to discover newer ones.
% \PFJL{This text needs some updates:} \AG{Updated:}
We perform a series of twin experiments based on flows past a ridge coupled with three-to-five component ecosystem models, including flows with chaotic advection. We quantify the learning skill, and evaluate convergence and the sensitivity to hyper-parameters. Our PDE framework successfully discriminates among functional forms and model complexities, and learns in the absence of prior knowledge by searching in dense function spaces. 
%Finally, performance is showcased in the presence of chaotic flows.
%
The probabilities of known, uncertain, and unknown
model formulations, and of biogeochemical-physical fields and parameters, are updated jointly using Bayes' law.
Non-Gaussian statistics, ambiguity, and biases are captured. The parameter values and the model formulations that best explain the noisy, sparse, and indirect data are identified.
When observations are sufficiently informative, model complexity and model functions are discovered.
\end{abstract}


\end{frontmatter}

\nolinenumbers
	
\section{Introduction} \label{sec:intro}

The ability to predict and understand marine ecosystems is essential for addressing many of the grand challenges faced by humanity, such as climate change, food security, and sustainability. 
% Biogeochemical-physical models are often used to model ocean ecosystems, which are a set of differential equations which describe the food-web interactions in the marine ecosystem, and also with the physical state of the ocean. 
In broad terms, marine ecosystems can be seen as food webs, or flows of food/energy from nutrients to phytoplankton, to zooplankton, to fish, and finally recycling back to the nutrients
\cite{lalli1997biological, fennel2014introduction}. However, there does not yet exist a single generic model that accurately represents all the components in marine food webs due to the presence of highly complex biological processes with many unknown interactions. %as well as of nonlinear physical forcing.
Therefore, many approximations are made in such ecosystem models. In addition, only parts of a food web are commonly modeled. The interactions of what is modeled with the portions of the food web that are not modeled are then either neglected or parameterized in terms of the modeled variables. Biology is also forced by complex nonlinear physics. Most  biogeochemical-physical modeling systems thus broadly categorize the nutrients and individual species,
%(plankton, fish, etc.) 
representing them as continuous state variable fields, defined as concentrations of nutrients, biomass, or number of organisms per unit volume of water.
% Being passive tracers, 
The dynamics of these fields consists of reaction terms representing biogeochemical processes such as nutrient uptake, grazing, death, etc., and of forcing by physical processes such as advection, diffusion, and sunlight. Each reaction term or physical process is commonly modeled mathematically, using functional forms or terms that contain multiple parameters and have different levels of accuracy.

%\PFJL{This para needs work, some sentences are off:
%- complexity definitions
%- examples of complexity in the lit
%} 
%\AG{As the previous para describes bio models, thus in this para could be about different types of models and how they are different.}

A plethora of biogeochemical modeling systems have been proposed, each of which with many model formulations \cite{hofmann2002predictive,fennel2022ocean}.
%by scientists. 
The models differ in their complexity, or ability to resolve different biological processes. % This ability is determined by the complexity of the model. 
Models of higher complexity have more biological components, functional terms, and parameters. However, process terms and parameters are often poorly known, which hampers the utility of highly complex models \cite{franks2002npz, ward2010parameter, denman2003modelling}. 
%Some prominent biogeochemical models of varying complexities are listed next. 
The simplest models are 3-component nutrient-phytoplankton-zooplankton (NPZ) models \citep{franks1986behavior,flierl2002mesoscale}. NPZ models are easily understood and serve an important role in ocean research. Including the intermediate state of detritus leads to four component NPZ-Detritus biological models \citep{davis1994biological}. Intermediate complexity models involve around 7 to 10 components, adding bacteria, nitrate, ammonium, and dissolved organic nitrogen \citep{fasham1990nitrogen}, or related state variables \citep{besiktepe_et_al_JMS2003}. 
One of the most complex lower-trophic-level marine ecosystem models is the European Regional Seas Ecosystem Model (ERSEM, \citep{baretta1995european,baretta1997preface,blackford2004ecosystem}), originally developed for the North Sea. Many choices of functional forms exist for each of the biological processes \cite{franks2002npz}, leading to application-specific variants of the above models.

Biogeochemical models are commonly developed semi-empirically, leading to uncertainty in their parameters, functional forms, and level of complexity. What is  adequate for a particular ocean region may not work elsewhere or may need to be updated due to seasonal or other variabilities \citep{lermusiaux_et_al_CS2004}. 
Such model uncertainties transfer to the state variables predicted, complicating model learning by direct comparisons of state variables with in situ data.
As a result, when observations are employed to develop models, it is often in an offline mode, fitting parameter values or functional forms to data in controlled experiments. 
With data assimilation, we could however use observations in a direct Bayesian sense, to jointly learn state variables, parameter values, and discriminate/discover functional forms
%of biogeochemical models 
with quantifiable uncertainty \citep{lermusiaux_PhysD2007}.
%for better estimation and prediction of ocean biology.
Most biogeochemical data assimilation \cite{robinson_lermusiaux_Sea2002,dowd2014statistical} can be categorized broadly into two categories. The first is parameter estimation, where model parameters are calibrated by minimizing misfits between model output fields and independent observations \citep{friedrichs2007assessment,losa2004weak,ward2010parameter,mattern2012estimating,toyoda2013improved}.
%lermusiaux_et_al_Oceanog2011}. 
The second is sequential estimation, where observations collected are used to update model states during the forward model integration \citep{besiktepe_et_al_JMS2003,mattern2010sequential,allen2003ensemble,natvik2003assimilation,hu2012data}. 
However, very few studies deal with the simultaneous estimation of parameters, state variables, and model equations. 
Doron et al.\ \citep{doron2011stochastic} used a Monte Carlo ensemble of 200 simulations lasting 30-days % during the spring bloom 
in the North Atlantic and conducted idealized twin experiments with surface observations of phytoplankton to estimate parameters and states with a Kalman filter-based scheme and state augmentation.
Jones et al.\ \citep{jones2010bayesian} performed state and parameter estimation in a nonlinear phytoplankton-zooplankton model using two Markov Chain Monte Carlo (MCMC) algorithms in an identical-twin setting. 
Mattern et al.\ \citep{mattern2013particle} used a nonlinear particle filter scheme to assimilate satellite sea surface color and jointly estimate the state and parameters of a three-dimensional biological ocean model.
Lately, along with state and parameter estimation, the selection of optimal complexity of biogeochemical models has become a new area of research \citep{lermusiaux_PhysD2007,ward2010parameter, giricheva2015aggregation, ward2013biogeochemical}. 
Because of the multiscale and intermittent variability of marine ecosystems, there is also a need for generalized and adaptive modeling, where models can learn and adapt during run-time \cite{lermusiaux_et_al_CS2004,evangelinos_et_al_ICCS2003,tian_etal_2004,lermusiaux_et_al_Oceanog2011}.

Several machine learning methods have been developed for the discovery of model equations. The sparse regression-based methods (SINDy; \cite{brunton2016discovering,rudy2019data}) are promising as they do not require prior knowledge, however, they often require large data sets. Variations of SINDy include weak SINDy to learn PDEs \cite{messenger2021weak}, adaptive generation of features to increase the library of models \cite{kulkarni_et_al_DDDAS2020}, and extensions to Bayesian identification \cite{niven2020bayesian}. Deep learning methods have been derived to obtain marine ecosystem closure models \cite{gupta_lermusiaux_PRSA2021,gupta_lermusiaux_SR2023}.
Genetic algorithms \cite{maslyaev1903data} and reinforcement learning \cite{bassenne2019computational, novati2021automating,wang2019learning} have been used to search the space of candidate models.
However, most of these approaches do not provide uncertainty estimates for the discovered models. 
Methods have also combined prior knowledge about underlying governing equations for model recovery and refinement. For example, Raissi and Karniadakis \cite{raissi2018hidden}  used Gaussian processes to learn the values of the parametric response of partially-known nonlinear differential equations. Unfortunately, data and knowledge of governing laws are luxuries in the case of realistic biogeochemical models.

It is clear that fundamental methods for
identifying dynamical models that best explain sparse data in accord with prior governing laws and uncertainties would be most useful. The Bayesian theory and schemes of Lu and Lermusiaux \cite{lu_lermusiaux_MSEAS2014,lu_lermusiaux_PhysD2021} address several of the above needs and drawbacks, using noisy, sparse, and indirect observations for joint Bayesian inference of states and parameters along with probabilistic discrimination among candidate models. 
However, several questions remain: 
Could we avoid assimilating observations independently in each candidate model when there are so many models to choose from? 
And if none of these models are that accurate, could the Bayesian machine find the elusive true formulations? 
%What to do in the case when none of the candidate models is exactly equal to the true model? 
%Or the functional form is yet completely elusive to scientists? 
%
Could it interpolate within and extrapolate out of known model spaces, while providing accurate joint probability distributions for model states, parameters, and formulations? 
Could such Bayesian learning be efficient and accurate with high-dimensional and multidisciplinary physical-biogeochemical stochastic PDEs?
The overall goal of the present paper is thus to extend and generalize the discrimination-based model learning developed in \cite{lu_lermusiaux_MSEAS2014,lu_lermusiaux_PhysD2021} to allow for interpolation in the space of candidate models and discovery of new models, in an efficient fashion. 
Our novel learning and discovery of differential models are achieved by introducing stochastic formulation parameters, stochastic complexity parameters, and piecewise function approximations assembled with stochastic expansion parameters.
% We address the challenges of simultaneous estimation of state variables, parameters, and model equations in dynamics-based Bayesian learning of high-dimensional coupled biogeochemical-physical models using sparse observations. 
We address the challenges of multidisciplinary dynamics and develop a rigorous PDE Bayesian learning framework
using state augmentation and the Gaussian Mixture Model - Dynamically Orthogonal (GMM-DO) filter \cite{sondergaard_lermusiaux_MWR2013_part1, sondergaard_lermusiaux_MWR2013_part2}.
The final estimates are notably joint probability distributions
for all learned quantities. 
To our knowledge, it is the first time that sequential Bayesian data assimilation is developed to predict and update the joint probability distributions of state variables, parameters, and known, uncertain, and unknown model formulations, 
enabling the Bayesian discovery of model functional forms and model complexities, with applications to high-dimensional ocean physical-biogeochemical dynamical systems.
%
%by combining the Dynamically Orthogonal (DO) methodology \cite{sapsis_lermusiaux_PD2009, sapsis_lermusiaux_PHYSD2012, feppon_lermusiaux_SIREV2018, feppon_lermusiaux_SIMAX2018a} %\AG{feppon-lermusiaux-PhysD2021 is missing} 
%for reduced dimension stochastic evolution, and Gaussian Mixture Model (GMM)-DO filtering algorithm 
% for the simultaneous nonlinear inference of the augmented state variables, parameters and model equations. 

In Sect.~\ref{sec:problem}, we present the problem statement. 
In Sect.~\ref{sec:method}, we develop the general Bayesian learning methodology with novel parameters for model learning and discovery. In Sect.~\ref{sec:exsetup}, we describe the stochastic biogeochemical-physical equations and simulated experiments. In Sect.~\ref{sec:results}, we apply our methodology to four sets of experiments of varying complexities and learning objectives. Conclusions are provided in Sect.~\ref{sec:conclusions}.

\section{Problem Statement} \label{sec:problem}

A single mathematical model that exactly captures all the 
physical and biological
processes occurring in the real world does not yet exist. 
Hence, there is inherent model uncertainty that manifests in many forms, including: initial and boundary condition uncertainties; unreliable parameter values; multiple competing candidate model functions; unknown functional forms; missing model terms; and, debatable complexity of the model. 
In this work, we consider discriminating among candidate models, learning among compatible models, and discovering new model formulations. Compatible models are models that can be related to a single dynamical system theoretically and that can also be combined numerically. Compatible models can nonetheless represent different dynamics, e.g., our goals include learning which dynamics are or are not present based on observations.

In general, we consider a stochastic dynamical modeling system defined on a domain $\mathcal{D}$, governing the uncertain spatiotemporal dynamics of $\mbs{\phi}(\mbs{x},t;\omega): \mathbb{R}^n\times [0, T]\rightarrow \mathbb{R}^{N_v}$, the stochastic state vector comprising $N_v$ dynamical state variable fields (e.g., physical fields, biogeochemical concentration fields, etc.). 
The realization index $\omega$  belongs to a measurable sample space $\Omega$ and the model depends on a vector $\mbs{\theta}(\omega)$ of $N_\theta$ uncertain parameters.
The main notation used is defined in Table~\ref{table: Notation compendium}.
To encompass the majority of scenarios, we write the general form of the uncertain dynamical modeling system as follows,
\begin{equation}
\label{eq:model uncertainty type 1}
\begin{split}
\frac{\Par \mbs{\phi}(\mbs{x},t; \omega)}{\Par t}
&= \mathcal{L}[\mbs{\phi}(\mbs{x},t; \omega),\mbs{\theta}(\omega), \mbs{x}, t] + \mathcal{\widehat{L}}[\mbs{\phi}(\mbs{x},t; \omega); \omega] + \widetilde{\mathcal{L}}[\mbs{\phi}(\mbs{x},t; \omega); \omega]\,, \\
& \hspace{0.5\textwidth} \mbs{x}\in \mathcal{D}, \: t\in [0, T], \: \omega\in\Omega \;, \\
\text{with} \quad & \mbs{\phi}(\mbs{x}, 0;\omega) = \mbs{\phi}_o(\mbs{x}; \omega)\,, \\
 \text{and} \quad & \mathcal{B}[\mbs{\phi}(\mbs{x}, t; \omega)] = \mbs{b}(\mbs{x}, t; \omega), \; \mbs{x} \in \partial\mathcal{D}, \: t\in [0, T], \: \omega\in\Omega \;,
\end{split}
\end{equation}
where $\mbs{\phi}_o(\mbs{x}; \omega)$, $\mathcal{B}$, and $\mbs{b}(\mbs{x}, t; \omega)$ are the stochastic initial conditions, boundary condition operators, and boundary values respectively.
The functional form of the first dynamics term
$\mathcal{L}[\mbs{\phi}(\mbs{x},t; \omega),\mbs{\theta}(\omega), \mbs{x}, t]$ is assumed to be known, but with uncertain parameters $\mbs{\theta}(\omega)$.
The second term $\mathcal{\widehat{L}}[\mbs{\phi}(\mbs{x},t; \omega); \omega] \in \{\widehat{\mathcal{L}}_1[\mbs{\phi}(\mbs{x},t; \omega); \omega], ..., \allowbreak \widehat{\mathcal{L}}_{N_m}[\mbs{\phi}(\mbs{x},t; \omega); \omega]\}$, represents a set of compatible candidate functional forms, where $N_m$ is the number of candidates. For example, for reaction terms, model functions are often from the polynomial, exponential, and/or sinusoidal families, and can be rational or irrational functions. The third term $\widetilde{\mathcal{L}}[\mbs{\phi}(\mbs{x},t; \omega); \omega]$ has a functional form completely unknown.
Each of these three functional terms has uncertainties, hence the $\omega$ dependence.
Their summation encompasses common scenarios, e.g., their multiplication simply absorbs the more known types into the most unknown type. 
The stochastic initial and boundary condition formulations can also have uncertain function forms, similar to the dynamical modeling system itself, i.e., they can be known, belonging to a family, or unknown.
% Along with this, we have time-variation of the uncertain parameters, the stochastic initial state of the system at $t_o$, and the stochastic boundary conditions as described by Equations \ref{eq: Parameter time-invariant app}, \ref{eq:stoch IC} and \ref{eq:stoch BC} respectively.

In some cases, candidate models have different complexities,
%\PFJL{Here, 1 in $u_1$ should become $u_{1,1}$ or some other way to make them model dependent.} \AG{Done}
\begin{eqnarray}
\label{eq:diff complexity models}
\mathcal{M}_i: \, \begin{cases}
\frac{\Par \phi^i_1(\mbs{x},t; \omega)}{\Par t}
=& \mathcal{L}^i_1[\phi^i_1(\mbs{x},t; \omega), ..., \phi^i_{N_v(i)}(\mbs{x},t; \omega), \mbs{\theta}^i(\omega), \mbs{x}, t; \omega] \\
&\vdots  \\
\frac{\Par \phi^i_{N_v(i)}(\mbs{x},t; \omega)}{\Par t}
=& \mathcal{L}^i_{N_v(i)}[\phi^i_1(\mbs{x},t; \omega), ..., \phi^i_{N_v(i)}(\mbs{x},t; \omega), \mbs{\theta}^i(\omega), \mbs{x}, t; \omega]
\end{cases}, \, i= 1, ..., N_m
\end{eqnarray}
where each model, $\mathcal{M}_i$, has $N_v(i)$ number of state variable fields ($\{\phi^i_1, ..., \phi^i_{N_v(i)} \}$) from a pool of candidates, and their aggregates.
In such situations, the candidate models can often remain compatible with each other, for example, low-complexity models are embedded in higher-complexity ones. We refer to such classes of candidate models as, \textit{compatible-embedded models}.
The number $N_v$ of dynamical state variable fields then denotes the number of state variables needed to encompass all models $\mathcal{M}_i$'s.
Of course, in general, uncertainty in parameter values, functional forms, and complexities occur simultaneously, thus, each term $\{\mathcal{L}^i_{1}, ..., \mathcal{L}^i_{N_v(i)}\}$ in equation \ref{eq:diff complexity models} can encompass the $\mathcal{\widehat{L}}$ and $\widetilde{\mathcal{L}}$ terms introduced in equation \ref{eq:model uncertainty type 1}.
Such scenarios are exemplified in our series of experiments in Sect.~\ref{sec:results}. For example, in Experiments-2, three- and four-component biogeochemical models are considered to be candidate $\mathcal{M}_i$'s; in Experiments-1 $\&$ 4, the zooplankton mortality function is considered to be either linear or quadratic, corresponding to $\mathcal{\widehat{L}}$; and in Experiments-3, the zooplankton mortality function is assumed completely unknown, corresponding to $\widetilde{\mathcal{L}}$.

Let $\mbs{\Phi}(t; \omega) \in \mathbb{R}^{N_vN_x}$ denote the spatially discretized state vector of the continuous field $\mbs{\phi}(\mbs{x}, t; \omega)$. where $N_x$ denotes the dimension of the discretized state space.  In the experiments, we assume that all observations $\mbs{\mathcal{Y}}(t;\omega)$ are noisy, sparse, and indirectly related to $\mbs{\Phi}(t;\omega)$ by a stochastic linear measurement model from the state to the data space,
\begin{equation}
\label{eq:observation model}
\mbs{\mathcal{Y}}(t;\omega) = \mbs{H}\mbs{\Phi}(t;\omega) + \mbs{V}(t;\omega), \qquad \mbs{V}(t;\omega) \sim \mathcal{N}(\mbs{0},\mbs{R})
\end{equation}
where $N_y$ is the number of available observations; $\mbs{H}\in \mathbb{R}^{N_y\times N_vN_x}$ the observation matrix; and $\mbs{V}\in \mathbb{R}^{N_y}$ a zero-mean, uncorrelated Gaussian observation noise with covariance matrix $\mbs{R}\in \mathbb{R}^{N_y\times N_y}$.
The latter noise in the observations is larger than the sensor noise as it also contains representation errors \cite{janjic2018representation}: in our experiments, it will be a fraction of the variability in the state variables \cite{lermusiaux_et_al_QJRMS2000,lermusiaux_JAOT2002}. The noisy observations are also sparse ($N_y \ll N_vN_x$) and indirect: they are available only at discrete time-instants, $t_k$ for $k=1,2,...,K$, at a very limited number of spatial locations, and only for one state variable. As in reality, the specific variable being measured will vary from experiment to experiment.

In summary, our specific objectives are thus two-fold, first to solve the stochastic forward-modeling system (Eqs.~\ref{eq:model uncertainty type 1}~\&~\ref{eq:diff complexity models}), taking into account all the associated uncertainties including compatible, compatible-embedded, and unknown model terms; and second to simultaneously learn, in the Bayesian sense, the state fields, parameters, and model equations based on the stochastic, sparse, and indirect observation model (Eq.~\ref{eq:observation model}). 
%
%As in \cite{lu_lermusiaux_PhysD2021},
Our Bayesian learning thus needs to evolve the prior and posterior joint probabilities of state fields, parameters, and model formulations, given the noisy observations available and all other uncertainties.
The overall goal is to accurately represent these probability density functions (pdfs), including the marginal probabilities of known, uncertain, and unknown model formulations. 
It is only if the noisy observations are sufficiently informative about either the state fields, parameters, and model formulations, that the Bayesian machine can identify the true state variables, true parameters, and true model.
If the noisy observations are not sufficiently informative, the perfect Bayesian machine will not lead to a unique identification, but provide the exact posterior probabilities of the models, parameter values, and state variable fields.

% In summary, we first propagate the stochastic dynamical system (Equation \ref{eq:model uncertainty type 1} or \ref{eq:diff complexity models}) forward in time, taking into account all the associated uncertainties. And when noisy observations are available, learn the state variables, parameters and the model equations based on the observation model (\autoref{eq:observation model}). 

\section{General Bayesian Learning Methodology} \label{sec:method}
%  \subsection{Model Learning}
% \label{sec:Model Learning}
% \PFJL{this should be in the intro or section 2:} The candidate models are based on theoretical or empirical prior knowledge about the dynamical system, or inspired from past observations. Thus, we utilize this prior knowledge and uncertainties while learning the true model governing the system from data. 
%\AG{An equivalent form of this para is already present in the intro, so I guess we can get rid of it altogether: "Observations are already an integral part in the formation of these models, and are in general only used for data fitting in order to find appropriate parameter values or functional form of these models in offline mode, or for interpolation/extrapolation of data using the models itself. However, with the availability of state-of-the-art data assimilation techniques, we should instead use these observations in a Bayesian sense to learn parameter values and discriminate / discover functional forms of biogeochemical models with quantifiable uncertainty for better estimation and prediction of ocean biology."}

Prior to developing our general methodology, we briefly review the Bayesian learning 
% of Lu and Lermusiaux 
for rigorous discrimination among candidate differential dynamical models \cite{lu_lermusiaux_MSEAS2014,lu_lermusiaux_PhysD2021}. In this Bayesian discrimination, each candidate model then evolves the joint pdf of its state variables and parameters, independently from other models, and provides probability distributions that are conditional on the candidate model. In other words, each model runs its own probabilistic forecast. When noisy observations are made, the model-conditional state variables and parameters (all contained in $\mbs{\Phi}$), and also the model pdfs themselves, are updated using Bayes' rules \cite{mr1763essay, bertsekas2008introduction},
\begin{equation}
\begin{split}
p_{\mbs{\Phi}|\mbs{\mathcal{Y}},\mathcal{M}} (\mbs{\Phi}|\mbs{y}, \mathcal{M}_i) &= \frac{p_{\mbs{\mathcal{Y}}|\mbs{\Phi},\mathcal{M}} (\mbs{y}|\mbs{\Phi}, \mathcal{M}_i)}{p_{\mbs{\mathcal{Y}}|\mathcal{M}} (\mbs{y}|\mathcal{M}_i)} p_{\mbs{\Phi}|\mathcal{M}} (\mbs{\Phi}|\mathcal{M}_i)\;, \quad \forall \; \mbs{\Phi}\in \mathbb{R}^{N_v N_x}, \forall \; i\in \{1,...,N_m\}\;, \\
\label{eq:post model dist}
p_{\mathcal{M}|\mbs{\mathcal{Y}}} (\mathcal{M}_n|\mbs{y}) &= \frac{p_{\mbs{\mathcal{Y}}|\mathcal{M}} (\mbs{y}|\mathcal{M}_i)}{p_{\mbs{\mathcal{Y}}} (\mbs{y})} p_{\mathcal{M}} (\mathcal{M}_i)\;, \quad  \forall \; i\in \{1,...,N_m\}\;,
\end{split}
\end{equation}
where $\mathcal{M}_i$ is the $i^{th}$ model candidate and the pdfs $p_{\mbs{\Phi}|\mathcal{M}} (\mbs{\Phi}|\mathcal{M}_i)$ and $p_{\mbs{\Phi}|\mbs{\mathcal{Y}},\mathcal{M}} (\mbs{\Phi}|\mbs{y}, \mathcal{M}_i)$ are the prior and posterior model-conditional state variable distributions, respectively. 
The model distribution $p_{\mathcal{M}} (\bullet)$ is the prior probability for each of the candidates being the true model and $p_{\mathcal{M}|\mbs{\mathcal{Y}}} (\bullet|\mbs{y})$ is the corresponding posterior model distribution. 
This pdf $p_{\mathcal{M}|\mbs{\mathcal{Y}}} (\bullet|\mbs{y})$ allows learning by exact Bayesian discrimination among candidate models and is one of the main novelty of \cite{lu_lermusiaux_MSEAS2014,lu_lermusiaux_PhysD2021}.
In particular, when the noisy observations are not sufficient to achieve unequivocally the ultimate learning objective, these posterior pdfs will adequately represent the ambiguity including possible multimodal distributions and the effects of biases in the candidate models \cite{lu_lermusiaux_PhysD2021}.

%\PFJL{I still need to work more on the para  below. We need to introduce this whole methodology and theory, outlining what is new, and then describe what is new in the 3.1, 3.2, etc. 
%It is likely that what had been done before which is most of what is in section 3 before 3.1 should be moved to section 1.2.}
%
The above Bayesian learning evolves each stochastic candidate model separately. To increase efficiency and allow the discrimination among many more models, all the way to a continuous space of models, this should be circumvented. For example, when models are compatible or compatible-embedded, the learning could interpolate in these model spaces, or even extrapolate out of them. New capabilities are also needed to enable Bayesian learning of unknown models.
Next, we thus develop new stochastic parameterizations that unify all such candidate models into a single general modeling system. We recast the model learning into new parameter estimation problems, using stochastic formulation and complexity parameters (Sect.\ \ref{Sec: Special Stochastic Parameters}) and piece-wise function approximation theory with stochastic expansion parameters (Sect.\ \ref{sec: Piece-wise Linear Function Approximations}). 
We then evolve the joint probabilities of the state fields, the regular parameters, and these new stochastic formulation, complexity, and expansion parameters, using stochastic DO equations
(Sect.\ \ref{sec: Bayesian Learning: DO and GMM-Do PDEs}).
%\AG{(\ref{app:DO} and Sect.\ \ref{sec: Stochastic Dynamically-Orthogonal PDEs in the bio-detailed paper})}. 
At each observation time, we perform Bayesian learning  using the GMM-DO filter %\AG{(\ref{app:GMM-DO})}
with state augmentation 
(Sect.\ \ref{sec: Bayesian Learning: DO and GMM-Do PDEs}).
%(\ref{app:state_aug}). 
%
Our methodology 
% (Sect.\ \ref{sec:method})
does not need to compute the discrete marginal likelihoods, $p_{\mbs{\mathcal{Y}}|\mathcal{M}} (\mbs{y}|\mathcal{M}_i)$; instead, it learns in a parameterized continuous model space. 
We thus extend learning among discrete model formulations to learning within a continuous infinite range of formulations as well as across models of different complexities and into unknown models.
% which represents the strength of the observational evidence of the $i^{th}$ candidate model, i.e. the likelihood of model $\mathcal{M}_i$ for all states $\mbs{U}$, and involves high-dimensional integrals. 
%Also, when the model functional form is unknown, one might be required to search in a large candidate space, thus making it completely computationally infeasible to evolve each stochastic candidate model separately. 
%In order to achieve the proposed unification of candidate models, we develop novel methods using special stochastic parameters and stochastic linear piece-wise linear function approximation theory. 
In other words,
we remain able to discriminate among existing models, but we can now also interpolate in or  extrapolate out of the space of models to discover new ones.
%\PFJL{Need to check changes made in defense slides and incorporate them at right places in this paper.} 
%AG{I think we cover all the changes already.}

\subsection{Stochastic Formulation and Complexity Parameters: Compatible and Compatible-embedded Models}
\label{Sec: Special Stochastic Parameters}

Let us first consider the case where, when according to prior scientific knowledge, the uncertain model belongs to a set of compatible candidate functional forms ($\mathcal{\hat{L}}[\bullet]$; Eq.~\ref{eq:model uncertainty type 1}). 
In order to recast this learning problem with multiple models into a learning problem with a single model and parameter estimation, the compatible candidate model functions are added to each other but only after being multiplied with novel stochastic parameters.
Each of the candidates is thus assigned a new stochastic formulation parameter that can take discrete or continuous values depending on the learning objectives and prior knowledge. 
For example, binary values would be utilized to discriminate between the presence or absence of certain functions, while other values would be utilized to allow some linear interpolation within the space defined by the compatible candidate models.
To complete Bayesian learning, when noisy observations are collected, 
the probability distributions of 
these \emph{stochastic formulation parameters}, $\alpha_k(\omega)$'s, $k=1,...,N_m$, are updated and their mean values estimated alongside these of other regular parameters $\mbs{\theta}(t; \omega)$, using state augmentation. 
%as described in \ref{app:state_aug}. 
Summarizing, 
the general model can thus be written as a stochastic linear combination of the candidates,
\begin{equation}
\mathcal{\hat{L}}[\mbs{\phi}(\mbs{x},t; \omega), t; \omega] = \sum_{k=1}^{N_m}\alpha_k(\omega)\mathcal{L}_k[\mbs{\phi}(\mbs{x},t; \omega), \mbs{x}, t; \omega] \;.
\end{equation}
where the distributions of the $\alpha_k(\omega)$'s are updated at each observation time. 
This new formulation can thus both help select active candidate functions and identify their linear combinations. It allows interpolating in the space of known candidate functions.
%\PFJL{Need to review names of stochastic parameters. 
%We also need to check notation (remove conflicts with DO $N_s$)
%For $N$'s, we could use: $N_s$ for DO subspace size, $N_r$ for realizations, $N_v$ and $N_v(i)$ for number of state variables, $N_m$ for number of models, and check to see if we don't have other $N$. You also use $N_{MC}$
%}
%\AG{We also have $N_{\alpha}$ for number of uncertain parameters; $N_{\phi}$ for number of biological variables; $N_{X}$ for number of grid points; $N_{Y}$ for number of observations. For the number of partitions of the range of state in function approx., we can use $N_H$.} 
%\PFJL{Finally, we should also use $i$, $j$, $k$, $\ell$, etc., for indexing - consistently - the same type of parameters to be learned.} 
%\AG{Would it be fine if I use $i$ and $j$ for miscellaneous indexing needs, while always index $a(\omega)$'s with $k$ and $\gamma$'s with $l$?}
%\PFJL{Maybe. What I would do is a list of everything, which we started above, and then we can think what is best.} \AG{Added the above list in the appendix.}

Next, we extend this approach to learn model complexity (Eq.\ \ref{eq:diff complexity models}). 
This is achieved by defining new states ${\phi}'_k$ that are the original states ${\phi}_k$ multiplied with new \emph{stochastic complexity parameters} $\beta_k(\omega)$. Hence, we define ${\phi}'_k = \beta_k(\omega) {\phi}_k$ and a new general model, $\mathcal{L}'_k$, which encompasses all the candidates in the class of compatible-embedded models,
\begin{equation}
\label{eq: compatible-embedded}
\frac{\Par {\phi}'_k(\mbs{x},t; \omega)}{\Par t}
= \mathcal{L}'_k[{\phi}'_1(\mbs{x},t; \omega), ..., {\phi}'_{N_v}(\mbs{x},t; \omega),\mbs{\theta}(t;\omega), \mbs{\beta}(\omega), \mbs{x}, t; \omega], \qquad k = 1,..., N_v
\end{equation}
where $N_v = \max\{N_v(i)\}_{i=1}^{N_m}$ or in general the number of state variables needed to encompass all models $\mathcal{M}_i$'s. By learning this vector of new complexity parameters $\mbs{\beta}(\omega)$, we can eliminate certain state variables or aggregate them to form new states, and determine the model of appropriate complexity that best explains the noisy observed data. 

To illustrate such combinations of compatible-embedded models into a general model, let us consider a case with only two candidate models ($N_m = 2$ in Eq.\ \ref{eq:diff complexity models}). Let us further assume that the set of states of the first model ($\{{\phi}_1, ..., {\phi}_{N_v(1)}\}$) are fully contained within the set of states of the second model ($\{{\phi}_1,\allowbreak  ...,\allowbreak {\phi}_{N_v(1)},\allowbreak ...,\allowbreak {\phi}_{N_v(2)}\}$), and the goal is to discriminate between the presence or absence of either of the models. 
Using a new complexity parameter $\beta(\omega)$ that is allowed to take only binary values and substituting new states variables defined as ${\phi}_{1}' = {\phi}_{1}$, $...$, ${\phi}_{N_v(1)}' = {\phi}_{N_v(1)}$, ${\phi}_{N_v(1)+1}' = \beta(\omega){\phi}_{N_v(1)+1}$, $...$, ${\phi}_{N_v(2)}' = \beta(\omega){\phi}_{N_v(2)}$, the general model can be written as (based on Eq.\ \ref{eq:diff complexity models} and omitting explicit dependence on $\mbs{x}$, $t$, \& $\omega$ for brevity),
\begin{equation}
\label{eq: example beta}
    \begin{split}
        \frac{\Par {\phi}_1'}{\Par t}
=& (1-\beta)\mathcal{L}^1_1[{\phi}_1', ..., {\phi}_{N_v(1)}', \mbs{\theta}^1]  + \beta\mathcal{L}^2_1[{\phi}_1', ..., {\phi}_{N_v(1)}', {\phi}_{N_v(1)+1}', ..., {\phi}_{N_v(2)}', \mbs{\theta}^2] \;, \\
& \vdots \\
\frac{\Par {\phi}_{N_v(1)}'}{\Par t}
=& (1-\beta)\mathcal{L}^1_{N_v(1)}[{\phi}_1', ..., {\phi}_{N_v(1)}', \mbs{\theta}^1]  + \beta\mathcal{L}^2_{N_v(1)}[{\phi}_1', ..., {\phi}_{N_v(1)}', {\phi}_{N_v(1)+1}', ..., {\phi}_{N_v(2)}', \mbs{\theta}^2] \;, \\
\frac{\Par {\phi}_{N_v(1)+1}'}{\Par t}
=&  \beta\mathcal{L}^2_{N_v(1)+1}[{\phi}_1', ..., {\phi}_{N_v(1)}', {\phi}_{N_v(1)+1}', ..., {\phi}_{N_v(2)}', \mbs{\theta}^2] \;, \\
& \vdots \\
\frac{\Par {\phi}_{N_v(2)}'}{\Par t}
=&  \beta\mathcal{L}^2_{N_v(2)}[{\phi}_1', ..., {\phi}_{N_v(1)}', {\phi}_{N_v(1)+1}', ..., {\phi}_{N_v(2)}', \mbs{\theta}^2] \;,
    \end{split}
\end{equation} 
where $\beta(\omega) = 0$ leads to the first candidate model, and $\beta(\omega) = 1$ to the second candidate model. In similar fashion, we can derive the general model for cases with more than two candidate models, with states in one model being aggregate of states in other models, etc.
%\PFJL{We likely need to show a generic equation for the betas in the RHS and add a few lines of text here to explain the embedding of complexity in the RHS? We could have a generic case. This is to balance section 3.2, and will help the DO eqs for beta and complete section 2 as well.} \AG{Done.}

%-------------
\subsection{Stochastic Piecewise Linear Function Approximations: Unknown Models}
\label{sec: Piece-wise Linear Function Approximations}

Model learning with the above new stochastic formulation and complexity parameters requires a set of candidate functional forms to choose from.
However, in some cases, there might be no such prior information\,/\,candidates available, hence there is an unknown part $\widetilde{\mathcal{L}}$ in the dynamical model (Eq.\ \ref{eq:model uncertainty type 1}).
Such dynamical model functions then need to be discovered. 
A way to achieve this is to parameterize the 
function space of the unknown dynamics, for example, using orthogonal polynomials. 
For the scalar right-hand-side of a single dynamical equation, 
if the polynomials are defined over the whole range of the scalar biogeochemical state variable, one would need high-order polynomials for the function approximation to achieve sufficient accuracy. The result is then susceptible to spurious oscillations. 
To remedy this, as in finite elements, one can divide the range of the biogeochemical variable into pieces and use low-order polynomials within each piece that are stitched together to approximate the unknown dynamical model function over its whole range. 
Generalizing, we thus propose to parameterize the unknown function space using stochastic piece-wise continuous functions, building on approximation theory \cite{trefethen2019approximation}.
In the present work, we consider dense piece-wise linear functions as this representation is both rich and simple,  and provides practical approximations of any unknown function. 
As we will showcase, it expands the functional space in which the Bayesian search is performed and enables searching outside of known models and the discovery of new learned functions. 

For brevity, let us only illustrate the scalar case, where $\widetilde{\mathcal{L}}[{\phi}(\mbs{x},t; \omega); \omega]$ is the unknown function (Eq.\ \ref{eq:model uncertainty type 1}) of a single scalar state variable. As biological concentration is finite, we assume that prior information about the range of values taken by the state variable is available, ${\phi}(\mbs{x},t; \omega) \in [{\phi}_L, {\phi}_R], ~ \forall \mbs{x}\in \mathcal{D}~\text{and}~t\in [0, T]$. 
Now, to define a parameterization using continuous piece-wise linear segments, this range $\mathcal{H} = [{\phi}_L, {\phi}_R]$ is divided into an indexed collection of $N_I$ number of intervals with non-zero measure, $\{I_i = [{\phi}_L^{i}, {\phi}_R^{i}]\,\}_{1\leq i \leq N_I}$, forming a partition of the range $\mathcal{H}$, i.e.,
\begin{equation}
\mathcal{H} = \bigcup_{i=1}^{N_I} I_i \quad \text{and} \quad {I_i} \cap {I_j} = \emptyset \quad \text{for} \; i\neq j \;,
\end{equation}
and we use $N_I+1$ points to discretize the range, such that,
\begin{equation}
{\phi}_L = {\phi}_L^1 < {\phi}_R^1 = {\phi}_L^2 < ... < {\phi}_R^{N_I-1} = {\phi}_L^{N_I} < {\phi}_R^{N_I} = {\phi}_R \;.
\end{equation}
Let $\{\Psi_1, ..., \Psi_{N_I+1}\}$ be the linear functions defined on these elements, 
\begin{equation}
\label{eq: linear basis functions}
\begin{split}
\Psi_1({\phi}) &= \begin{cases} 
\frac{1}{({\phi}_R^{1} - {\phi}_L)}({\phi}_R^{1} - \phi) & \text{if} \; {\phi}\in I_{1} \;, \\
0 & \text{otherwise} 
\end{cases} \\
\Psi_k({\phi}) &= \begin{cases} \frac{1}{({\phi}_R^{k-1} - {\phi}_L^{k-1})}({\phi}-{\phi}_L^{k-1}) & \text{if} \; {\phi}\in I_{k-1} \;, \\
\frac{1}{({\phi}_R^{k} - {\phi}_L^{k})}({\phi}_R^{k} - {\phi}) & \text{if} \; {\phi}\in I_{k} \;, \\
0 & \text{otherwise} 
\end{cases} \quad \text{for} \; k \in \{2, ..., N_I\} \;, \\
\Psi_{N_I+1}({\phi}) &= \begin{cases} \frac{1}{({\phi}_R - {\phi}_L^{N_I})}({\phi}-{\phi}_L^{N_I}) & \text{if} \; {\phi}\in I_{N_I} \;, \\
0 & \text{otherwise} 
\end{cases}
\end{split}
\end{equation}
and $\gamma_k (\omega)'s, \; k\in{1, ..., N_I+1}$ be $N_I+1$ \textit{stochastic expansion parameters} that parameterize the unknown function space by taking a linear combination of the functions defined on each element. 
Hence, all together we obtain:
\begin{equation}
\label{eq: stochastic expansion}
\widetilde{\mathcal{L}}[{\phi}(\mbs{x},t; \omega); \omega] = \sum_{k=1}^{N_I+1} \gamma_k(\omega) \Psi_k({\phi}(\mbs{x},t; \omega)) \;.
\end{equation}
Thus, estimating the stochastic expansion parameters $\gamma_k$'s, in turn, leads to the learning of the unknown model function. 
The above formulation ensures $C^0$ continuity in the functional space 
and is equivalent to linear interpolating splines \cite{jim2023mat}. 
The prior pdf of these parameters defines the functional space in which the search is performed. By construction, this parameterized space can be made as dense as desired, by increasing the number of realization $\omega$ of $\gamma_k (\omega)$'s.
Throughout this paper, we only consider linear segments as the basis, however, this formulation can be extended to any other basis, such as higher-degree splines.

%\PFJL{For the paper, should we move section 4.2, and 4.3 here, and perhaps merge with 3.3, and then have 4.9 after 3.3?. All of 4.2, 4.3 and 4.9 are general method subsections I think.} \AG{Thinking more about this after our meeting, moving 4.2, 4.3, and 4.9 here feels a little weird in terms of flow of the paper. I agree that they are general sections, however, still specific to the application shown in the paper. In the current section 3, we have not really talked anything about biogeochemical models.}

%-------------
\subsection{Bayesian Learning: stochastic DO PDEs, GMM-DO Filter, and Learning Skill}
\label{sec: Bayesian Learning: DO and GMM-Do PDEs}

To provide an accurate and informative prior for our new Bayesian learning paradigm with uncertain and unknown nonlinear dynamics and PDEs, we employ Dynamically Orthogonal (DO) equations \cite{sapsis_lermusiaux_PD2009, sapsis_lermusiaux_PHYSD2012, feppon_lermusiaux_SIMAX2018a}. 
% The high dimensionality associated with ocean models, render this problem intractable. 
The DO equations are instantaneously optimal reduced-order differential equations that evolve, based on the governing nonlinear dynamics, the dominant probabilistic subspace.
Their derivation with uncertain parameters is outlined in \ref{app:DO} and in Sect.\ \ref{sec: Stochastic Dynamically-Orthogonal PDEs in the bio-detailed paper} for biogeochemical specifics. 
%\PFJL{In this section, we could list the parameters: $N_s$, $N_{GMM}$,
%$N_{MC}$, etc. Note we have conflicting symbols. We may need to do a table in an appendix as I did with Pete.} \AG{Done.}

For the Bayesian learning at each observation time, the GMM-DO filter \cite{sondergaard_lermusiaux_MWR2013_part1, sondergaard_lermusiaux_MWR2013_part2} is used to perform nonlinear, non-Gaussian updates of the probability distribution of all quantities estimated, as detailed in \ref{app:GMM-DO}. 
For the joint Bayesian learning of state variables and parameters, 
we combine the GMM-DO filter with state augmentation \cite{gelb1974applied,lu_lermusiaux_PhysD2021} ($\mbs{\Phi}_{aug}$; \ref{app:state_aug}).
%
%This approach is extended to stochastic dynamical models featuring uncertain parameters using the technique of state augmentation \cite{gelb1974applied,lu_lermusiaux_PhysD2021} (\ref{app:state_aug}), enabling joint Bayesian learning of state variables and parameters. 
%
%
In essence, the DO equations evolve the initial joint probability distribution for the augmented state variable, $p_{\mbs{\Phi}_{aug}}(\mbs{\Phi}_{aug}(0; \omega))$, to obtain the forecast/prior joint distribution at the observation time $t$, $p_{\mbs{\Phi}_{aug}^f} (\mbs{\Phi}_{aug}^f(t; \omega))$.
This prior pdf is approximated with multivariate Gaussian Mixture Models (GMMs) 
\begin{eqnarray}
p_{\mbs{\Phi}_{aug}^f} (\mbs{\Phi}_{aug}^f(t; \omega)) \approx \sum_{j=1}^{N_{\text{GMM}}} \pi^f_{\mbs{\Phi}_{aug},j} \times \mathcal{N}(\mbs{\Phi}_{aug}^f(t; \omega);\boldsymbol{\mu}^f_{\mbs{\Phi}_{aug},j}, \mbs{\Sigma}^f_{\mbs{\Phi}_{aug},j}) 
\end{eqnarray}
using an efficient fit in the DO subspace (\ref{app:GMM-DO}).
Given the property that GMMs are conjugate priors to Gaussian observation models (Eq.~\ref{eq:observation model}), their Bayesian update remains a GMM \cite{sondergaard_lermusiaux_MWR2013_part1,casella2021statistical}, providing the posterior pdf, $p_{\mbs{\Phi}_{aug}^a} (\mbs{\Phi}_{aug}^a(t; \omega)) $,
\begin{eqnarray}
p_{\mbs{\Phi}_{aug}^a} (\mbs{\Phi}_{aug}^a(t; \omega)) \approx \sum_{j=1}^{N_{\text{GMM}}} \pi^a_{\mbs{\Phi}_{aug},j} \times \mathcal{N}(\mbs{\Phi}_{aug}^a(t; \omega);\boldsymbol{\mu}^a_{\mbs{\Phi}_{aug},j}, \mbs{\Sigma}^a_{\mbs{\Phi}_{aug},j})
\end{eqnarray}
where, $\forall j\in \{1,...,N_{\text{GMM}}\}$,
\begin{eqnarray}
\begin{split}
\pi_{\mbs{\Phi}_{aug},j}^a &= \frac{\pi_{\mbs{\Phi}_{aug},j}^f \times \mathcal{N}(\mbs{{y}};\mbs{{H}}_{aug}\boldsymbol{\mu}^f_{\mbs{\Phi}_{aug},j}, \mbs{{H}}_{aug}\mbs{\Sigma}^f_{\mbs{\Phi}_{aug},j}\mbs{{H}}_{aug}^T + \mbs{R})}{\sum_{m=1}^{N_{\text{GMM}}}\pi_{\mbs{\Phi}_{aug},m}^f \times \mathcal{N}(\mbs{{y}};\mbs{{H}}_{aug}\boldsymbol{\mu}^f_{\mbs{\Phi}_{aug},m}, \mbs{{H}}_{aug}\mbs{\Sigma}^f_{\mbs{\Phi}_{aug},m}\mbs{{H}}_{aug}^T + \mbs{R})} \;, 
%\quad \forall j\in \{1,...,N_{\text{GMM}}\} \;, 
\\
\boldsymbol{{\mu}}^a_{\mbs{\Phi}_{aug},j} &= \boldsymbol{\mu}^f_{\mbs{\Phi}_{aug},j} + \mbs{{K}}_j(\mbs{{y}} - \mbs{{H}}_{aug}\boldsymbol{\mu}^f_{\mbs{\Phi}_{aug},j}) \;, 
%\qquad \forall j\in \{1,...,N_{\text{GMM}}\} \;, 
\\
\mbs{\Sigma}^a_{\mbs{\Phi}_{aug},j} &= (\mbs{I} - \mbs{{K}}_j\mbs{{H}}_{aug})\mbs{\Sigma}^f_{\mbs{\Phi}_{aug},j} \;, 
%\qquad \forall j\in \{1,...,N_{\text{GMM}}\} \;, 
\\
\mbs{{K}}_j &= \mbs{\Sigma}^f_{\mbs{\Phi}_{aug},j}\mbs{{H}}_{aug}^T(\mbs{{H}}_{aug}\mbs{\Sigma}^f_{\mbs{\Phi}_{aug},j}\mbs{{H}}_{aug}^T + \mbs{R})^{-1} \;. 
%\;, \qquad \forall j\in \{1,...,N_{\text{GMM}}\} \;.
\end{split}
\end{eqnarray}
Further, using the properties of affine transformation, the above Bayesian update is performed in the DO subspace (\ref{app:GMM-DO}), thus rendering it computationally tractable.
%\AG{Based on what we keep, we might need to update the notation table.}
%\PFJL{Nice job.  I think it is fine, it will help the new Fig 1. I moved the for-all to make it easier to read. I think you can update the notation table.} 
%\AG{Done!}

Our novel schemes allow for efficient simultaneous Bayesian estimation of state variable fields, parameters, and model equations themselves, all while using a single modeling system. 
They recast the learning of compatible and compatible-embedded models into formulation and complexity parameter estimations  (Sect.\ \ref{Sec: Special Stochastic Parameters}) and,
to allow the discovery of formulations,
parameterize the space of unknown model functions using piece-wise linear continuous functions (Sect.\ \ref{sec: Piece-wise Linear Function Approximations}).
%
%Then, combining the GMM-DO filter with state augmentation, our novel schemes of recasting the learning of compatible and compatible-embedded models into special parameter estimations and of parameterizing the space of unknown model functions using piece-wise linear continuous functions,
%
%allow for efficient simultaneous Bayesian estimation of state variable fields, parameters, and model equations themselves, all while using a single modeling system. 
%
For the former, the learning occurs
within the space of candidate models while for the latter, it occurs outside of that space and into the space of unknown model functions, hence providing the capability for model discovery.
Importantly, this discovery is interpretable as it is in the form of piece-wise continuous functions. In addition, all of our Bayesian estimations provide much more than maximum likelihood estimates: they predict and update the complete joint probability distribution of states, parameters, and models.
If the noisy, sparse, and indirect observations are not sufficiently informative to learn and eliminate all but one model, parameter value, or state variable field, our Bayesian learning estimates the correct multi-modal pdfs. 
Our learning can indeed represent ambiguity, e.g.\ multiple options are possible, or even equifinality \cite{hart2000pattern}, e.g.\ a set of model estimates have the same likelihood.
It can also signal the presence of bias in competing model formulations. 
Such capabilities will be showcased in Sect.\ (\ref{sec:results}).
%
%\PFJL{Pierre to use/rephrase some of the above in abstract, intro and conclusion.}

To evaluate the learning skill, we first compare the mean fields and parameters with the noisy observations, using several error metrics. We also analyze the evolution of the pdfs of fields and parameters, as well as the convergence of these pdfs with stochastic resolution.

\begin{figure}[h]
 	\centering
 	\includegraphics[width=0.9\textwidth]{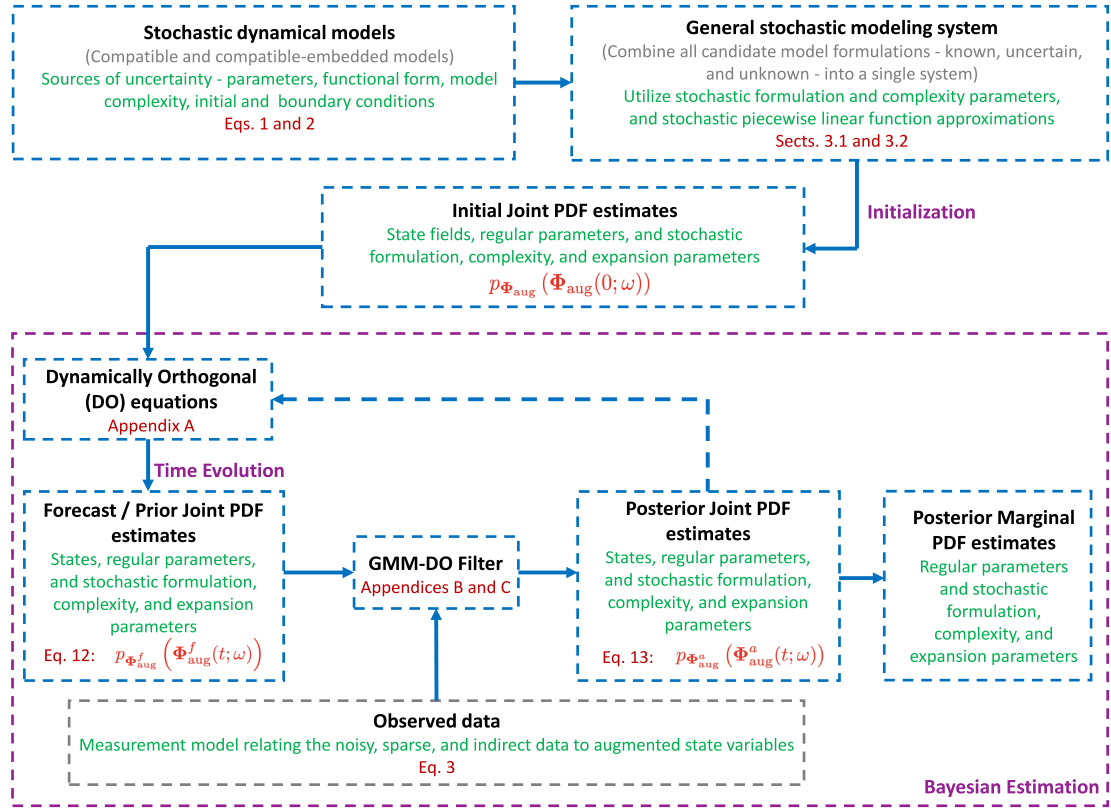}
 	\vspace*{-3mm}
 	\caption{
 	Overview of the general Bayesian model learning methodology.
  }
    
 	\label{fig: overview general Bayesian model learning methodology}
 \end{figure}

The definitions and notation for the hyper-parameters used in the DO methodology and the GMM-DO filter are provided in Table~\ref{table: Notation compendium}. To summarize, in Fig.~\ref{fig: overview general Bayesian model learning methodology}, we provide an overview of our general Bayesian model learning methodology with references to relevant equations, sections, and appendices.

\section{Biogeochemical-Physical Equations and Simulated Experiments Setup} \label{sec:exsetup}
\label{sec: bio detailed paper ExSetup}

In this section, we describe the specifics of our simulated Bayesian learning experiments. We start with the biogeochemical differential equations, their coupling with the physics PDEs, and the stochastic DO decomposition with uncertain and unknown terms. This is followed by details of the modeling domain, numerical methods, initialization of the stochastic simulations, true solution generation, simulated noisy, sparse, and indirect observations, and learning metrics.

%--------------------------
%--------------------------
\subsection{Biogeochemical Models}
\label{sec:biogeochemical models}

The biogeochemical differential equations that we employ are adapted from \cite{tian_etal_2004,lermusiaux_PhysD2007} and references therein, and from Newberger et.\;al., \cite{newberger2003analysis}.
% The latter models were used to simulate the ecosystem in the Oregon coastal upwelling zone in a two-dimensional approximation. 
%
%These models allow generalized and adaptive biogeochemical modeling \cite{lermusiaux_et_al_CS2004,evangelinos_et_al_ICCS2003,tian_etal_2004,lermusiaux_PhysD2007}. 
They meet the criterion of being compatible with each other, with low complexity models being embedded in higher complexity models (compatible-embedded-models).
We will utilize three reaction models: the three-component NPZ model, i.e., nutrients ($N$), phytoplankton ($P$), and zooplankton ($Z$); the four-component NPZD model that adds Detritus ($D$); and, the five-component NNPZD model that adds a second nutrient, simulating ammonia ($NH_4$), nitrate ($NO_3$), $P$, $Z$, and $D$. 

The NPZ biogeochemical reaction model is given by,
\begin{equation}
\label{eq:NPZ model}
\begin{split}
\frac{dN}{dt} &= -\underbrace{G\frac{PN}{N+K_u}}_{Nutrient~Uptake}+ \underbrace{\Xi P}_{Phyt.~Mortality} + \underbrace{\Gamma Z}_{Zoo.~Mortality} + \underbrace{R_m \gamma Z(1-\exp^{-\Lambda P})}_{Zoo.~Egestion} \;, \\
\frac{dP}{dt} &= \underbrace{G\frac{PN}{N+K_u}}_{Nutrient~Uptake} - \underbrace{\Xi P}_{Phyt.~Mortality}  - \underbrace{R_m Z(1-\exp^{-\Lambda P})}_{Zoo.~Grazing} \;, \\
\frac{dZ}{dt} &=   \underbrace{R_m (1-\gamma) Z(1-\exp^{-\Lambda P})}_{Zoo.~Ingestion} - \underbrace{\Gamma Z}_{Zoo.~Mortality} \;,
\end{split} 
\end{equation}
where $G$ representing the optical model,
\begin{equation}
\label{eq: optical model}
    G = V_m\frac{\alpha I}{(V_m^2 + \alpha^2 I^2)^{1/2}} \;, \quad \text{and} \quad I(z) = I_0 \exp^{k_w z}\;,
\end{equation}
$z$ is depth, and $I(z)$ models the availability of sunlight for photo-chemical reactions. 
The parameters in Eqs.~\ref{eq:NPZ model}~\&~\ref{eq: optical model} are: $k_w$, light attenuation by sea water; $\alpha$, initial slope of the $P$-$I$ curve; $I_0$, surface photosynthetically available radiation; $V_m$, phytoplankton maximum uptake rate; $K_u$, half-saturation constant for phytoplankton uptake of nutrients; $\Xi$, phytoplankton specific mortality rate; $R_m$, zooplankton maximum grazing rate; $\Lambda$, Ivlev grazing constant; $\gamma$, fraction of zooplankton grazing egested; and $\Gamma$, zooplankton specific excretion/mortality rate. 
%and $T_{bio}$, total biomass concentration (Sect.\ \ref{sec:initialization}).
%\AG{$T_{bio}$ is only used in sect. 4.6 for the first time, i.e. why I had removed it from here}. 
In this NPZ model (Eq.\ \ref{eq:NPZ model}), the nutrient uptake by phytoplankton is governed by a Michaelis-Menten formulation, which amounts to a linear uptake relationship at low nutrient concentrations that saturates to a constant at high concentrations. 
The grazing of phytoplankton by zooplankton follows a similar behavior: their growth rate becomes independent of $P$ in case of abundance, but proportional to available $P$ when resources are scarce; hence, zooplankton grazing is modeled by an Ivlev function. The death rates of both $P$ and $Z$ are linear, and a portion of zooplankton grazing in the form of excretion goes directly to nutrients.

For the NPZD model, the only change is in the addition of detritus, which is the intermediate state through which dead plankton is converted to nutrients,
\begin{equation}
\label{eq:NPZD model}
\begin{split}
\frac{dN}{dt} &= -G\frac{PN}{N+K_u}+ \underbrace{\Phi D}_{Remineralization} + \Gamma Z  \;, \\
\frac{dD}{dt} &= R_m \gamma Z(1-\exp^{-\Lambda P}) + \Xi P - \underbrace{\Phi D}_{Remineralization} \;.
\end{split}
\end{equation}

However, for the NNPZD model, the nutrients are divided into ammonia and nitrates, which are the two most important forms of nitrogen in the ocean \cite{lalli1997biological, fennel2014introduction}. This helps to capture new processes such as phytoplankton cells preferentially taking up ammonia over nitrates because the presence of ammonia inhibits the activity of the enzyme nitrate reductase essential for the uptake kinetics, the pool of ammonia coming from remineralization of detritus, and part of this ammonia pool getting oxidized to become a source of nitrates referred to as nitrification, etc.\ \cite{lalli1997biological, fennel2014introduction,besiktepe_et_al_JMS2003}. The NNPZD model is given by,
\begin{equation}
\label{eq:NNPZD Model}
\begin{split}
\frac{dNO_3}{dt} =& ~\Omega NH_4 - \underbrace{G\left[\frac{NO_3}{NO_3+K_u}\exp^{-\Psi_I NH_4}\right]P}_{Nitrate~Uptake}  \,, \\
\frac{dNH_4}{dt} =& ~-\Omega NH_4 + \Phi D +\Gamma Z  - \underbrace{G\left[\frac{NH_4}{NH_4+K_u}\right]P}_{Ammonia~Uptake}  \,, \\
\frac{dP}{dt} =& ~\underbrace{G\left[\frac{NO_3}{NO_3+K_u}\exp^{-\Psi_I NH_4} + \frac{NH_4}{NH_4+K_u}\right]P}_{Nitrate~+~Ammonia~Uptake} - \Xi P - R_m Z(1-\exp^{-\Lambda P}) \,, \\
\frac{dZ}{dt} =&  ~ R_m (1-\gamma) Z(1-\exp^{-\Lambda P}) - \Gamma Z \,,  \\
\frac{dD}{dt} =&~ R_m \gamma Z(1-\exp^{-\Lambda P}) + \Xi P - \Phi D \,.
\end{split}
\end{equation} 

The above three reaction models aim to capture the lower-trophic-level (LTL) interactions in the ocean ecosystem. 
They are the Lagrangian or ordinary differential equation (ODE) versions of these models. For realistic ocean field simulations, the above rates of change are material derivatives of dynamic tracers that are coupled with the physics using advection-diffusion-reaction PDEs. Of course, these models are not directly applicable in every ocean region without parameter tuning or modifying the functional form of the reaction terms. 
Regional diversity is one of the reasons for parameter and functional form (model) uncertainties. 

%--------------------------
%--------------------------
\subsection{Coupling with the Physics}

In biogeochemical-physical models, the physics is provided by solving PDEs for the conservation of mass and momentum (Navier-Stokes), internal energy, and salt, e.g., the ocean primitive equations \citep{haley_lermusiaux_OD2010,
haley_et_al_OM2015}.
These models often contain parameterizations to represent subgrid-scale processes \citep{mcwilliams2008nature, hecht2013ocean}. 
In the present work, we employ the incompressible nonhydrostatic Reynolds-averaged Navier-Stokes (RANS) PDEs \cite{ferziger2002computational},
\begin{equation}
\label{eq:deterministic NVs} 
\begin{split}
\nabla\cdot\mbs{u}(\mbs{x},t) &= 0, \qquad \mbs{x} \in \mathcal{D} \;, \\
\pder{\mbs{u} (\mbs{x},t)}{t} + \mbs{u}(\mbs{x},t) \cdot \nabla\mbs{u}(\mbs{x},t)  &=  -\nabla p(\mbs{x},t) + \nu_E \nabla^2 \mbs{u}(\mbs{x},t) \;, \qquad \mbs{x} \in \mathcal{D} \;, 
\end{split}
\end{equation}
where $\mbs{u}(\mbs{x},t)$ is the velocity field, $p(\mbs{x},t)$  the  pressure field, and $\nu_E$ the turbulent eddy viscosity. 
% In the biogeochemical-physical models, the physics part is either provided by directly solving the Navier-Stokes equations or one of its variants, such as the primitive equations \citep{haley_et_al_OM2015}. The physics models contains their own sources of uncertainties, especially due to the presence of various parameterizations needed to capture the sub-mesoscale processes in the ocean \citep{mcwilliams2008nature, hecht2013ocean}.    
% For the physics part, we make use of the stochastic Navier-Stokes equations,
% \begin{equation}
% \label{eq:stoch NVs} 
% \begin{split}
% \nabla.\mbs{u}(\mbs{x},t;\omega) &= 0, \qquad \mbs{x} \in \mathcal{D}, \qquad \omega \in \Omega \\
% \pder{\mbs{u} (\mbs{x},t;\omega)}{t} + \nabla.(\mbs{u}(\mbs{x},t;\omega) \mbs{u}(\mbs{x},t;\omega))  &=  -\nabla p(\mbs{x},t;\omega) + \Lambda_{Re}(\omega)\nabla^2 \mbs{u}(\mbs{x},t;\omega) \;, \;\;\; \mbs{x} \in \mathcal{D}, \;\;\; \omega \in \Omega \;, 
% \end{split}
% \end{equation}
% where $\mbs{u}(\mbs{x},t;\omega)$ is the stochastic velocity field, $p(\mbs{x},t;\omega)$ is the stochastic pressure field, while $\Lambda_{Re}(\omega)$ is an uncertain parameter representing the inverse of eddy viscosity based Reynolds number. As described earlier, $\omega$ is a realization index belonging to measurable sample space $\Omega$.

The Lagrangian biogeochemical models (Sect.\ \ref{sec:biogeochemical models}) are coupled with the physics using stochastic advection-diffusion-reaction (ADR) PDEs. 
For $N_\phi$ stochastic biogeochemical tracers, $\phi^i(\mbs{x},t; \omega)$'s (concentration per unit volume, \cite{kulkarni_lermusiaux_JCP2019}), we obtain,
%these ADR PDEs are given by,
\begin{equation}
\label{eq:Generic ADR equations}
\begin{split}
\pder{\phi^i(\mbs{x},t; \omega)}{t} + \underbrace{\mbs{u}(\mbs{x},t) \cdot \nabla \phi^i(\mbs{x},t; \omega)}_{\text{Advection}} -\underbrace{\mathcal{K}_E\nabla^2 \phi^i(\mbs{x},t; \omega)}_{\text{Diffusion}} &= \underbrace{S^{\phi^i}(\phi^1, ..., \phi^{N_\phi}, {\theta}^1(\omega), ...,{\theta}^{N_{\theta}}(\omega), \mbs{x},t; \omega)}_{\text{Reaction}} \;, \\
& \hspace{0.1\textwidth} {\rm for\;} i \in \{1,...,N_\phi\},
\end{split}
\end{equation}
where $\mbs{u}(\mbs{x},t)$ is the deterministic velocity field governed by (\ref{eq:deterministic NVs}), 
% \begin{equation}
% \label{eq:Generic ADR equations}
% \begin{split}
% \pder{\phi^i(\mbs{x},t; \omega)}{t} + \underbrace{\nabla.(\mbs{u}(\mbs{x},t; \omega) \phi^i(\mbs{x},t; \omega))}_{Advection} -\underbrace{\frac{1}{Pe}\nabla^2 \phi^i(\mbs{x},t; \omega)}_{Diffusion} = \underbrace{S^{\phi^i}(\phi^1, ..., \phi^{N_\phi},\mbs{x},t; \omega)}_{Reaction} \;, \; \forall i = \{1,...,N_\phi\},
% \end{split}
% \end{equation}
% where $\mbs{u}(\mbs{x},t; \omega)$ is the stochastic velocity field which is provided by the physics model, 
$\mathcal{K}_E$ is the eddy diffusivity, 
% $Pe = \frac{UD}{\mathcal{K}}$ is the turbulent Peclet number (with $\mathcal{K}$ as the eddy diffusivity; $U$ and $D$ are representative velocity and length scales, respectively), 
$S^{\phi^i}(\phi^1, ..., \phi^{N_\phi},\allowbreak{\theta}^1(\omega), ...,{\theta}^{N_{\theta}}(\omega),\mbs{x},t; \omega)$ are the reaction or source terms defined by the right-hand-side of the ODEs of Sect.\ \ref{sec:biogeochemical models}, and the ${\theta}^{l}(\omega)$'s, $l = \{1,...,N_\theta\}$, are the uncertain biogeochemical parameters. 
Biogeochemical reactions are nonlinear in nature, hence, the PDEs (\ref{eq:Generic ADR equations}) form a set of strongly nonlinear, stiff, and coupled PDEs.

%\PFJL{Here, since we have the Re and Pe, the physics and bio equations are and variables are already non-dimensional. We need to indicate this before PDEs (15). Then, we can say at the end}.
%\AG{I think it it would be better to use just $\nu$ and $\kappa$ in the above equations, and later on we introduce Re and Pe. This is because for example, later in the boundary conditions we use $u=U$, where $U$ is actually dimensional.}
%
%
% In the later Sect. \ref{sec: Modeling Domain}, we will provide actual dimensions so as to connect the above dimensionalized variables and PDEs to real ocean conditions. 
%non-dimensionalize all state variables and use the non-dimensionalized versions of the above equations. 

%\PFJL{Abhinav, how you refer to equations and section numbers is not really right or efficient. It should be as [\ref{eq:deterministic NVs}] or (\ref{eq:deterministic NVs}). To avoid adding brackets everywhere, perhaps we can define a macro and/or use eref or something like this.} \AG{I understand and agree. Does cleveref: https://texblog.org/2013/05/06/cleveref-a-clever-way-to-reference-in-latex/ work? I could use that throughout if that looks good to you. Also, do you prefer using "Eq. (1)" or just "(1)"?} \AG{Ensured consistency throughout the paper. I have explicitly mentioned "Eq." or "Eqs." at all the places where the word "equation/s" did not precede the needed reference. }

%--------------------------
%--------------------------
\subsection{Biogeochemical-Physical Stochastic Dynamically-Orthogonal PDEs}
\label{sec: Stochastic Dynamically-Orthogonal PDEs in the bio-detailed paper}

To solve the 
%stochastic coupled physical-biogeochemical modeling 
system of Eqs.~(\ref{eq:deterministic NVs}~\&~\ref{eq:Generic ADR equations}) efficiently, we
now
% employ the DO methodology (\ref{app:DO}), and derive
develop the DO equations for the stochastic ADR PDEs (\ref{eq:Generic ADR equations}) with model and parameter uncertainty.
%similarly to a general stochastic nonlinear dynamical model.
%
We first separate the reactions into known, uncertain, and unknown terms, and write 
(\ref{eq:Generic ADR equations}) in vector form,
\begin{equation}
\begin{split}
\pder{\mbs{\phi}(\mbs{x},t; \omega)}{t} + & \nabla\cdot(\mbs{u}(\mbs{x},t) \mbs{\phi}(\mbs{x},t; \omega)) - \mathcal{K}_E\nabla^2 \mbs{\phi}(\mbs{x},t; \omega) = 
\mbs{S}^{\mbs{\phi}}(\mbs{\phi}(\mbs{x},t; \omega), \mbs{\theta}(\omega), \mbs{\beta}(\omega), \mbs{x},t; \omega) \\
&
+ \mbs{\widehat{S}}^{\mbs{\phi}}(\mbs{\phi}(\mbs{x},t; \omega), \mbs{\theta}(\omega), \mbs{\alpha}(\omega), \mbs{\beta}(\omega), \mbs{x},t; \omega) 
+ \mbs{\widetilde{S}}^{\mbs{\phi}}(\mbs{\phi}(\mbs{x},t; \omega), \mbs{\gamma}(\omega), \mbs{x},t; \omega) \;,
\end{split}
\end{equation}
where 
% $\mbs{\phi} = \begin{bmatrix} \phi^1 \\ \vdots \\ \phi^{N_{\phi}} \end{bmatrix}$
$\mbs{\phi} = \left[ \phi^i\right]_{i=1}^{N_{\phi}}$. The functional form of the first reaction term
% $\mbs{S}^{\mbs{\phi}}(\bullet) = \begin{bmatrix} S^{\phi^1}(\bullet) \\ \vdots \\ S^{\phi^{N_{\phi}}}(\bullet) \end{bmatrix}$ 
$\mbs{S}^{\mbs{\phi}}(\bullet) = \left[ S^{\phi^i}(\bullet) \right]_{i=1}^{N_{\phi}}$ is assumed to be known, however it contains $N_\theta$ uncertain regular parameters 
% $\mbs{\theta}(\omega) = \begin{bmatrix} \theta^1 \\ \vdots \\ \theta^{N_{\theta}} \end{bmatrix}$
$\mbs{\theta}(\omega) = \left[ \theta^k \right]_{k=1}^{N_{\theta}}$.
% The second term $\mbs{\widehat{S}}^{\mbs{\phi}}(\bullet) = \begin{bmatrix} \widehat{S}^{\phi^1}(\bullet) \\ \vdots \\ \widehat{S}^{\phi^{N_{\phi}}}(\bullet) \end{bmatrix}$ 
The second term $\mbs{\widehat{S}}^{\mbs{\phi}}(\bullet) = \left[ \widehat{S}^{\phi^i}(\bullet) \right]_{i=1}^{N_{\phi}}$ is uncertain: it belongs to a family of candidate functions, parameterized using $N_\alpha$ stochastic formulation parameters 
% $\mbs{\alpha}(\omega) = \begin{bmatrix} \alpha^1 \\ \vdots \\ \alpha^{N_{\alpha}} \end{bmatrix}$
$\mbs{\alpha}(\omega) = \left[ \alpha^k \right]_{k=1}^{N_{\alpha}}$, and may %$\mbs{\widehat{S}}^{\mbs{\phi}}(\bullet)$ could also
contain uncertain regular parameters $\mbs{\theta}(\omega)$. 
The candidate models of different complexities are combined using $N_\beta$  stochastic complexity parameters 
% $\mbs{\beta}(\omega) = \begin{bmatrix} \beta^1 \\ \vdots \\ \beta^{N_{\beta}} \end{bmatrix}$
$\mbs{\beta}(\omega) = \left[ \beta^k \right]_{k=1}^{N_{\beta}}$. 
The $\beta_k(\omega)$'s multiplied with the original biological tracer fields (as described in Sect.\ \ref{Sec: Special Stochastic Parameters}) are absorbed into $\phi_i$'s and not explicitly shown; however, $\beta_k(\omega)$'s usually appear on the right-hand-side (RHS) in $\mbs{S}^{\mbs{\phi}}(\bullet)$ and $\mbs{\widehat{S}}^{\mbs{\phi}}(\bullet)$.
The third term 
% $\widetilde{\mbs{S}}^{\mbs{\phi}}(\bullet) = \begin{bmatrix} \widetilde{S}^{\phi^1}(\bullet) \\ \vdots \\ \widetilde{S}^{\phi^{N_{\phi}}}(\bullet) \end{bmatrix}$ 
$\widetilde{\mbs{S}}^{\mbs{\phi}}(\bullet) = \left[ \widetilde{S}^{\phi^i}(\bullet) \right]_{i=1}^{N_{\phi}}$ has a functional form completely unknown, and is parameterized using $N_\gamma$ stochastic expansion parameters 
% $\mbs{\gamma}(\omega) = \begin{bmatrix} \gamma^1 \\ \vdots \\ \gamma^{N_{\gamma}} \end{bmatrix}$
$\mbs{\gamma}(\omega) = \left[ \gamma^k \right]_{k=1}^{N_{\gamma}}$.

The DO decomposition for the biogeochemical fields consists of the sum of the statistical mean $\bar{\mbs{\phi}}$ with the sum of the $N_s$ modes $\tilde{\mbs{\phi}}_i$ multiplied by their stochastic coefficients $Y_i$, all of which will be evolved dynamically using DO differential equations,
\begin{equation}
    \begin{split}
        \mbs{\phi}(\mbs{x}, t; \omega) = \bar{\mbs{\phi}}(\mbs{x}, t) + \sum_{i=1}^{N_s} \tilde{\mbs{\phi}}_i(\mbs{x}, t) Y_i(t; \omega) \,.
    \end{split}
     \label{eq:DO_decomp}
\end{equation}

The uncertain regular and formulation and complexity parameters are split into means and deviations, $\mbs{\theta}(\omega) = \bar{\mbs{\theta}} + \mbs{\mathfrak{D}}^{\theta}(\omega)$, $\mbs{\alpha}(\omega) = \bar{\mbs{\alpha}} + \mbs{\mathfrak{D}}^{\alpha}(\omega)$, and $\mbs{\beta}(\omega) = \bar{\mbs{\beta}} + \mbs{\mathfrak{D}}^{\beta}(\omega)$.
%
%\PFJL{Need to discuss use of $\mbs{E}^{\theta}(\omega)$ for deviations since $E$ is mean operator.} 
%\AG{We can change it, however, the mean operator is $\mathbb{E}$, which is different that what used for deviations.}
%\PFJL{We can use $\mbs{\mathcal{D}}$ or $\mbs{\mathfrak{D}}$, e.g.\  $\mbs{\mathfrak{D}}^{\theta}(\omega)$, for all these deviations.} 
%\AG{Done. Used $\mathfrak{D}$ throughout.}
%
For the nonlinear reaction terms in ${\mbs{S}}^{\mbs{\phi}}(\bullet)$ and $\widehat{\mbs{S}}^{\mbs{\phi}}(\bullet)$, as for the nonlinear path planning optimal propulsion term \cite{subramani_lermusiaux_OM2016,subramani_et_al_CMAME2018},
we utilize a local Taylor series expansion around the statistical means, $\bar{\mbs{\phi}}(\mbs{x},t)$, $\bar{\mbs{\theta}}$, $\bar{\mbs{\alpha}}$, and $\bar{\mbs{\beta}}$, to locally represent the nonlinear stochastic effects in the reaction equations as nonlinear mean terms plus stochastic deviations.
As we will exemplify, for most uncertainties, such stochastic approximation is efficient for Bayesian learning as it maintains the significant computational advantages of DO with respect to the other methods 
\cite{branicki_majda_limit_PCE_CMS2013}.
%
%Handling the $\mbs{\widetilde{S}}[\bullet]$ term is less straightforward because of the need to evaluate the interval in which each state realization value lies at every spatial location and time (see Sect.\ \ref{sec: Piece-wise Linear Function Approximations}). 
Finally, for maximum accuracy, we evaluate the $\mbs{\widetilde{S}}[\bullet]$ terms for every state realization in a Monte-Carlo fashion. 
%To increase efficiency without much loss of accuracy, recent techniques such as dynamic clustering \cite{humara_MSThesis2020,humara_et_al_Oceans2022,charous_clustering} can be used.
%\AG{Please check if the arron-clustering citation is the correct one.} 
%
Details on DO schemes are provided in \ref{app:DO}.
Next, we directly provide the DO differential equations for the mean and for $N_s$ modes and stochastic coefficients with $i \in \{1,...,N_s\}$
(omitting function arguments and using $j$, $n$, and $m$ as summation indices),
%
%\PFJL{I tried to allow separation for the equations below, but it didn't work. Perhaps you can fix this?} 
%\AG{Not clear what you want to achieve?}
%\PFJL{With group, equations in align can normally split on two pages (to avoid the latex big white spaces before/after).}
%\AG{I doubt if this is possible, with split, and also it would look weird if equations referenced by the same number gets separated between pages?}
\begingroup
\allowdisplaybreaks
\begin{align}
% \begin{equation}
    \begin{split}
        \frac{\partial \bar{\mbs{\phi}}}{\partial t} &= -\nabla\cdot (\mbs{u} \bar{\mbs{\phi}}) + \mathcal{K}_E \nabla^2\bar{\mbs{\phi}} + \mbs{S}^{\mbs{\phi}}|_{\begin{smallmatrix}\mbs{\phi} = \mbs{\bar{\phi}}, \\\mbs{\theta} = \mbs{\bar{\theta}}, \\\mbs{\beta} = \mbs{\bar{\beta}}\end{smallmatrix}} 
        +
        \mbs{\widehat{S}}^{\mbs{\phi}}|_{\begin{smallmatrix}\mbs{\phi} = \mbs{\bar{\phi}}, \\
        \mbs{\theta} = \mbs{\bar{\theta}}, \\\mbs{\alpha} = \mbs{\bar{\alpha}}, \\
        \mbs{\beta} = \mbs{\bar{\beta}}\end{smallmatrix}}
        +
        \mathbb{E}[\mbs{\widetilde{S}}^{\mbs{\phi}}]\,, \\
        \frac{\partial \tilde{\mbs{\phi}}_i}{\partial t} &= \mbs{Q}_i - \sum_{j=1}^{N_s}\langle \mbs{Q}_i, \tilde{\mbs{\phi}}_j \rangle \tilde{{\mbs{\phi}}}_j \,, \\
        \frac{d Y_i}{dt} &= \sum_{m=1}^{N_s} \langle \mbs{F}_m, \tilde{\mbs{\phi}}_i \rangle Y_m 
        + \sum_{m=1}^{N_{\theta}} \left\langle \frac{\partial\mbs{S}^{\mbs{\phi}}}{\partial \theta_i}\bigg|_{\begin{smallmatrix}\mbs{\phi} = \mbs{\bar{\phi}}, \\\mbs{\theta} = \mbs{\bar{\theta}}, \\\mbs{\beta} = \mbs{\bar{\beta}}\end{smallmatrix}} , \tilde{\mbs{\phi}}_i \right\rangle \mathfrak{D}_m^{\theta} 
        + \sum_{m=1}^{N_{\beta}} \left\langle \frac{\partial\mbs{S}^{\mbs{\phi}}}{\partial \beta}\bigg|_{\begin{smallmatrix}\mbs{\phi} = \mbs{\bar{\phi}}, \\\mbs{\theta} = \mbs{\bar{\theta}}, \\\mbs{\beta} = \mbs{\bar{\beta}}\end{smallmatrix}}  , \tilde{\mbs{\phi}}_i \right\rangle \mathfrak{D}_m^{\beta}
        \\
        & + \sum_{m=1}^{N_{\theta}} \left\langle \frac{\partial\mbs{\widehat{S}}^{\mbs{\phi}}}{\partial \theta_i}\bigg|_{\begin{smallmatrix}\mbs{\phi} = \mbs{\bar{\phi}}, \\
        \mbs{\theta} = \mbs{\bar{\theta}}, \\\mbs{\alpha} = \mbs{\bar{\alpha}}, \\
        \mbs{\beta} = \mbs{\bar{\beta}}\end{smallmatrix}} , \tilde{\mbs{\phi}}_i \right\rangle \mathfrak{D}_m^{\theta}
        + \sum_{m=1}^{N_{\alpha}} \left\langle \frac{\partial\mbs{\widehat{S}}^{\mbs{\phi}}}{\partial \alpha_i}\bigg|_{\begin{smallmatrix}\mbs{\phi} = \mbs{\bar{\phi}}, \\
        \mbs{\theta} = \mbs{\bar{\theta}}, \\\mbs{\alpha} = \mbs{\bar{\alpha}}, \\
        \mbs{\beta} = \mbs{\bar{\beta}}\end{smallmatrix}} , \tilde{\mbs{\phi}}_i \right\rangle \mathfrak{D}_m^{\alpha} 
        + \sum_{m=1}^{N_{\beta}} \left\langle \frac{\partial\mbs{\widehat{S}}^{\mbs{\phi}}}{\partial \beta_i}\bigg|_{\begin{smallmatrix}\mbs{\phi} = \mbs{\bar{\phi}}, \\
        \mbs{\theta} = \mbs{\bar{\theta}}, \\\mbs{\alpha} = \mbs{\bar{\alpha}}, \\
        \mbs{\beta} = \mbs{\bar{\beta}}\end{smallmatrix}} , \tilde{\mbs{\phi}}_i \right\rangle \mathfrak{D}_m^{\beta} 
        \\
        & +
        \left\langle \mbs{\widetilde{S}}^{\mbs{\phi}} - \mathbb{E}[\mbs{\widetilde{S}}^{\mbs{\phi}}], \tilde{\mbs{\phi}}_i \right\rangle
        \,,
    \end{split}
    \label{eq:DO_Eq1}
% \end{equation}
\end{align}
\endgroup
where,
\begin{equation}
    \begin{split}
       \mbs{Q}_i &= -\nabla\cdot(\mbs{u}\tilde{\mbs{\phi}}_i) +  \mathcal{K}_E \nabla^2\tilde{\mbs{\phi}}_i 
       + \frac{\partial\mbs{S}^{\mbs{\phi}}}{\partial \mbs{\phi}}\bigg|_{\begin{smallmatrix}\mbs{\phi} = \mbs{\bar{\phi}}, \\\mbs{\theta} = \mbs{\bar{\theta}}, \\\mbs{\beta} = \mbs{\bar{\beta}}\end{smallmatrix}} \mbs{\tilde{\phi}}_i 
       + \sum_{j=1}^{N_s}\sum_{n=1}^{N_{\theta}}C_{Y_iY_j}^{-1}C_{\mathfrak{D}_n^{\theta}Y_j}\frac{\partial \mbs{S}^{\mbs{\phi}}}{\partial \theta_n}\bigg|_{\begin{smallmatrix}\mbs{\phi} = \mbs{\bar{\phi}}, \\\mbs{\theta} = \mbs{\bar{\theta}}, \\\mbs{\beta} = \mbs{\bar{\beta}}\end{smallmatrix}} \\
       & + \sum_{j=1}^{N_s}\sum_{n=1}^{N_{\beta}}C_{Y_iY_j}^{-1}C_{\mathfrak{D}_n^{\beta}Y_j}\frac{\partial \mbs{S}^{\mbs{\phi}}}{\partial \beta_n}\bigg|_{\begin{smallmatrix}\mbs{\phi} = \mbs{\bar{\phi}}, \\\mbs{\theta} = \mbs{\bar{\theta}}, \\\mbs{\beta} = \mbs{\bar{\beta}}\end{smallmatrix}} 
       +
       \frac{\partial\mbs{\widehat{S}}^{\mbs{\phi}}}{\partial \mbs{\phi}}\bigg|_{\begin{smallmatrix}\mbs{\phi} = \mbs{\bar{\phi}}, \\
        \mbs{\theta} = \mbs{\bar{\theta}}, \\\mbs{\alpha} = \mbs{\bar{\alpha}}, \\
        \mbs{\beta} = \mbs{\bar{\beta}}\end{smallmatrix}} \mbs{\tilde{\phi}}_i 
       +
       \sum_{j=1}^{N_s}\sum_{n=1}^{N_{\theta}}C_{Y_iY_j}^{-1}C_{\mathfrak{D}_n^{\theta}Y_j}\frac{\partial \mbs{\widehat{S}}^{\mbs{\phi}}}{\partial \theta_n}\bigg|_{\begin{smallmatrix}\mbs{\phi} = \mbs{\bar{\phi}}, \\
        \mbs{\theta} = \mbs{\bar{\theta}}, \\\mbs{\alpha} = \mbs{\bar{\alpha}}, \\
        \mbs{\beta} = \mbs{\bar{\beta}}\end{smallmatrix}} \\
       & + \sum_{j=1}^{N_s}\sum_{n=1}^{N_{\alpha}}C_{Y_iY_j}^{-1}C_{\mathfrak{D}_n^{\alpha}Y_j}\frac{\partial \mbs{\widehat{S}}^{\mbs{\phi}}}{\partial \alpha_n}\bigg|_{\begin{smallmatrix}\mbs{\phi} = \mbs{\bar{\phi}}, \\
        \mbs{\theta} = \mbs{\bar{\theta}}, \\\mbs{\alpha} = \mbs{\bar{\alpha}}, \\
        \mbs{\beta} = \mbs{\bar{\beta}}\end{smallmatrix}}
        + \sum_{j=1}^{N_s}\sum_{n=1}^{N_{\beta}}C_{Y_iY_j}^{-1}C_{\mathfrak{D}_n^{\beta}Y_j}\frac{\partial \mbs{\widehat{S}}^{\mbs{\phi}}}{\partial \beta_n}\bigg|_{\begin{smallmatrix}\mbs{\phi} = \mbs{\bar{\phi}}, \\
        \mbs{\theta} = \mbs{\bar{\theta}}, \\\mbs{\alpha} = \mbs{\bar{\alpha}}, \\
        \mbs{\beta} = \mbs{\bar{\beta}}\end{smallmatrix}}
        +
       \sum_{j=1}^{N_s} C_{Y_iY_j}^{-1} \mathbb{E}[Y_j\mbs{\widetilde{S}}^{\mbs{\phi}}]
       \,, \\
       \mbs{F}_m &= -\nabla\cdot(\mbs{u}\tilde{\mbs{\phi}}_m) +  \mathcal{K}_E \nabla^2\tilde{\mbs{\phi}}_m
       + \frac{\partial\mbs{S}^{\mbs{\phi}}}{\partial \mbs{\phi}}\bigg|_{\begin{smallmatrix}\mbs{\phi} = \mbs{\bar{\phi}}, \\\mbs{\theta} = \mbs{\bar{\theta}}, \\\mbs{\beta} = \mbs{\bar{\beta}}\end{smallmatrix}} \mbs{\tilde{\phi}}_m  + \frac{\partial\mbs{\widehat{S}}^{\mbs{\phi}}}{\partial \mbs{\phi}}\bigg|_{\begin{smallmatrix}\mbs{\phi} = \mbs{\bar{\phi}}, \\
        \mbs{\theta} = \mbs{\bar{\theta}}, \\\mbs{\alpha} = \mbs{\bar{\alpha}}, \\
        \mbs{\beta} = \mbs{\bar{\beta}}\end{smallmatrix}} \mbs{\tilde{\phi}}_m \,,
    \end{split}
    \label{eq:DO_Eq2}
\end{equation}
with $C_{\bu, \bu}$ representing cross-covariances, $\mathbb{E}[\bullet]$ expectations, and $\langle \bu, \bu \rangle$ spatial inner-products operators.

%------------------------
\subsection{Modeling Domain and Boundary Conditions}
\label{sec: Modeling Domain}

\begin{figure}
 	\centering
 	\includegraphics[width=0.75\textwidth]{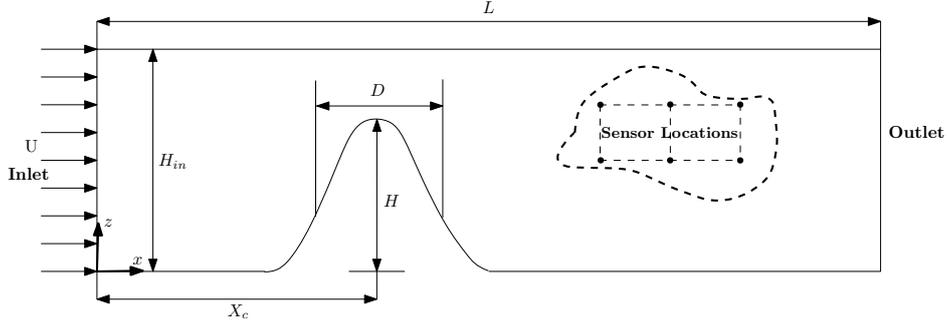}
 	\vspace*{-3mm}
 	\caption{
 	Two-dimensional spatial domain of the flow past a ridge. 
 	%\PFJL{Here, I would simply draw and define the dimensions: $H$ is the ocean height at the inlet, etc. Adding these h, 1, etc is confusing. Also, should we show at least 6 sensor locations, perhaps a little closer to each other} \AG{Done.}
 	The ridge is defined by $He^{-(x-X_c)^2/D^2}$, where $D$ is the characteristic width, $H$ the height, and $X_c$ the distance between the inlet and the center of the ridge. 
  Noisy and sparse observations of state variables are collected downstream of the ridge (see example sensor locations inset). 
  The exact observation locations and the state variables being measured vary with the particular experiment.
%  	All dimensions are given in terms of length scales (D \& H) of the seamount defined by $He^{-(x-X_c)^2/D^2}$ \PFJL{where $D$ is its characteristic width}. Observations are collected downstream of the seamount (sensor locations), with exact observation locations dependent on the particular experiment.}
    }
 	\label{domain BGC flows}
 \end{figure}

Our modeling domain is inspired by Stellwagen Bank at the edge of Massachusetts Bay, which is a whale feeding ground \cite{mcgillicuddy1998adjoint,lermusiaux_JMS2001,besiktepe_et_al_JMS2003,pineda2015whales,tian2015model,pershing_GoM_temp2050_2019,silva_SB_Sanct_2021}, as well as by many other coastal banks and ridges. 
The experimental setup consists of a two-dimensional domain with a bathymetry obstacle (Fig.~\ref{domain BGC flows}), which could be considered a slice of a wide seamount or bank, an idealized sill, or a cross-section of a ridge. In the rest of the text, we will consider our experimental setup as flow over an idealized ridge. 
The mean flow occurs from left to right in the positive $x$-direction over the ridge. Such flows can create an upwelling of nutrients, leading to phytoplankton blooms, zooplankton responses, and nutrient uptake and recycling.
% The velocity field is governed by the two-dimensional, incompressible RANS PDEs (\ref{eq:deterministic NVs}).
% The turbulent Reynolds number is $Re = \frac{UD}{\nu_E}$, where $\nu_E$ is the eddy viscosity, $U$ the inlet free-stream velocity, and $D$ the seamount horizontal length scale. 
% The biogeochemical models are coupled with this RANS flow using the ADR PDEs (\ref{eq:Generic ADR equations}).

% The turbulent Reynolds number is $Re = \frac{UD}{\nu_E}$, where $\nu_E$ is the eddy viscosity, $U$ the inlet free-stream velocity, and $D$ the seamount horizontal length scale. 

% The experimental setup consists of a 2-dimensional domain with a seamount representing an idealized sill or strait (\autoref{domain BGC flows}), which can create an upwelling of the nutrients, leading to phytoplankton blooms. Flow occurs from left to right in the positive $x$-direction over the seamount. This domain is inspired from the Stellwagen Bank at the edge of Massachusetts Bay, which is a whale feeding ground. The velocity field is governed by two-dimensional, incompressible Navier-Stokes equations (\autoref{eq:stoch NVs}). Reynolds number of this flow is based on eddy viscosity, given by $Re = \frac{UD}{\nu_E}$, where $\nu_E$ is the eddy viscosity, $U$ is the inlet free-stream velocity, and $D$ is the seamount horizontal length scale. This parameter is considered to be uncertain, with its probability distribution given in \autoref{table:DA Setup information}. The biogeochemical models are coupled with the flow using the ADR equations (\autoref{eq:Generic ADR equations}).

A horizontal length scale of $D\approx 1~km$ is chosen for the ridge, while the vertical height scale is $H\approx 50~m$. The overall transverse height of the domain is $H_{in} = 100~m$. 
The cross-ridge length of the domain is $L = 20~km$, with center of the ridge at $X_c = 7.5~km$. 

Further, we only consider deterministic boundary conditions (BCs) models. The inlet at the left boundary  has Dirichlet BCs for velocity, and zero Neumann for biological tracers,
\begin{equation}
u=U, ~ v=0  \qquad \text{and} ~ \pder{\mbs{\phi}^i}{x} = 0,\qquad \text{at} ~ x = 0 \;, ~\text{for}~ i \in \{1, ..., N_{\phi}\} \;.
\end{equation} 
On the top and bottom boundary, free slip for velocity and again zero Neumann for tracers are applied,
\begin{equation}
\pder{u}{z} = 0, ~ v=0  ~ \text{and} ~ \pder{\mbs{\phi}^i}{z} = 0,\qquad \text{at} ~ z = 0~\text{and}~z = h \;, ~\text{for}~ i \in \{1, ..., N_{\phi}\} \;.
\end{equation} 
At the outlet on the right boundary, we have open BCs with zero Neumann for all the state variables,
\begin{equation}
\pder{u}{x} = 0, ~ \pder{v}{x}=0
\qquad\text{and} \qquad
\pder{\mbs{\phi}^i}{x} = 0,\qquad \text{at} ~ x = L \;, ~\text{for}~ i \in \{1, ..., N_{\phi}\} \;.
\end{equation} 
Finally, on the obstacle surface, no-slip for velocity and zero Neumann for tracers are used,
\begin{equation}
u=0, ~v= 0 \qquad\text{and} \qquad \pder{\mbs{\phi}^i}{x} = \pder{\mbs{\phi}^i}{z} = 0  ,\qquad \text{at} ~ z = H e^{-(x-X_c)^2 / D^2} \;, ~\text{for}~ i \in \{1, ..., N_{\phi}\} \;.
\end{equation}
 
% (\textcolor{red}{more need to clarify non-dimensionalization?})

\subsection{Numerical Schemes}
\label{sec: Numerical Schemes}

%\PFJL{This was a bandit in the prior section:} 

The velocity and pressure fields are governed by the incompressible nonhydrostatic RANS PDEs (\ref{eq:deterministic NVs}). The stochastic biogeochemical fields are coupled with this dynamic RANS flow and governed by a dynamic reduced-order representation of the original stochastic ADR PDEs (\ref{eq:Generic ADR equations}), the DO ADR PDEs we derived (Eqs.~\ref{eq:DO_Eq1}~\&~\ref{eq:DO_Eq2}).
%
%For numerical implementation, 
%\PFJL{You should read the other papers again and try to catch these things. } 
%\AG{I assume you mean non-dimensionalization. I will have a look at others too.} 
%\PFJL{No, I mean numerics. It was at the wrong place.}
%
%
In all experiments Sect.\ \ref{sec: Numerical Schemes},
we solve the deterministic RANS-biogeochemical PDEs for the true solution as well as
the RANS-biogeochemical DO equations for the predicted pdfs using our modular finite-volume framework \cite{ueckermann_and_lermusiaux_MSEAS2012}.
The physical domain (Sect.\ \ref{sec: Modeling Domain}) is discretized using a uniform finite-volume staggered C-grid, for both the flow and stochastic biogeochemical fields. 
The size of finite volumes in each $x-$ and $z-$ direction is equal to $\Delta x = \frac{1}{15}$ and $\Delta z = \frac{1}{15}$ (non-dimensional) respectively, thus, a grid-size of $300 \times 30$. 
Advection is computed explicitly, using a total variation diminishing (TVD) scheme with a monotonized flux limiter \cite{van1977towards}. 
Diffusion is treated implicitly, with a second-order central difference scheme. All the reaction terms are computed explicitly. To handle the complex boundaries with the structured Cartesian grid, a ghost cell immersed boundary method is adopted for accurate enforcement of the boundary conditions \cite{gupta_PhDThesis2022}. For time-marching of the PDEs (RANS, DO mean, and DO modes), we use a first-order forward Euler method, while for the stochastic DO coefficient ODEs, we use a four-stage Runge-Kutta scheme.
%
% \PFJL{For diffusion, we employ
%..., and for the reaction terms, we ...}. \PFJL{For the RANS time-marching, we use the first-order forward Euler method.} \AG{Done.}
%
% For the stochastic DO, first-order forward Euler is also used to evolve the mean and DO modes, and a four-stage Runge-Kutta scheme is employed for the stochastic DO coefficients. 
A non-dimensional time-step of $\Delta t =\frac{1}{240}$ is used in all the experiments. It is also ensured that we satisfy the Courant-Friedrichs-Lewy (CFL) condition at all times. We refer to  (\cite{ueckermann_et_al_JCP2013}) and (\cite{feppon_lermusiaux_SIREV2018}) for more details on the numerical schemes we employ. 

%In all experiments Sect.\ \ref{sec: Numerical Schemes}, the deterministic equations for the true solution and DO equations for the estimate pdfs are solved using the modular finite-volume framework. 

\subsection{Stochastic Balanced Initialization: Parameters, State Variable Fields, and Probabilities}
\label{sec:initialization}

The values of the parameters for the physics are chosen such that the flow emulates some coastal ocean dynamics. The dimensional barotropic velocity at the inlet is chosen to be $U\approx 10^{-2}$ to $10^{-1}~m/s$. The subgridscale eddy-viscosity is $\nu_E \approx 0.01$ to $0.5~m^2/s$. Considering the vertical length scale of $H\approx 50$~m for the ridge, we obtain an eddy-viscosity Reynolds number of $Re = \frac{U H}{\nu_E} \approx 1$ to $500$. 
Further, we do not consider any wind-forcing explicitly. 
%, thus, helping us implement Large Eddy Simulations (LES) in the simplest form. 
%
%\PFJL{We need to add values of H and of - all -  other physics parameters as needed, say no winds, etc.}. \AG{Done.}
%
For the initial velocity, we use a divergence free velocity field that satisfies the inlet and outlet boundary conditions, and so mass conservation in the given domain. The pressure field is initialized to be zero throughout the domain.
%\PFJL{The initial pressure field is obtained ...} \AG{Done.}

% The parameters for the physics are chosen such that the flow emulates the large scale ocean dynamics. The dimensional velocity at the inlet is chosen to be $U\approx 10^{-2}~m/s$, while the eddy viscosity to be of the order, $\nu_E\sim \mathcal{O}(10)~m^2/s$. Considering the horizontal length scale of $D\approx 1~km$ for the seamount, the typical value of eddy viscosity based Reynolds number ($Re$) turns out to be of the order of, $Re \sim \mathcal{O}(1)$, thus helping us implement Large Eddy Simulations (LES) in the simplest form. For the mean velocity, we use a divergence free velocity field, conforming to the given domain. For experiments with uncertainty in the physics, the velocity initial uncertainty is prescribed using the same method as in Lermusiaux, 2013 (\cite{ueckermann_et_al_JCP2013}), while the inverse of the Reynolds number ($\Lambda_{Re}(\omega)$) is sampled from the distribution prescribed in \autoref{table:DA Setup information}.

The biological parameters 
% used in the biological models 
are either deterministic or stochastic. 
The values of the deterministic parameters are kept fixed for every realization. The stochastic parameters are sampled from their respective initial pdf or joint pdfs, if available; when measurements are made, these pdfs are updated along with all other state variables, using the GMM-DO filter. The stochastic parameters are divided into two categories, regular ones that were originally present in the biogeochemical models and have biological origins, and new parameters introduced for the unification of candidate models and parameterizations of unknown functions. 
The deterministic values and initial pdfs of the biological parameters used in the main experiments are given in Table \ref{table:DA Setup information}. 
The initial pdfs of all the stochastic parameters are set to be uniform and independent of each other unless otherwise specified. 
Several but not all regular parameters are assumed stochastic, aiming to showcase parameter uncertainties that are commonly significant or of different mathematical types (e.g., nonlinear functions) or learning types (e.g., parameters in equations not directly related to the sparse measurements).
%\PFJL{and uniform or Gaussian? Need to describe special and regular.} \AG{Done.}
%
In the experiments presented in this paper, advection-reaction dominates and the eddy-diffusivity for the biological tracers can be taken as negligible, $\mathcal{K}_E \approx 0$, such that the eddy-diffusivity Peclet number $Pe = \frac{UH}{\mathcal{K}_E} \rightarrow \infty$. Other experiments (not shown) were also successful however with non-negligible diffusivity, e.g.\ \cite{lu_lermusiaux_PhysD2021}.
In all our simulations, a biological time-scale of the order of $1~day$ is used for all non-dimensionalization purposes.
%\PFJL{Need to discuss values of initial parameters here and types of initial probabilities for the parameters. May also need to discuss biological non-dimensional times and Peclet number}. \AG{Done.}
%\PFJL{Why did you use $\mathcal{K}_E$ negligible? This means the tracers have no diffusion, in all experiments?} \AG{Yes, the tracers have no diffusion, but, I don't remember the exact rationale behind it. It has been this case from the starting.} 
%\AG{Added the other experiments line.}

Following \cite{lermusiaux_et_al_QJRMS2000,lermusiaux_JAOT2002,besiktepe_et_al_JMS2003,lermusiaux_PhysD2007,lermusiaux_et_al_Oceanog2011}, in all the subsequent experiments,
biogeochemical fields are initialized in approximate dynamical balance, in accord with their stochastic model PDEs (\ref{eq:Generic ADR equations}) and their sampled parameter values and local total biomass field.
Specifically,
the initial concentration fields for every sampled realization are obtained by finding a biogeochemical equilibrium solution corresponding to their sampled parameter values and vertical biomass profile. These equilibrium fields are found by solving the ODE  nonlinear biogeochemical models of Sect.\ \ref{sec:biogeochemical models} at all depths.
Dynamical equilibrium for initialization is reached when temporal variations become negligible, or when the system reaches a limit cycle or time-variations that are without unrealistic transients \cite{lermusiaux_JAOT2002,haley_et_al_OM2015}. However, in the current setup, the approximate equilibrium solutions at each spatial location are found by using MATLAB's \textit{fsolve} \cite{fsolve} to find the roots of the nonlinear system obtained with time derivatives set to zero \cite{newberger2003analysis}. 
Further, we also impose the initial total biomass profile, $\sum_{i=1}^{N_\phi}\phi^i(z; \omega) = T_{bio}(z)$, to be given, with $T_{bio}$ to be linearly increasing from $10~mmol~N~m^{-3}$ at the surface to $30~mmol~N~m^{-3}$ at the depth of $100~m$, for all the biogeochemical models. 
% 
% In  \autoref{fig:equilibrium solutions}, as an example, we provide equilibrium states for a particular set of parameter values. Further, to initialize the whole domain, 
% 
%  \begin{figure}
% 	\centering
% 	\subfloat[NPZ equilibrium solution]{\label{subfig:NPZ equib }\includegraphics[width=.4\textwidth]{Figures/equib_sol/NPZ.png}}\quad
% 	\subfloat[NPZD equilibrium solution]{\label{subfig:NPZD equib }\includegraphics[width=.4\textwidth]{Figures/equib_sol/NPZD.png}} \\
% 	\subfloat[NNPZD equilibrium solution]{\label{subfig:NNPZD equib }\includegraphics[width=.4\textwidth]{Figures/equib_sol/NNPZD.png}}
% 	\caption{Sample equilibrium solution along depth for the NPZ, NPZD and NNPZD biogeochemical models.}
% 	\label{fig:equilibrium solutions}
% \end{figure}
% 
This depth-dependent equilibrium solution for each realization of the biogeochemical state variables is used to initialize the corresponding fields in space, with the ridge masked at every $x$ location. 
We also ensure that none of the realizations of the stochastic parameters lead to nonphysical equilibrium solutions, such as negative tracer values. 
The value of $30~mmol~N~m^{-3}$ is used to non-dimensionalize all the biogeochemical fields and parameter values. For the non-dimensionalization of parameters, when needed, we additionally use a length-scale of $50~m$ (the height $H$ of the ridge) and a time-scale of $1~day$.
%\PFJL{Likely need to add a bit more for the non-dimen of the parameter values. Also, the non-dim of the biological variables by 30 seems a bit off, but maybe this is ok if we non-dim already in prior section.} \AG{Done.}

To initialize the DO decomposition of the biogeochemical fields, after generating the initial fields for each realization, we compute their statistical average and use it to initialize the mean biogeochemical fields.
To initialize the DO modes and stochastic coefficients, we take the singular value decomposition (SVD) of the ensemble of mean-removed concatenated fields, keeping the dominant singular values and vectors. We account for the differences in the magnitude of the variability of individual biogeochemical tracers before taking the SVD, by appropriate normalization based on their standard deviations (\ref{app:DO}). 

%----------
\subsection{True Solution Generation}
\label{sec:True Solution Generation}

In the present work, twin experiments \cite{bengtsson1981dynamic,ide1998extended1, ide1998extended2,lermusiaux_MWR1999} are conducted and the noisy observations are extracted from a simulated truth. 
To obtain the simulated truth fields for each experiment,
a set of parameters and initial state fields are sampled. 
Starting from these initial conditions, the Navier-Stokes PDEs (\ref{eq:deterministic NVs}) and the deterministic version of the ADR PDEs (\ref{eq:Generic ADR equations}) with the true biogeochemical model are numerically integrated. The result is the simulated truth solution.
% with  sampled from the initial realization space. 
In each experiment, all the remaining deterministic parameters, modeling domain, and numerical schemes are as these of the stochastic simulation using the DO equations.

\subsection{Observations and Inference}
\label{sec:Observations and Inference}

Noisy, sparse, and indirect observations are taken from the simulated true solution (Sect.\ \ref{sec:True Solution Generation}). 
The observation locations are kept in or near the euphotic zone because deeper depths have limited biological variability (Fig.~\ref{domain BGC flows}).
In each experiment, only one of the biological tracer fields is observed at 6 to 9 locations.
What is measured thus varies from experiment to experiment, as is common in real sea experiments.
The observation schedule is also experiment dependent, however, it is not more frequent than once every non-dimensional time:
data are collected at far-apart discrete time-instants, $t_k$ for $k=1,2,...,K$.
This use of a few, infrequent, and indirect observations allows us to showcase the capabilities of multivariate GMM-DO Bayesian learning: it can use 
noisy and sparse measurements of only one state variable to jointly update all pdfs, and thus indirectly estimate multiple state variable fields, parameters, and model functions.

The linear observation matrix $\mbs{H}$ (Eq.\ \ref{eq:observation model}) is specified such that it predicts the concentration of the observed tracer field at the observation locations by interpolating the concatenated state fields at the observation locations. 
The observation error standard deviation matrix ($\sqrt{\mbs{R}}$ in Eq.\ \ref{eq:observation model}) 
%represents the confidence in the sensors, and in all the experiments, the sensor errors are considered independent of each other, such that $\mbs{R}$ 
is assumed diagonal.
%The chosen standard deviation is kept comparable to that expected at the measurement locations before the first observation. 
It models both sensor noise and representation errors \cite{janjic2018representation}. The latter representation errors include subgrid-scale processes and variability in the data not simulated by the dynamical model; the representation errors are thus often much larger than sensor errors. As a result, in our experiments, the observation error standard deviation values contained in $\sqrt{\mbs{R}}$ are set to be a fraction of the local variability in the state variables \cite{lermusiaux_et_al_QJRMS2000,lermusiaux_JAOT2002}. In most of our at-sea experiments, observation errors are significantly smaller than errors in biogeochemical model equations and variability of their fields, and the relative observation error is here set at five percent.
We note that as long as these 
%observation error 
standard deviations and the actual simulated observation noise are consistent and not larger than the natural variability, their values mainly affect the learning rate achieved by data assimilation 
as well as the amount of uncertainty remaining in the learned states and parameters (i.e., smaller observation errors accelerate the learning and reduce the uncertainty  remaining at the end).

Further, the hyper-parameters related to the DO equations and the GMM-DO filter were chosen based on numerical tests and experience \cite{sondergaard_lermusiaux_MWR2013_part2,lolla_et_al_OD2014_part2,lu_lermusiaux_MSEAS2014,gupta_et_al_Oceans2019}, for each of the experiments.
For the DO equations, the number of modes, number of Monte-Carlo coefficient samples, time-step, etc., were selected so as to be sufficient to capture the dominant uncertainty and evolving probability distribution for each of the state vector fields, parameters, and model equations themselves.
For Bayesian learning with the GMM-DO filter, the expectation-maximization (EM) algorithm \cite{bilmes1998gentle} and Bayesian Information Criterion (BIC) \cite{stoica2004model,wornell2013} were employed to select the optimal number of GMM components at each data time. Typical BIC-optimized values for $N_{\mathrm{GMM}}$ were found to be 10 for the
present experiments.

\subsection{Learning Metrics}
\label{sec:Learning Metrics}

We evaluate the performance of our Bayesian learning framework by comparing the learned solution with the true solution from which noisy observations were collected and by examining the posterior joint state-parameter-model probability distributions.
For the former solution evaluations, we compare the true fields to the DO mean fields, and the true parameter values to the most probable DO pdf values of the parameters.
To quantify performance, we examine
the evolution %with time 
of the Root Mean Square Error (RMSE) of the biogeochemical tracer fields, uncertain regular parameters $\mbs{\theta}(\omega)$, and novel formulation $\mbs{\alpha}(\omega)$, complexity $\mbs{\beta}(\omega)$, and/or expansion $\mbs{\gamma}(\omega)$ parameters. 
%\PFJL{PL and AG to discuss this:} \AG{Refer section 5.1 of Jing's thesis to help with the discussion} 
%
%\PFJL{The two of us should discuss RMSE between mean and true fields.} \AG{Done and added RMSE formula.}
%
The RMSE between an evolved stochastic state field/parameter estimate $\phi(\mbs{x}, t;\omega)$ and its corresponding true field/parameter $\phi^{true}(\mbs{x}, t)$, is given by, $\sqrt{\frac{1}{|\mathcal{D}|}\int_{\mathcal{D}}\mathbb{E}[(\phi(\mbs{x}, t;\omega) - \phi^{true}(\mbs{x}, t))^2] d\mbs{x}}$.
The square of RMSE hence consists of two contributions \cite{lin_PhDThesis2020}, one is the square of the $L_2$ distance between the mean of the variable in the stochastic run and the simulated truth, while the other is the variance of the variable. 
In every experiment, the RMSE values of each variable are normalized by the corresponding RMSE value just before the first assimilation step.
%\PFJL{Explain how we examine the posterior state-parameter-model probability distributions}. \AG{Done.}
%
For the latter pdf evaluations, we analyze the evolution of the posterior pdfs of the stochastic DO coefficients, and of the regular and new stochastic parameters. For example, for the DO coefficient realizations, we employ 2-D scatter plots. For the stochastic parameters, we use marginals and kernel-density fits. We also evaluate the convergence of pdf estimates with stochastic resolution, i.e.\ increasing/decreasing stochastic numerical parameters ($N_s$, $N_{r}$, etc.), see Sect.\ \ref{sec: Bayesian Learning: DO and GMM-Do PDEs}.
%\PFJL{Need to check symbols and use right ones here.} \AG{Done.}

\section{Application Results and Discussion} \label{sec:results}

In order to demonstrate the capabilities of our Bayesian learning
% framework 
we utilize four sets of twin experiments with different coupled biogeochemical-physical dynamics and learning objectives. We perform simultaneous Bayesian estimation of state variables, parameters, and model equations, using noisy observations that are sparse in both space and time.
% Four sets different identical twin experiments using the NPZ, NPZD and NNPZD models are conducted, with different learning objectives. 
%
To quantify performance, we evaluate several learning metrics, emphasizing the sharpness of the inference and the accuracy of probability distributions.
For each of the four sets of experiments, we conduct multiple studies so as to evaluate the sensitivity to hyper-parameters. %associated with the setup.
However, for each set, we present detailed results for only one experiment and summarize the other results and sensitivity studies.
%of the sensitivity studies.}
%to hyper-parameters.}
%
% With the sparse observations of biological fields, we wish to learn the high-dimensional and multidisciplinary system, starting only with initial uncertainties on states, parameters and model formulations.
The main parameters of the physical-biogeochemical models, the hyper-parameters for the DO equations, and the observation and assimilation parameters
% and the GMM-DO filter 
are provided in Table \ref{table:DA Setup information}.

\begin{footnotesize}
\begin{longtable}{|p{0.4\textwidth}|p{0.12\textwidth}|p{0.12\textwidth}|p{0.12\textwidth}|p{0.12\textwidth}|}
	\caption{Deterministic or pdf values of the various domain-related, biological, physical, and hyper- parameters used in the four sets of experiments. The values
	$H=50~m$, $max\{T_{bio}(z)\} = 30~mmol~N~m^{-3}$, and time-scale of $1~day$ are the characteristic scales used for non-dimensionalization. 
 %\PFJL{Need to clearly say what is non-dim: I added one non-dim, but please check all and add non-dim where missing.} \AG{Checked and that one was only missing.}
 }
	\label{table:DA Setup information}
% 	\centering
% 	\begin{tabular}
\\
		\hline
		\textbf{Parameters}     &   \textbf{Exp. 1}     &   \textbf{Exp. 2}     &   \textbf{Exp. 3}     &   \textbf{Exp. 4}\\\hline
		Biogeochemical model    &   NPZ     &   NPZ \& NPZD    &  NPZ     &   NNPZD \\\hline
		\multicolumn{5}{|c|}{\textbf{Biological Parameters}} \\\hline
		Light attenuation due to sea water, $k_w$ ($m^{-1}$)     &   0.067   &   0.067   &   0.067   &   0.067 \\\hline
		Initial slope of the P-I curve, $\alpha$ ($(W~m^{-2}~day)^{-1}$)   &  0.025   &   0.025   &   0.025   &   0.025 \\\hline
		Surface photosynthetically available radiation, $I_o$ ($W~m^{-2}$)    & 158.075   &   158.075     &   158.075      &   158.075   \\\hline
		Phytoplankton maximum uptake rate, $V_m$ ($day^{-1}$)     &   1.5     &     1.5  &   1.5     &   1.5 \\\hline
		Half-saturation for phytoplankton uptake of nutrients, $K_u^*$ ($mmol~N ~m^{-3}$)  &   1   &   1   &   1   & 1 \\\hline
		$NH_4$ inhibition parameter,  $\Psi_I$ ($(mmol ~N ~m^{-3})^{-1}$)     &   --    &   --    &   --      &   1.46  \\\hline
		$NH_4$ oxidation coefficient,  $\Omega$ ($day^{-1}$)   &   --  &     --     &   --      &   0.25 \\\hline
		Phytoplankton specific mortality rate,  $\Xi$ ($day^{-1}$)    &   0.1     &  0.1    &   0.1     &   unif(0.01, 0.08) \\\hline
		Zooplankton specific excretion and mortality rate, $\Gamma$ ($day^{-1}$)  &  0.145   &   0.145  &   0.145   &   unif(0.125, 0.150) \\\hline
		Presence/absence of quadratic zooplankton term, $\alpha$   &         unif\{0, 1\}    &   unif\{0, 1\}    &     --     &   unif\{0, 1\}    \\\hline
		Quadratic zooplankton specific excretion and mortality rate, $\tilde{\Gamma}$ ($day^{-1}$)   &   0.2     &   0.2     &   0.2     &     0.2 \\\hline
		Zooplankton maximum grazing rate, $R_m$ ($day^{-1}$)  &   0.52    &   0.52    &   0.52    &   unif(0.52, 0.72)    \\\hline
		Ivlev constant, $\Lambda$ ($(mmol ~N ~m^{-3})^{-1}$)     &         unif(0.1, 0.2)      &       unif(0.1, 0.2)      &       0.13   &   unif(0.052, 0.072) \\\hline
		Fraction of zooplankton grazing egested, $\gamma$   &   0.3     &   0.3     &   0.2        &   0.3  \\\hline
		Detritus decomposition rate, $\Phi$ ($day^{-1}$)   &   1.03    &     1.03     &   1.03    &   1.03    \\\hline
		Diffusion constants -- horizontal \& vertical, ($\mathcal{K}_E$)     &   0   &	0	&   0   &   0 \\\hline
		\multicolumn{5}{|c|}{\textbf{Modeling Domain}}  \\\hline
		Height of the ridge, $H$ ($m$)  &   50	    &   50  &   50      &	50\\ \hline
		Characteristic width of the ridge, $D$ ($km$)  &   1	    &   1  &   1      &	1\\ \hline
		Distance between inlet and center of ridge, $X_c$ ($km$)  &   7.5	    &   7.5  &   7.5      &	7.5\\ \hline
		Domain height, $H_{in}$ ($m$)  &   100	    &   100  &   100      &	100\\ \hline
		Domain length, $L$ ($km$)  &   20	    &   20  &   20      &	20\\ \hline
		\multicolumn{5}{|c|}{\textbf{Physical Parameters}}  \\\hline
		Inverse of Eddy-viscosity Reynolds nb., ($\Lambda_{Re}$)    &   1	&   1   &   1    &	1/500   \\\hline
		\multicolumn{5}{|c|}{\textbf{DO Parameters}}  \\\hline
		Number of Modes, $N_s$  &   20	    &   40  &   20      &	15\\ \hline
		Number of Monte-Carlo samples, $N_{r}$  &   10,000      &	    10,000	    &   1,000      &       10,000\\ \hline
		\multicolumn{5}{|c|}{\textbf{GMM-DO Filtering -- Observation and Assimilation Parameters}}  \\\hline
		State variables being observed &      $Z$     &	$Z$	    &       $N$     &      $P$ \\\hline
		Observation error standard deviation, ($\sqrt{\mbs{R}}$) (non-dim.)  &   0.05	&   0.05    &   0.035     &	    0.04\\ \hline
		Size of Observation vector, $N_y$   &   6	&   6 &     8    &	9\\ \hline
		Observation start time (non-dim.)  & 5	&   5   &   1      &	2\\\hline
		Time interval between assimilations (non-dim.)   &   2	&   2  &   2    &	1\\\hline
		Observation end time (non-dim.)   &   25      &   25		&   25  &   25 \\\hline
% 	\end{tabular}
\end{longtable}
\end{footnotesize}

\subsection{Experiments 1: Discriminating among candidate functional forms}
\label{sec: Exp_NPZ_Parameter_Quad_Mort_Discriminate}

Biologically, mortality is a linear rate process. The mortality terms of phytoplankton and zooplankton however commonly act 
as ``closure'' parameterizations in models because
as they allow for recycling of nutrients directly from plankton. As a result, due to the missing intermediate states in the recycling model, the zooplankton mortality and recycling processes are often modeled nonlinearly, with a concentration-dependent loss rate \cite{franks2002npz}, which allows representing unresolved processes such as predation by larger predators, diseases, etc. 
In this first set of experiments, we use the NPZ model with uncertainty introduced by the ambiguity in the presence or absence of a quadratic zooplankton mortality function, along with the uncertainty in the value of the Ivlev grazing parameter ($\Lambda$). 
This ambiguity in the zooplankton mortality function corresponds to the $\mathcal{\widehat{L}}$ term introduced in Eq.~\ref{eq:model uncertainty type 1}.
Uncertainties in the initial biogeochemical conditions are set in balance with the uncertain parameters and model equations, as explained in Sect.\ \ref{sec:initialization}. 
The learning objective is to simultaneously learn all the biological states, regular parameter $\Lambda$, and functional form of zooplankton mortality using a stochastic formulatiom parameter, by assimilating sparse noisy observations.

The right-hand-side of the NPZ model (Eq.\ \ref{eq:NPZ model}) with the quadratic zooplankton mortality are the reaction or source terms in the ADR PDEs (\ref{eq:Generic ADR equations})). These NPZ source terms $S^{\phi^i}$ are given by,
\begin{equation}
\label{eq:Modified NPZ model}
\begin{split}
S^N &= -G\frac{PN}{N+K_u}+\Xi P + \Gamma Z + \underbrace{\alpha(\omega)(\tilde{\Gamma} Z^2) }_{\text{Quad.~Z~Mort.}} + R_m \gamma Z(1-\exp^{-\Lambda(\omega) P})  \\
S^P &= G\frac{PN}{N+K_u} - \Xi P  - R_m Z(1-\exp^{-\Lambda(\omega) P})  \\
S^Z &=   R_m (1-\gamma) Z(1-\exp^{-\Lambda(\omega) P}) - \Gamma Z - \underbrace{\alpha(\omega)(\tilde{\Gamma} Z^2) }_{\text{Quad.~Z~Mort.}} \;.
\end{split} 
\end{equation}
The stochastic parameters are explicitly shown using the realization index ($\omega$), and the ambiguous quadratic mortality term is pointed out. The stochastic formulation parameter, $\alpha(\omega)$, is restricted to binary values, i.e., 0 or 1, corresponding to the absence or presence of the quadratic mortality term, respectively.
$\Lambda(\omega)$ is sampled from a uniform probability distribution between the non-dimensional values of 3 and 6, and $\alpha(\omega)$ is assumed to have an initial 50\%-50\% probability of being 0 or 1.
% Thus, this helps us achieve model learning by parameter estimation, using state augmentation with the GMM-DO filter. 
%
%The above stochastic NPZ reactions (Eq.~\ref{eq:Modified NPZ model}) are  coupled with the RANS flow PDEs and used in the stochastic ADR PDEs that are solved with the DO methodology (Sects.\ \ref{sec: Stochastic Dynamically-Orthogonal PDEs in the bio-detailed paper}--\ref{sec: Numerical Schemes}).
%
The stochastic ADR PDEs with the above stochastic NPZ reactions (Eq.\ \ref{eq:Modified NPZ model}) are coupled with the RANS flow PDEs, and solved with the DO methodology (Sects.\ \ref{sec: Stochastic Dynamically-Orthogonal PDEs in the bio-detailed paper}--\ref{sec: Numerical Schemes}).
The other known physical-biogeochemical model parameters, the hyper-parameters for the DO equations, and the observation and assimilation parameters
% and the GMM-DO filter 
are given in Table \ref{table:DA Setup information}.

%\PFJL{Make table fonts smaller.} \AG{Done.}
%\PFJL{Can you make them even smaller so the table fits on a page?} \AG{Done.}

%\PFJL{I recommend we follow the exact same organization for the description of each experiment (order/style of what we say). You could use something similar to Pete's paper, I had tried to do this (Model formulation - dynamics, geometry, BCs, ICs, True solution generation, Observations and learning parameter, Numerical method).
% For example, you should define the truth here before the obs.}
%

\emph{True solution generation:}
The true solution corresponds to the non-dimensional values, 3.6 for $\Lambda$, and 1 for $\alpha$, i.e., the quadratic mortality term present. 
The true state fields are initialized and evolved as described in Sect.\ \ref{sec:True Solution Generation}, and noisy observations are extracted from this simulation.

\begin{figure}
	\centering
	\includegraphics[width=\textwidth]{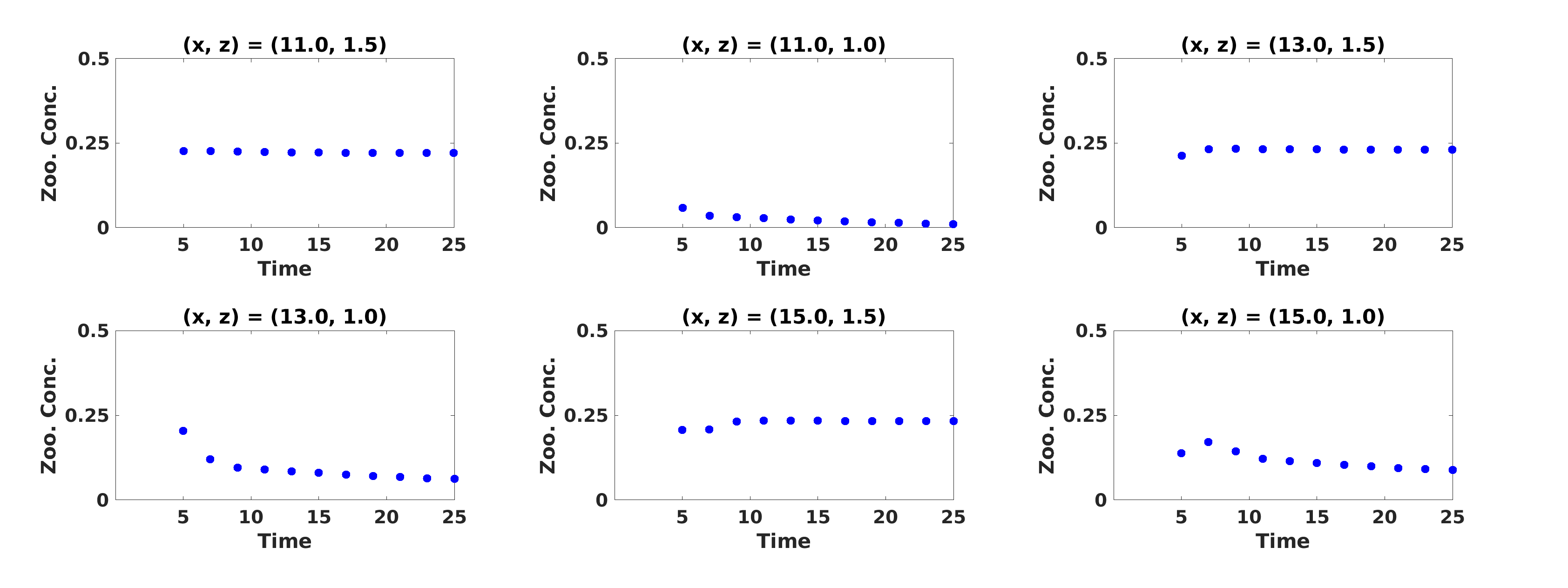}
	\vspace*{-10mm}
	\caption{
	Experiments-1. Time series of non-dimensional zooplankton concentration collected at six observation locations (their coordinates are given in the respective titles). For all experiments-1, the non-dimensional data error standard deviation is 0.05 (see Table \ref{table:DA Setup information}).}
	\label{fig: NPZ_Parameter_Quad_Mort data time-series}
\end{figure}

\emph{Observations and learning parameters:}
The simulated observations are sparse in both space and time. They consist of noisy zooplankton measurements at six locations downstream of the ridge, only at every two non-dimensional times, starting at $t=5$. 
The non-dimensional $Z$-data error standard deviation is 0.05.
The data shown in Fig.~\ref{fig: NPZ_Parameter_Quad_Mort data time-series} is all that the Bayesian learning framework gets to assimilate over the course of the experiment. 
Other hyper-parameters related to the GMM-DO filtering are provided in Table \ref{table:DA Setup information}.
%\AG{Numerical method:}
%The deterministic equations for the true solution and DO equations for the estimate pdfs are solved using the modular finite-volume framework described in Sect.\ \ref{sec: Numerical Schemes}. 

\emph{Learning metrics:} 
As time advances, the sparse noisy data are assimilated using the Bayesian GMM-DO filter in the augmented state space.
We compare the true fields and paramaters to their DO estimates (mean and most probable values). 
% To both quantitatively and qualitatively analyse the performance of our learning framework, we make use of the normalized RMSE and posterior pdf of the stochastic parameters as described in Sect. \ref{sec:Learning Metrics}.
To quantify performance, we examine the evolution of the normalized RMSEs (Sect.\ \ref{sec:Learning Metrics}) for the N, P, and Z fields, and for the $\Lambda(\omega)$ and $\alpha(\omega)$ parameters, as well as 
the pdfs of the stochastic parameters, DO coefficients, and  biological states. 
%, and examine their evolution over time. 
%
%The RMSEs are normalized using the corresponding RMSE value just before the first assimilation step, as mentioned in Sect.\ \ref{sec:Learning Metrics}. 
%Along with this, we also analyze the evolution of the pdfs of the stochastic parameters, DO coefficients, and  biological states. 
%\AG{The headings in this para need to be removed.}
%The square of RMSE consists of two contributions, one is the square of the $L_2$ distance between the mean of the state variables in the stochastic run and the simulated truth, while other is the variance of the state variables (similarly for parameters also). 
%In every experiment, the RMSE value for each of the variables at different times, is normalized by the corresponding RMSE value just before the first assimilation step.

\subsubsection{Learning results}

\begin{figure}
	\centering
	\includegraphics[width=\textwidth]{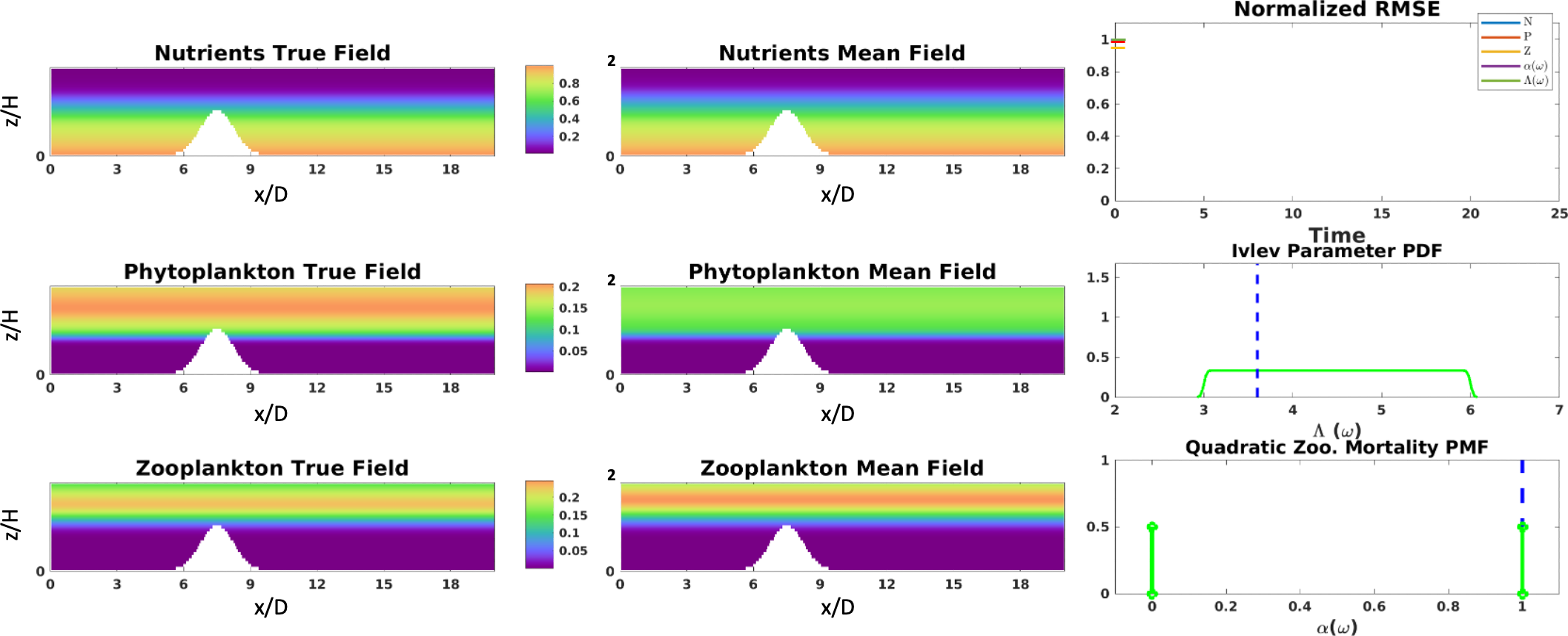}
	\vspace*{-6mm}
	\caption{
	Experiments-1: State of the true and estimate NPZ fields and parameters at $t = 0$ (i.e.\ initial conditions). The first two columns consist of the non-dimensionalized true (left) and mean estimate (right) tracer fields of N, P, and Z. 
    In the third column, the top panel shows the variation of normalized root-mean-square-error (RMSE) with time for the stochastic state variables and parameters. We explicitly mark the RMSE values at $t=0$ using short horizontal lines for better visibility. All the RMSE values are concentrated around $1$, and for some, the lines overlap.
    %\PFJL{We should make sure the RMSEs at t=0 are visible. Perhaps we should reduce the time duration (I am not sure)?} 
    %\AG{We can have a short line segment protruding to the left of the y-axis for each of the variables corresponding to the RMSE values and the color?} 
    The next two panels contain the pdfs of the non-dimensional $\Lambda(\omega)$ and $a(\omega)$ (to learn the presence or absence of quadratic zooplankton mortality), each
    marked with solid green lines, with the true unknown parameter values marked with blue dotted lines. The velocity field is deterministic with $Re = 1$.}
	\label{fig: NPZ_Parameter_Quad_Mort T=0 Prior}
\end{figure}

\begin{figure}
	\centering
	\includegraphics[width=\textwidth]{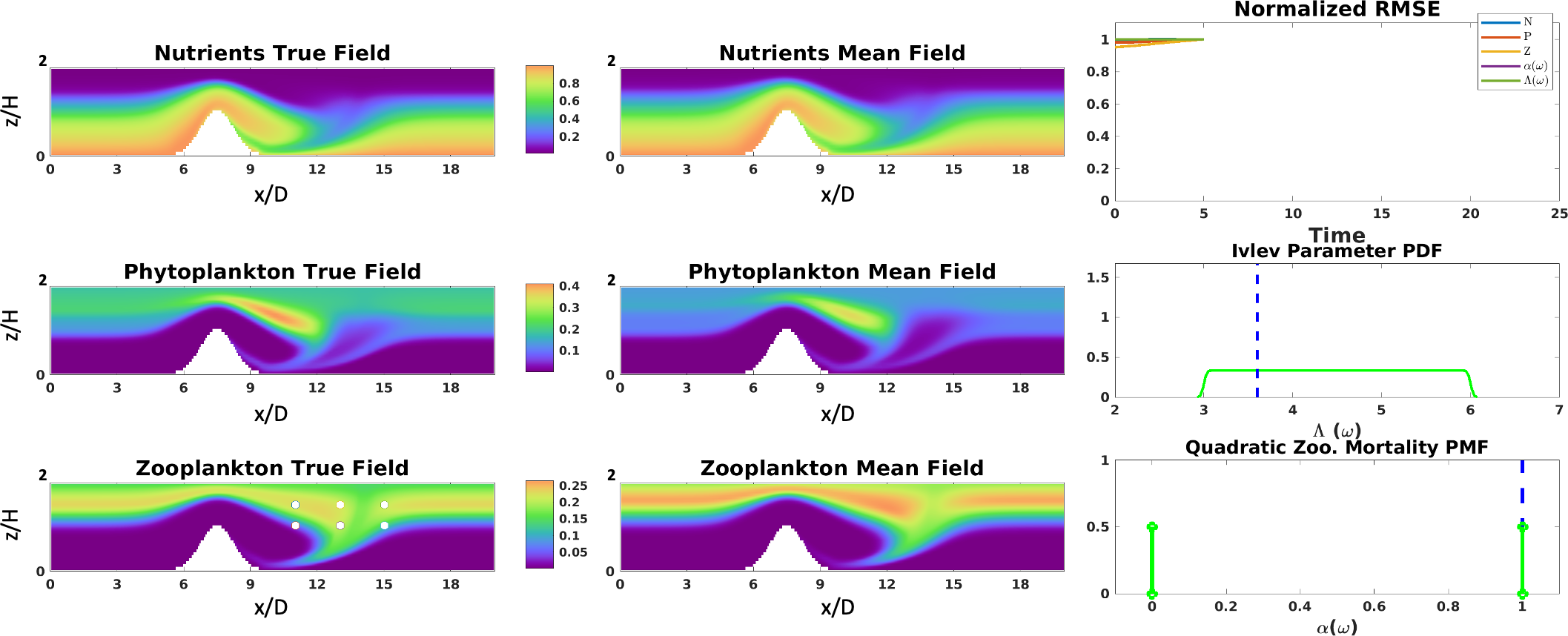}
	\vspace*{-6mm}
	\caption{
	Experiments-1: As Fig.\ \ref{fig: NPZ_Parameter_Quad_Mort T=0 Prior}, but for the prior fields and parameters at $t = 5$ (i.e.\ just before the 1st assimilation). 
    The white circles on the zooplankton true field mark the six observation locations.
    In the first two columns, the axis limits for the state variables have changed so as to follow the bloom, but in the third column, they remain as in Fig.\ \ref{fig: NPZ_Parameter_Quad_Mort T=0 Prior} so as to directly highlight the uncertainty evolution.}
	\label{fig: NPZ_Parameter_Quad_Mort T=5 Prior}
\end{figure}

Figure \ref{fig: NPZ_Parameter_Quad_Mort T=0 Prior} shows the initial state and parameters of the system (at $t=0$), while Fig.\ \ref{fig: NPZ_Parameter_Quad_Mort T=5 Prior} shows the evolved prior state and parameters of the system at $t=5$ (i.e.\ just before the 1st observational episode).
There are significant differences between the true and prior DO mean fields of the biogeochemical tracers. 
% The blue dotted line in the probability distribution plots of the parameters marks their true non-dimensional values, 3.6 for $\Lambda(\omega)$, and 1 for $a(\omega)$. 
% The probability of $\Lambda(\omega)$ is uniform throughout its range, and for $a(\omega)$ there is 50\%-50\% probability of it being either 0 or 1.
During these first five non-dimensional time units, a phytoplankton bloom develops just downstream (top-right) of the ridge: upwelling of nutrients above the ridge within the euphotic zone feeds the growth in phytoplankton biomass in the wake. 

\begin{figure}
	\centering
	\subfloat[Phytoplankton mean, standard deviation and top three DO modes, at $t=0$.]{\label{subfig:P modes t=0}\includegraphics[width=.425\textwidth]{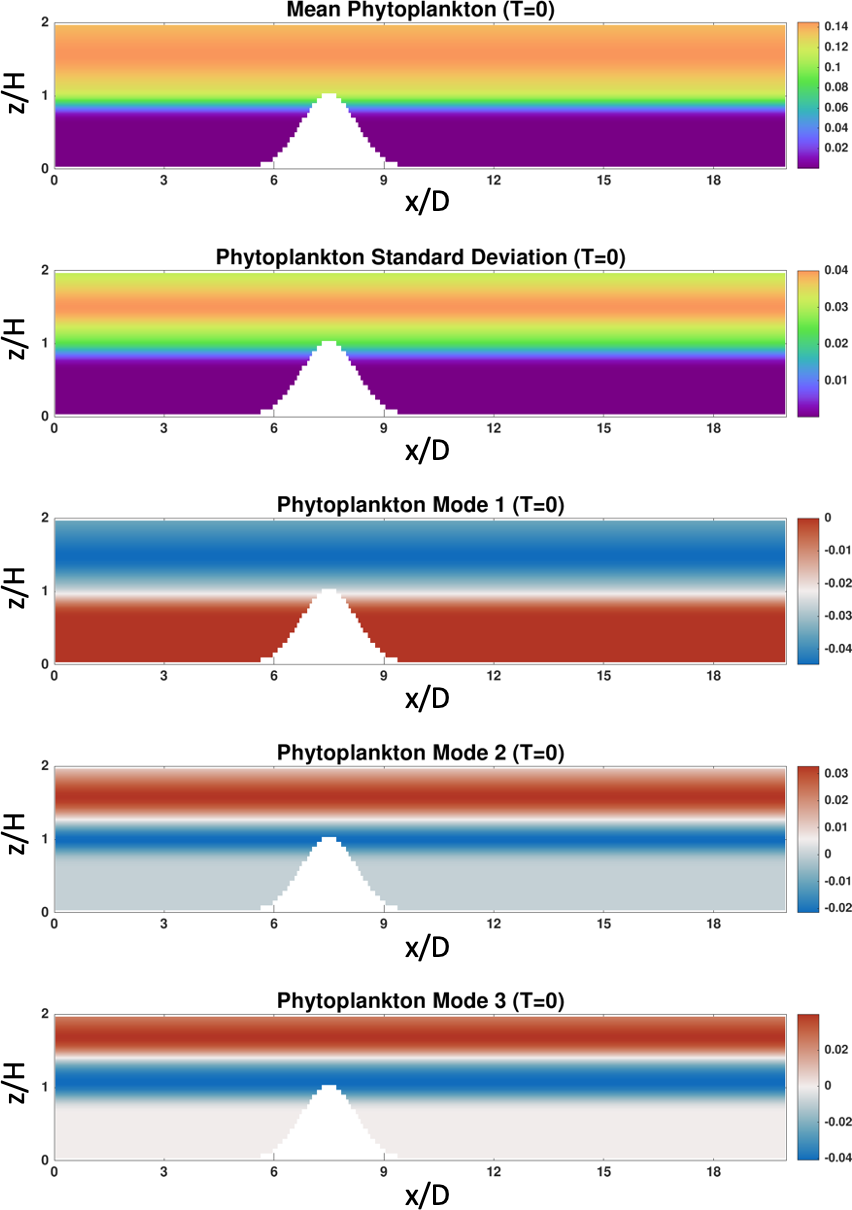}}\quad
	\subfloat[Phytoplankton mean, standard deviation and top three DO modes, at $t=5$ (prior).]{\label{subfig:P modes t=5}\includegraphics[width=.425\textwidth]{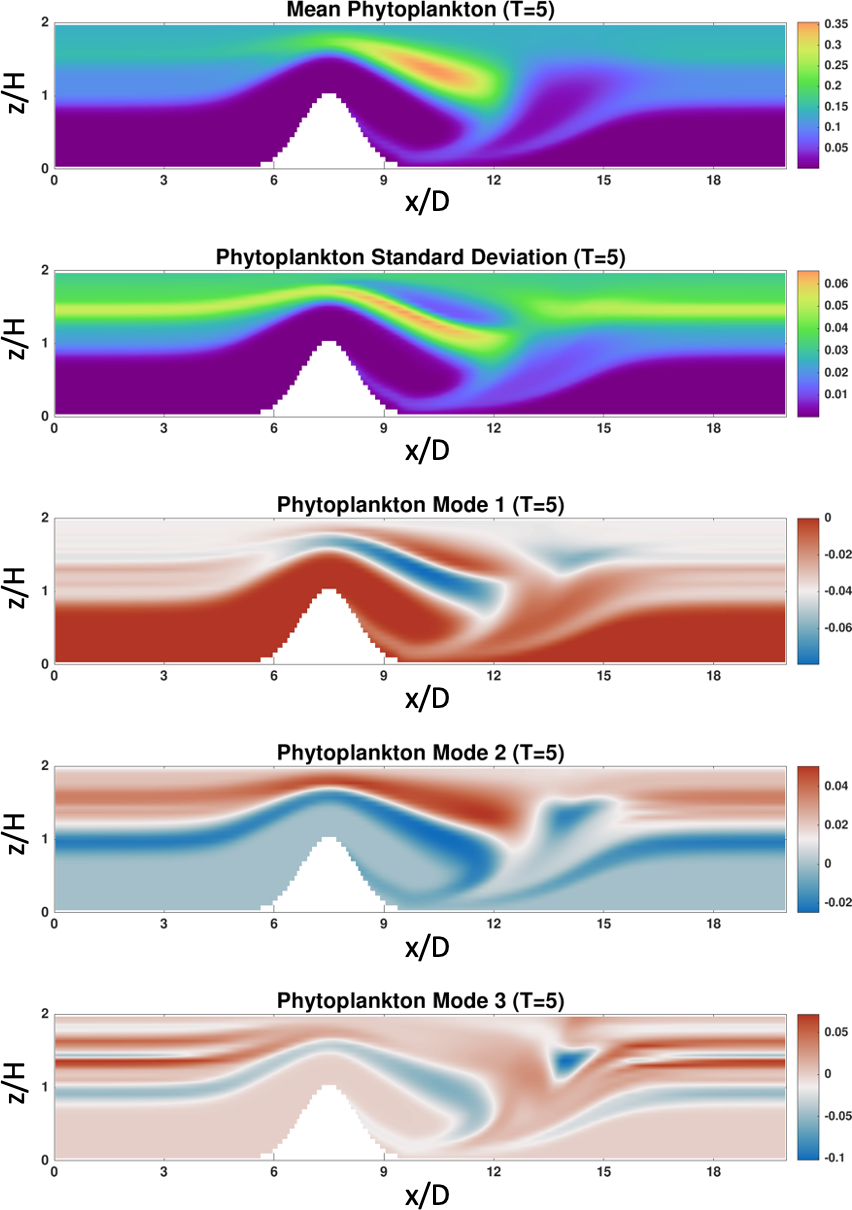}}\\
	\subfloat[Joint distributions and respective marginals of the top four stochastic DO coefficients, at $t=0$.]{\label{subfig:NPZ Yi t=0}\includegraphics[width=.425\textwidth]{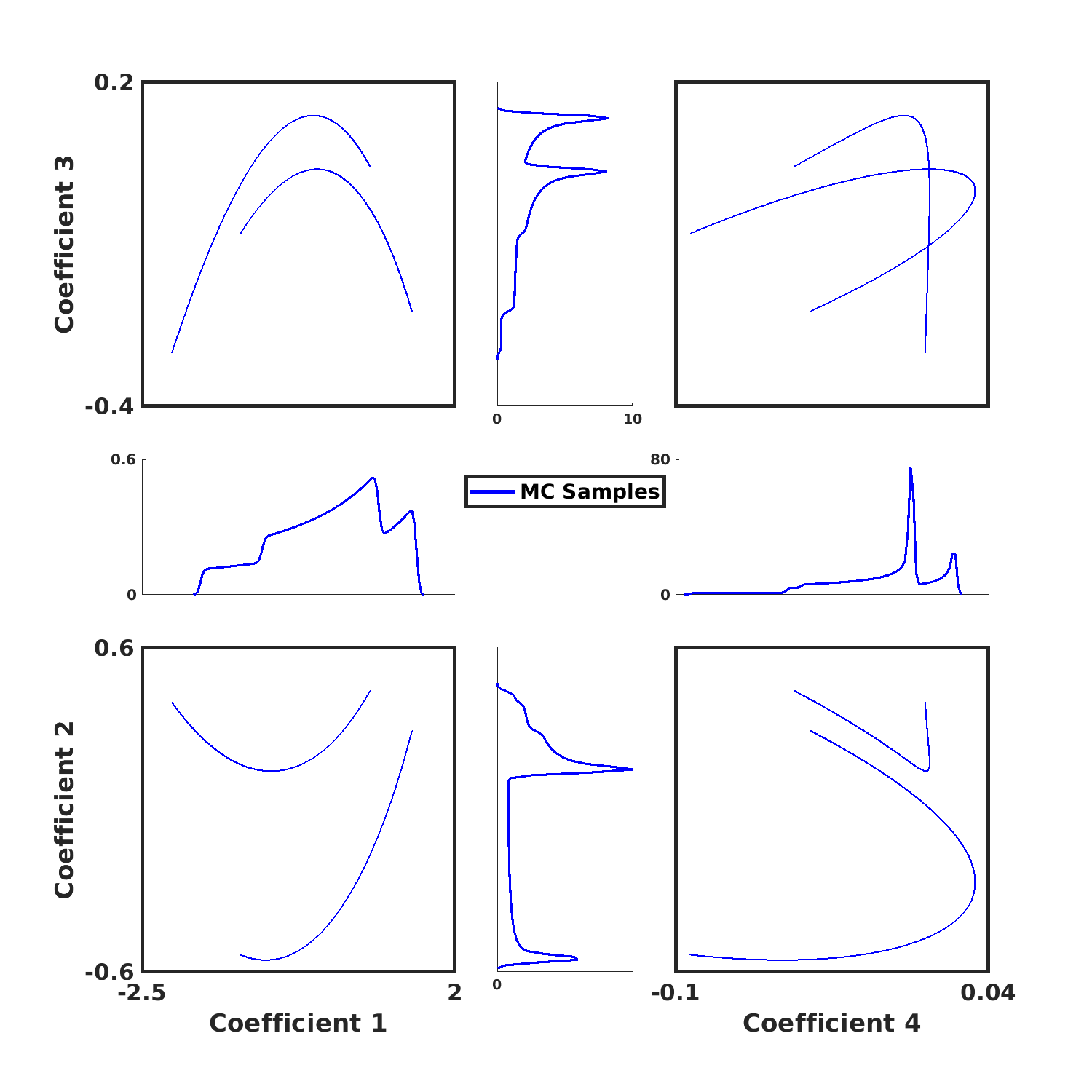}} \quad
	\subfloat[Joint distributions and respective marginals of the top four stochastic DO coefficients, along with the GMM fit, at $t=5$ (prior). 
    In the joint distribution plots, one standard deviation contours of each member of the GMM are marked with solid-line ovals (at the 1-sigma level) colored according to their respective weights (colorbar to the right).]{\label{subfig:NPZ Yi t=5}\includegraphics[width=.425\textwidth]{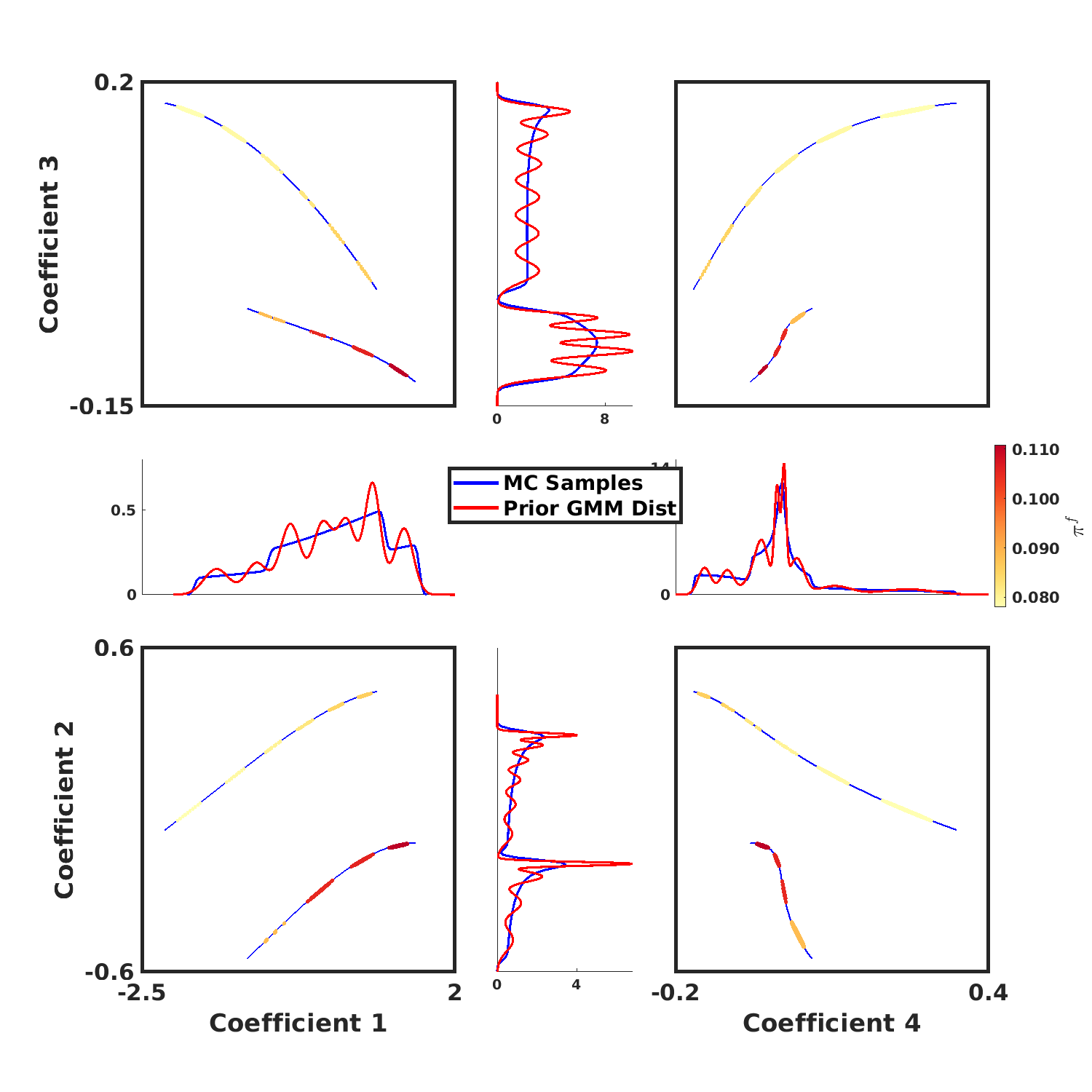}}
	\caption{
	Experiments-1: Statistics for the initial ($t=0$) and prior ($t=5$, just before the 1st assimilation) states of the stochastic NPZ ADR dynamical system.}
	\label{fig: NPZ_Parameter_Quad_Mort modes and Yi T=0 & 5}
\end{figure}

In Fig.\ \ref{fig: NPZ_Parameter_Quad_Mort modes and Yi T=0 & 5}, we illustrate the evolving statistics of the stochastic dynamical system from $t=0$ to $t=5$ just before data assimilation. 
We show fields of the phytoplankton standard deviation and dominant three DO modes (Panels \ref{subfig:P modes t=0}~\&~\ref{subfig:P modes t=5}). 
The standard deviation fields clearly highlight the significant uncertainty around the phytoplankton subsurface maxima and bloom, reaching 30 percent of the mean field maxima. The uncertain subsurface maxima and bloom also clearly affect the DO modes. 
In Panels \ref{subfig:NPZ Yi t=0}~\&~\ref{subfig:NPZ Yi t=5}, we show the joint distribution of the top four stochastic coefficients, $Y_i(t; \omega)$ with $i=1,2,3,4$ in (Eq.\ \ref{eq:DO_decomp}) at $t=0$ and $t=5$, respectively. 
In Panel \ref{subfig:NPZ Yi t=5}, we also show the corresponding prior GMM fits at $t=5$, using $N_{\text{GMM}}=10$ components (Eq.\ \ref{eq:DO-coef-GMM-prior} in \ref{app:GMM-DO}).
We use the BIC to find this optimal number of components required \cite{sondergaard_lermusiaux_MWR2013_part1}.
The marginalized distributions shown are those of the joint-double and single coefficient spaces for $i=1,2,3,4$. They demonstrate the highly non-Gaussian nature of the stochastic DO coefficients, which the DO equations evolve, and the GMM-DO filter account for. 
The strong parametric uncertainties are reflected by the thin joint-double coefficient distributions.
In addition, the realizations of the stochastic coefficients are clearly divided into two groups, corresponding to the presence or absence of the quadratic mortality term. 
This is already the case at $t=0$ (Panel \ref{subfig:NPZ Yi t=0}) because our uncertainty initialization scheme (Sect.\ \ref{sec:initialization}) respects the stochastic dynamical balances.

\begin{figure}
	\centering
	\includegraphics[width=\textwidth]{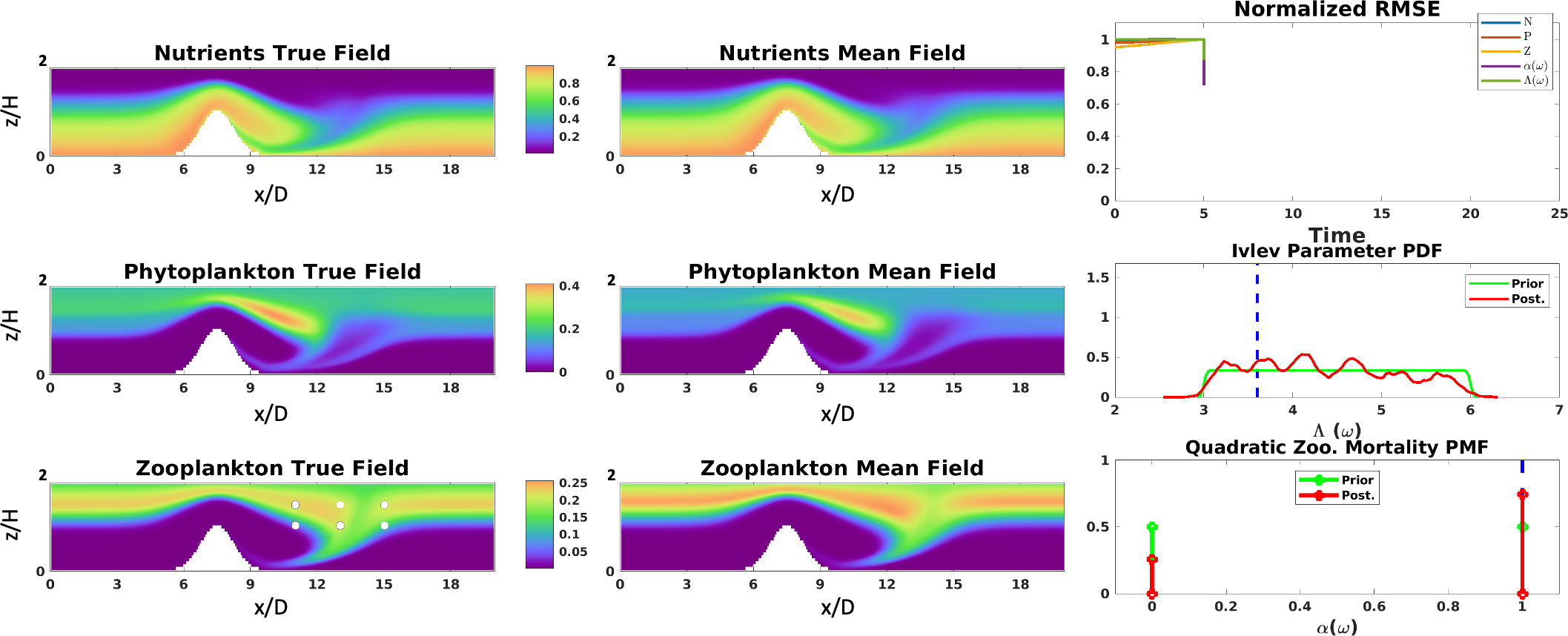}
	\vspace*{-6mm}
	\caption{
	Experiments-1: As Figs.\ \ref{fig: NPZ_Parameter_Quad_Mort T=0 Prior}~\&~\ref{fig: NPZ_Parameter_Quad_Mort T=5 Prior}, but for posterior fields and parameters at $t=5$ (i.e.\ just after the 1st assimilation). 
    In the last two panels of the third column, the prior pdfs associated with the non-dimensional $\Lambda(\omega)$ and $a(\omega)$ at $t=5$ are marked with solid green lines, while the posterior pdfs are marked with solid red lines.}
	\label{fig: NPZ_Parameter_Quad_Mort T=5 Post}
\end{figure}

At $t=5$, the first sparse noisy data are assimilated. Fig.\ \ref{fig: NPZ_Parameter_Quad_Mort T=5 Post} shows the posterior mean fields, prior and posterior parametric distributions, and the normalized RMSE values for the mean fields and two stochastic parameters. 
By only observing noisy zooplankton at six locations, the GMM-DO filter simultaneously updates all the biological fields and parameters. This is evident from the mean fields getting aligned with the true fields and quantified by the RMSE reductions of about 20 to 30 percent. Also visible is the slight change in the pdf for $\Lambda(\omega)$ and a higher probability value for $\alpha(\omega)$ being one. The six data are so far much more informative about the mortality term than the Ivlev parameter.

\begin{figure}
	\centering
	\includegraphics[width=\textwidth]{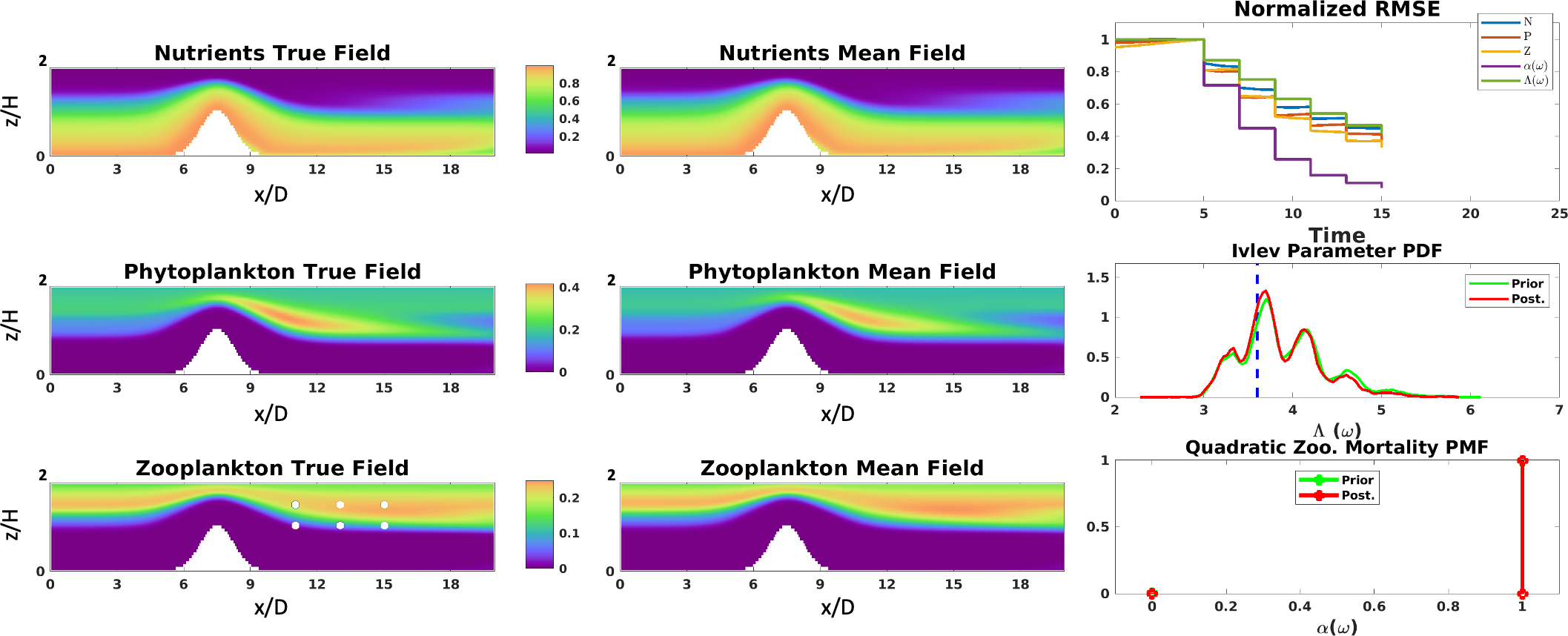}
	\vspace*{-6mm}
	\caption{
	Experiments-1: As Figs.\ \ref{fig: NPZ_Parameter_Quad_Mort T=0 Prior}~,~\ref{fig: NPZ_Parameter_Quad_Mort T=5 Prior}~\&~ \ref{fig: NPZ_Parameter_Quad_Mort T=5 Post} but for posterior fields and parameters at $t=15$ (i.e.\ just after the 6th assimilation). 
    In the first two columns, the axis limits for the state variables have changed so as to follow the bloom, but in the third column, they remain as in Fig.\ \ref{fig: NPZ_Parameter_Quad_Mort T=0 Prior} so as to directly highlight the uncertainty evolution.
    %In the last two panels of the third column, the prior pdfs associated with the non-dimensional $\Lambda(\omega)$ and $a(\omega)$ at $t=15$ are marked with solid green lines, while the posterior pdfs are marked with solid red lines.
    }
	\label{fig: NPZ_Parameter_Quad_Mort T=15 Post}
\end{figure}

Next, in Fig.\	\ref{fig: NPZ_Parameter_Quad_Mort T=15 Post}, we illustrate the same posterior mean fields, prior and posterior parameters, and normalized RMSE values, but at $t=15$, i.e., at the sixth data assimilation. 
The flow is fully developed with the biogeochemical fields well learned, as quantified by the normalized RMSEs. 
The GMM-DO filter unambiguously detects the presence of quadratic mortality of $Z$, as confirmed by the RMSE
of $\alpha(\omega)$ at 10 percent.
The other RMSEs are higher at about 40-50 percent, with that of $Z$ a bit lower (reflecting that $Z$ is the only variable measured at six locations).
The pdf of $\Lambda(\omega)$ is also accumulated around its true value, but is now clearly multi-modal, indicating nonlinearities and remaining ambiguity. In other words, with the sparse noisy data assimilated so far, several values of the Ivlev constant $\Lambda(\omega)$ remain very likely, indicating possible biases and equifinality \cite{duda2006pattern,lu_lermusiaux_PhysD2021}.

\begin{figure}
	\centering
	\includegraphics[width=\textwidth]{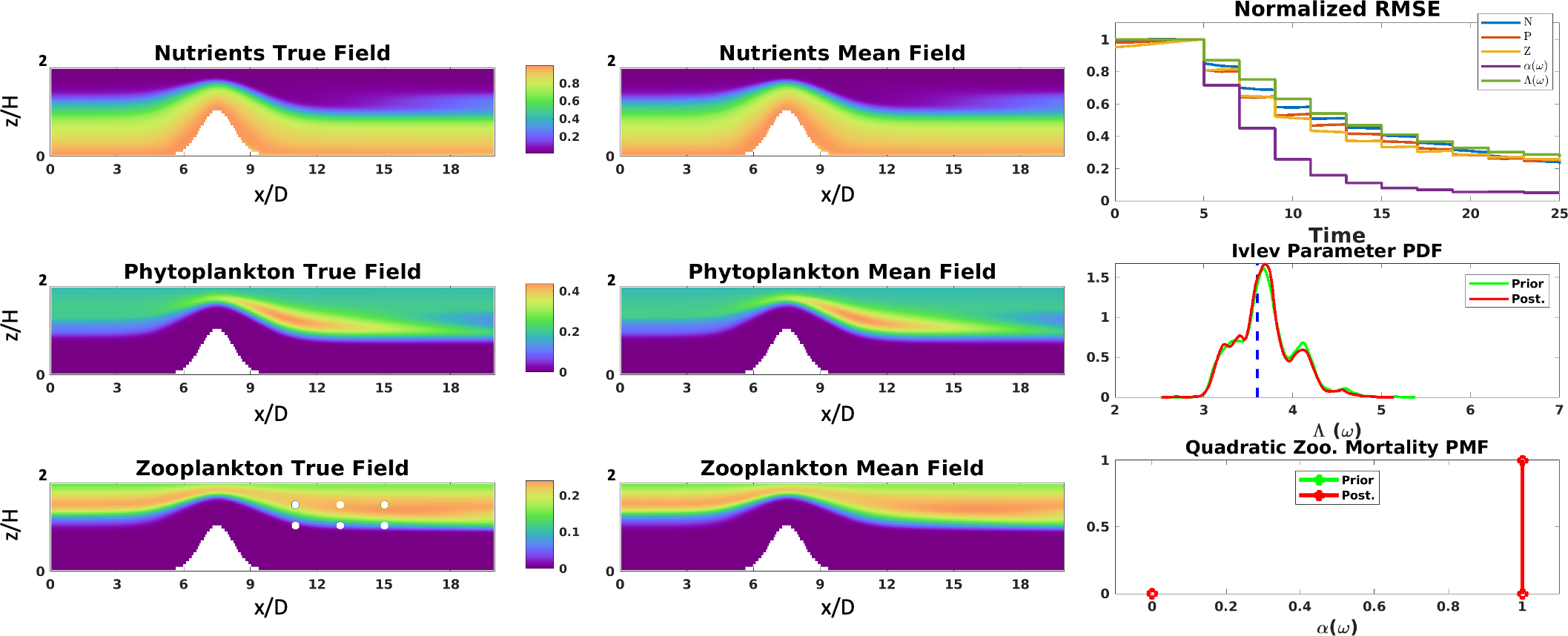}
	\vspace*{-6mm}
	\caption{
	Experiments-1: As Figs.\ \ref{fig: NPZ_Parameter_Quad_Mort T=0 Prior}~,~\ref{fig: NPZ_Parameter_Quad_Mort T=5 Prior}~,~ \ref{fig: NPZ_Parameter_Quad_Mort T=5 Post}~\&~ \ref{fig: NPZ_Parameter_Quad_Mort T=15 Post} but for posterior fields and parameters 
	at $t=25$ (i.e.\ just after the 11th assimilation). 
    In the first two columns, the axis limits for the state variables have changed so as to follow the bloom, but in the third column, they remain as in Fig.\ \ref{fig: NPZ_Parameter_Quad_Mort T=0 Prior} so as to directly highlight the uncertainty evolution.
    %In the last two panels of the third column, the prior pdfs associated with the non-dimensional $\Lambda(\omega)$ and $a(\omega)$ at $t=25$ are marked with solid green lines, while the posterior pdfs are marked with solid red lines.
    }
	\label{fig: NPZ_Parameter_Quad_Mort T=25 Post}
\end{figure}

Finally, at $t=25$, after 11 assimilation events, the same quantities are shown in Fig.\ \ref{fig: NPZ_Parameter_Quad_Mort T=25 Post}. All the biogeochemical mean and true fields match with each other with RMSEs around 20 percent or less.
The probability of the presence of the quadratic mortality term is now almost one, while the $\Lambda(\omega)$ pdf has a clear peak near 3.6 with a couple other much lower biased peaks around it.
In general, the presence of lower peaks in pdfs of parameters indicate alternative combinations that could explain the data, and also the ability of the GMM-DO filter to capture non-Gaussian pdfs. 
The learning is also evident from the sustained decrease in the normalized RMSEs at every assimilation step for all the biogeochemical fields and parameters. 

%\PFJL{In addition to RMSE, we weed to describe pdf evolution and how do we know the pdf we get are correct (you have some of this above, so perhaps here it is a summary or we need to organize a bit more, I need to check the above text first), Need to talk about bias. 

%Then, we need to summarize results of other experiments. Check what I did in Pete's paper}

\emph{Sensitivity Studies.}
Many similar experiments were completed, changing various hyperparameters related to the GMM-DO filter, such as the biological variable being observed, observation locations, frequency, start-time, etc. 
Noisy observations from simulated truths with different combinations of $\Lambda (\omega)$ and $\alpha(\omega)$ were also used. We found that the biological variable being observed has an impact on the sharpness of the inference or learnability of the given learning objectives. 
For example, observing $N$ led to the learning of two distinct combinations of $\Lambda(\omega)$ \& $\alpha(\omega)$, 3.1 \& 0, and 3.6 \& 1, respectively with nearly equal amount of confidence \cite{gupta_MSThesis2016}. 
We also varied the location of observations (e.g., surface versus subsurface) and confirmed that more informative locations \cite{lermusiaux_et_al_TheSea2017} improve the learning rate.
Decreasing the amount of observation data, or increasing the value of the observation error standard deviation, led to a slower learning rate and larger uncertainty in the learned states and parameters. 
We also confirmed the convergence of our GMM-DO Bayesian posteriors by repeating learning experiments with an increasing number of DO modes and coefficients (not shown), until the results converged to those shown.
This convergence of the pdfs of the parameters and DO coefficients, and of the DO modes and mean, suggests that our Bayesian GMM-DO filter provides accurate pdf estimates: it thus shows what has been learned without ambiguity or with some ambiguity remaining. For the latter case, the multi-model posterior pdfs show that additional observations are needed to sharpen the inference further.

%%%%%%%%%%%%%%%%%%%%%%%%%%%%%%%%%%%%%%%%%%%%%%%%%%%%%%%%%%%%%%%%%%%%%%%%%%%%%%%%%%%%%%%%%%%%%%%%%%%%%%%%%%%%%%%%%%%%%%%%%%%%%%%%%%%%%%%%%%%%%%

\subsection{Experiments 2: Discriminating among models of different complexities}

In the second set of experiments, the primary goal is to learn the complexity of the biogeochemical model, e.g., its state variables, along with the biogeochemical fields and Ivlev grazing parameter. Two candidate hierarchical model classes, NPZ and NPZD, are considered possible. 
These NPZ and NPZD models correspond to the candidates $\mathcal{M}_i$'s introduced in Eq.~\ref{eq:diff complexity models}.
To represent them with a single modeling system, we embed the former into the latter using the stochastic complexity parameter, $\beta(\omega)$. We multiply the detritus state variable ($D$) and other appropriate terms with $\beta(\omega)$, such that, the value of 1 derives the NPZD model, while the value of 0 derives the NPZ model (see Eq.\ \ref{eq: example beta}). Thus, the RHS of the general stochastic model which encompasses both NPZ and NPZD models is given by,
\begin{equation}
\label{eq:Modified NPZD model}
\begin{split}
S^N &= -G\frac{PN}{N+K_u} + \Phi D' + \Gamma Z + (1-\beta(\omega))\Xi P   \\
S^P &= G\frac{PN}{N+K_u} - \Xi P  - R_m Z(1-\exp^{-\Lambda(\omega) P})  \\
S^Z &=   R_m (1-\beta(\omega)\gamma) Z(1-\exp^{-\Lambda(\omega) P}) - \Gamma Z  \\
S^{D'} &=  \beta(\omega)R_m \gamma Z(1-\exp^{-\Lambda(\omega) P}) + \beta(\omega)\Xi P - \Phi D' \\
D' &= \beta(\omega) D \;,
\end{split} 
\end{equation}
where $D'$ is the modified detritus state.
In Experiments-2, the formulation uncertainty is thus within the class of compatible models $\mathcal{\widehat{L}}$ introduced in Eq.~\ref{eq:model uncertainty type 1} and of the compatible-embedded model type defined by Eq.~\ref{eq: compatible-embedded} in 
Sect.\ \ref{Sec: Special Stochastic Parameters}.
Once again, $\Lambda(\omega)$ is sampled from a uniform probability distribution between the non-dimensional values of 3 and 6, and $\beta(\omega)$ is assumed to have 50\%-50\% probability of being 0 or 1.
% Thus, this helps us achieve model learning by parameter estimation, using state augmentation with the GMM-DO filter. 
%
The stochastic ADR PDEs with the stochastic NPZD$'$ reactions (Eq.~\ref{eq:Modified NPZD model}) are coupled with the RANS flow PDEs, and solved with the DO methodology (Sects.\ \ref{sec: Stochastic Dynamically-Orthogonal PDEs in the bio-detailed paper}--\ref{sec: Numerical Schemes}).
The other known physical-biogeochemical  parameters as well as the hyper-parameters for the DO equations 
% and the GMM-DO filter 
are given in Table \ref{table:DA Setup information}.

% \PFJL{Need to select and follow consistent template of describing things, and describe each key thing.}
\emph{True solution generation:}
The true solution corresponds to the NPZ model with a non-dimensional value of 3.6 for the $\Lambda$ parameter. The state fields are initialized and evolved as described in Sect.\ \ref{sec:True Solution Generation}.

\emph{Observations and learning parameters:}
The simulated observations are sparse in both space and time, and again consist of noisy zooplankton measurements at six locations downstream of the ridge, only at every two non-dimensional times, starting at $t=5$. 
The non-dimensional $Z$-data error standard deviation is 0.05.
Other hyper-parameters related to the GMM-DO filtering are provided in Table \ref{table:DA Setup information}.
%\AG{Numerical method:}
%Similar to the last set of experiments, the DO equations and the deterministic governing equations for the true solution are solved using the modular finite-volume framework described in Sect. \ref{sec: Numerical Schemes}. 

\emph{Learning metrics:} 
As time advances and sparse noisy data are assimilated, we compare the true fields and parameters to their DO estimates.
To quantify performance, we examine the evolution of the normalized RMSEs of state fields and parameters, pdfs of stochastic parameters, and variances of DO coefficients.
%
% \AG{The headings in this para need to be removed.}
% We start with equally likely probability of detritus field being present or absent, and an uniform $\Lambda(\omega)$ distribution. Table \ref{table:DA Setup information} provides the values of other relevant model and hyper- parameters. In this experiment as well, the zooplankton field is being observed at six locations every two non-dimensional time-steps, starting at $t=5$.

\subsubsection{Learning results}

\begin{figure}[h]
	\centering
	\includegraphics[width=\textwidth]{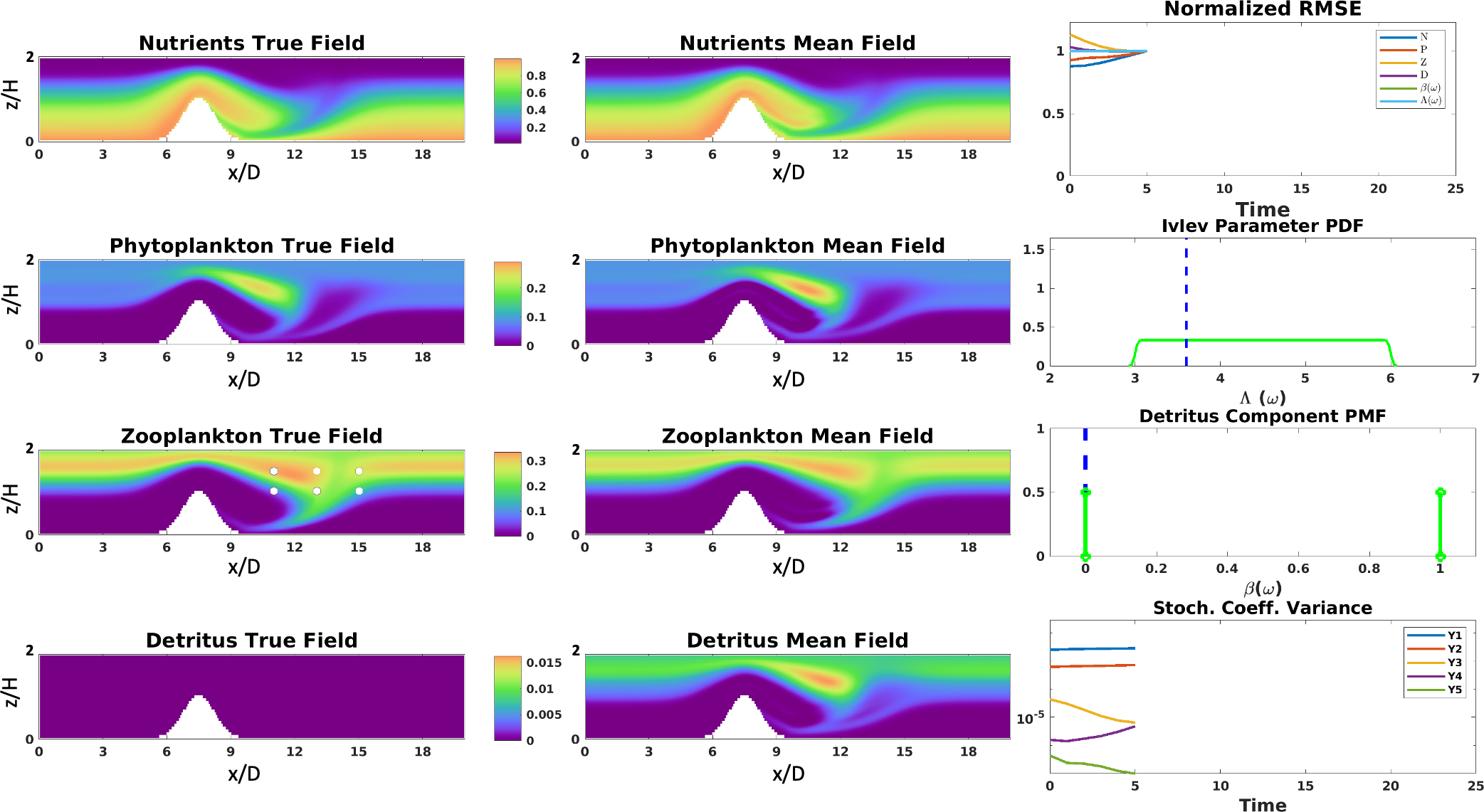}
	\vspace*{-4mm}
	\caption{
	Experiments-2: State of the true and prior estimate NPZD fields and parameters at $t = 5$ (i.e.\ just before the 1st assimilation). The first two columns consist of the non-dimensionalized true (left) and mean estimate (right) tracer fields of $N$, $P$, $Z$, and $D$. In the third column, the first panel shows the variation of normalized RMSE with time for all the stochastic state variables and parameters. The next two panels contain the pdfs of the non-dimensional $\Lambda(\omega)$ and $\beta(\omega)$ (to learn the complexity, NPZ vs.\ NPZD), each marked with solid green lines, with the true unknown parameter values marked with blue dotted lines. The last panel shows the evolution with time of the variance (log scale) of the top five modes. The velocity field is deterministic with $Re=1$. Additionally, the white circles on the zooplankton true field mark the six observation locations.}
	\label{fig: NPZD_Complexity T=5 Prior}
\end{figure}

Figure \ref{fig: NPZD_Complexity T=5 Prior} shows the state and parameters of the system at $t=5$, just before the first observational episode. 
The most distinctive difference is between the true and  mean detritus fields. Since the true model is NPZ, the true detritus field is equal to zero, while the mean detritus field is non-zero because half of the realizations correspond to the NPZD model.
The RMSEs of all the variables exactly equal 1, because their respective values just before the first assimilation were used for normalization.
The pdf of $\Lambda(\omega)$ is uniform in the main range, and $\beta(\omega)$ has 0.5 probability of being 0 or 1. 
The variances of the top five modes show a rapid decay with mode number, with the top two variances orders of magnitude larger. The variances of modes 3 and 4 differ initially but become similar over time, indicating a potential cross-over at $t=5$. 
% \PFJL{Explain why since true model is NPZD. Add some more descriptions here on RMSE (why all equal to 1), pdf, variance, etc.} \AG{Done.}

\begin{figure}[h]
	\centering
	\includegraphics[width=\textwidth]{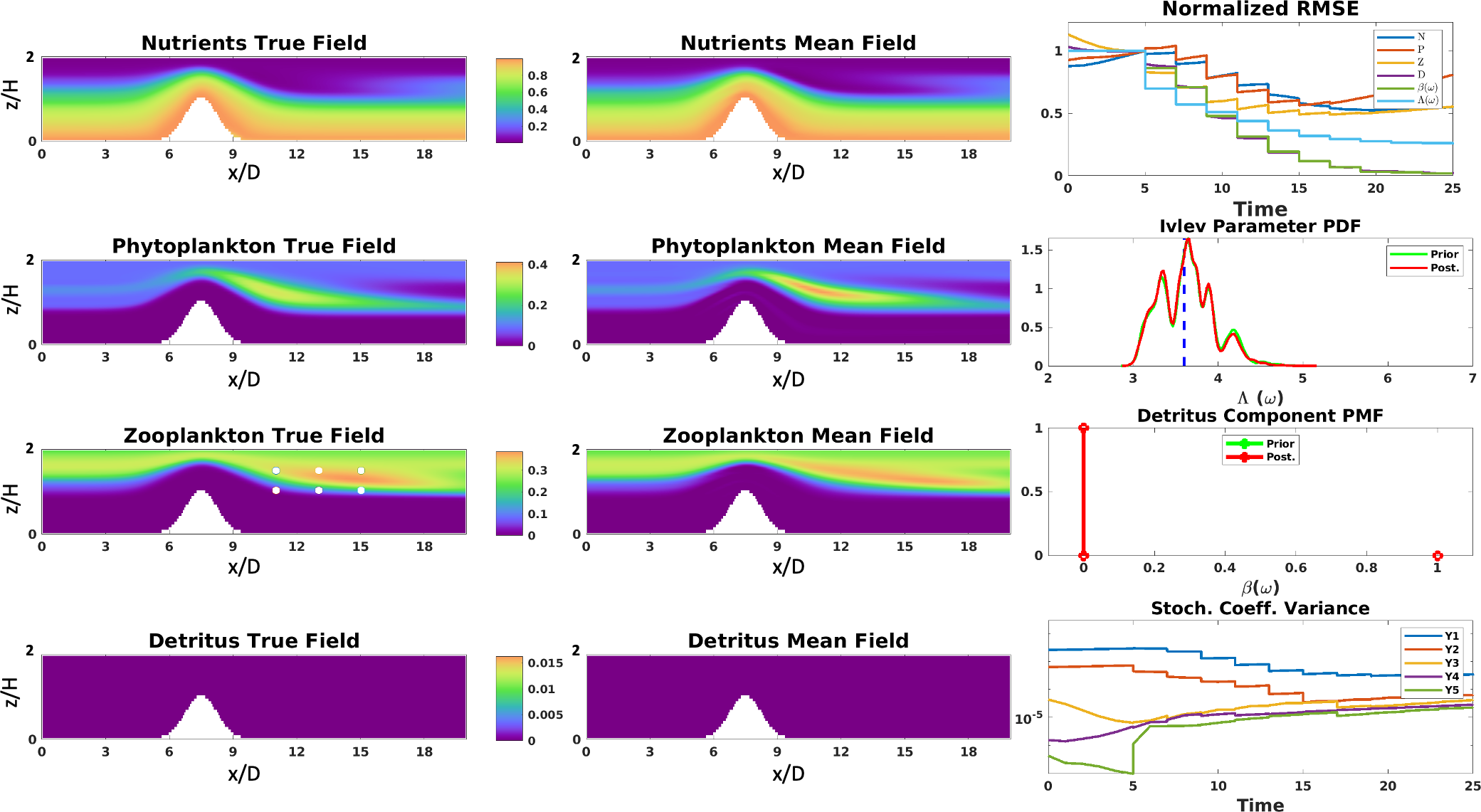}
	\vspace*{-4mm}
	\caption{
	Experiments-2: As Fig.\ \ref{fig: NPZD_Complexity T=5 Prior} but for posterior fields and parameters 
	at $t=25$ (i.e.\ just after the 11th assimilation). 
    In the middle two panels of the third column, the prior pdfs associated with the non-dimensional $\Lambda(\omega)$ and $\beta(\omega)$ at $t=25$ are marked with solid green lines, while the posterior pdfs are marked with solid red lines.
    In the first two columns, the axis limits for the state variables have changed so as to follow the bloom, but in the third column, they remain as in Fig.\ \ref{fig: NPZD_Complexity T=5 Prior} so as to directly highlight the uncertainty evolution.}
    \label{fig: NPZD_Complexity T=25 Post}
\end{figure}

In Fig.\ \ref{fig: NPZD_Complexity T=25 Post}, we directly show the state of the system at time $t=25$, after eleven GMM-DO data assimilation (six zooplankton values every two non-dimensional times). 
We find that our Bayesian learning framework is able to learn the true model to be NPZ, along with the posterior pdf of $\Lambda(\omega)$ concentrated around the true value of 3.6. 
The mean fields also match the true fields, especially the detritus mean field becoming very close to 0 at all the spatial locations. 
The
RMSEs for all the variables decrease over time, up to about $t=15$. At that time, the RMSE for the phytoplankton field increases due to a mismatch in the strength of the bloom, thus showing that the zooplankton data are not sufficiently informative for the same.
The pdf of $\Lambda(\omega)$ features multiple peaks and thus still indicates that competing hypotheses remain for different pairs of parameter values; this was already the case in the intermediate assimilation steps (not shown).
The evolution of the variances of the top five modes shows that these variances can increase and cross-over, for example, lower modes become more important as learning progresses. 
As the bloom develops, more complex nonlinear dynamics are activated, leading to the growth of some uncertainty modes. Results show that our Bayesian filter captures this as well as biases and non-Gaussian behaviors in the pdfs.
%
%\PFJL{The above is perhaps a bit short, we may need to describe figures and intermediate steps (that are not shown) a bit more. I also need to make sure Fig 9 should be in this section.} \AG{Done.}

%\PFJL{Need to describe pdf evolution and how do we know the pdf we get are correct, Then, need to summarize results of other experiments.} 
%

\emph{Sensitivity Studies.}
We performed other experiments with parameter sensitivity studies similar to those of Experiments-1; similar trends were found.

%%%%%%%%%%%%%%%%%%%%%%%%%%%%%%%%%%%%%%%%%%%%%%%%%%%%%%%%%%%%%%%%%%%%%%%%%%%%%%%%%%%%%%%%%%%%%%%%%%%%%%%%%%%%%%%%%%%%%%%%%%%%%%%%%%%%%%%%%%%%%%

\subsection{Experiments 3: Learning unknown functional forms}

In our third set of experiments, the primary goal is to learn the functional form of the zooplankton mortality without any prior knowledge of candidate forms, along with the uncertain biological tracer fields. 
This completely unknown zooplankton mortality function corresponds to the $\mathcal{\widetilde{L}}$ term introduced in Eq.~\ref{eq:model uncertainty type 1}.
We utilize stochastic piece-wise linear functions to parameterize a large set of possible functional forms within a specified range, as explained in Sect.\ \ref{sec: Piece-wise Linear Function Approximations} and Eq.~\ref{eq: stochastic expansion}. Such a parameterization encompasses many different classes of functions, for example, polynomial, exponential, logarithmic, sinusoidal, etc. The right-hand-side of the stochastic NPZ model with the unknown function is given by,
\begin{equation}
\label{eq:Modified NPZ model functional learning}
\begin{split}
S^N &= -G\frac{PN}{N+K_u}+\Xi P + \Gamma Z + \underbrace{F(Z; \omega) }_{\text{Unknown~Function}} + R_m \gamma Z(1-\exp^{-\Lambda P})  \\
S^P &= G\frac{PN}{N+K_u} - \Xi P  - R_m Z(1-\exp^{-\Lambda P})  \\
S^Z &=   R_m (1-\gamma) Z(1-\exp^{-\Lambda P}) - \Gamma Z - \underbrace{F(Z; \omega)}_{\text{Unknown~Function}}
\end{split} 
\end{equation}

\begin{figure}[h]
	\centering
	\includegraphics[width=0.65\textwidth]{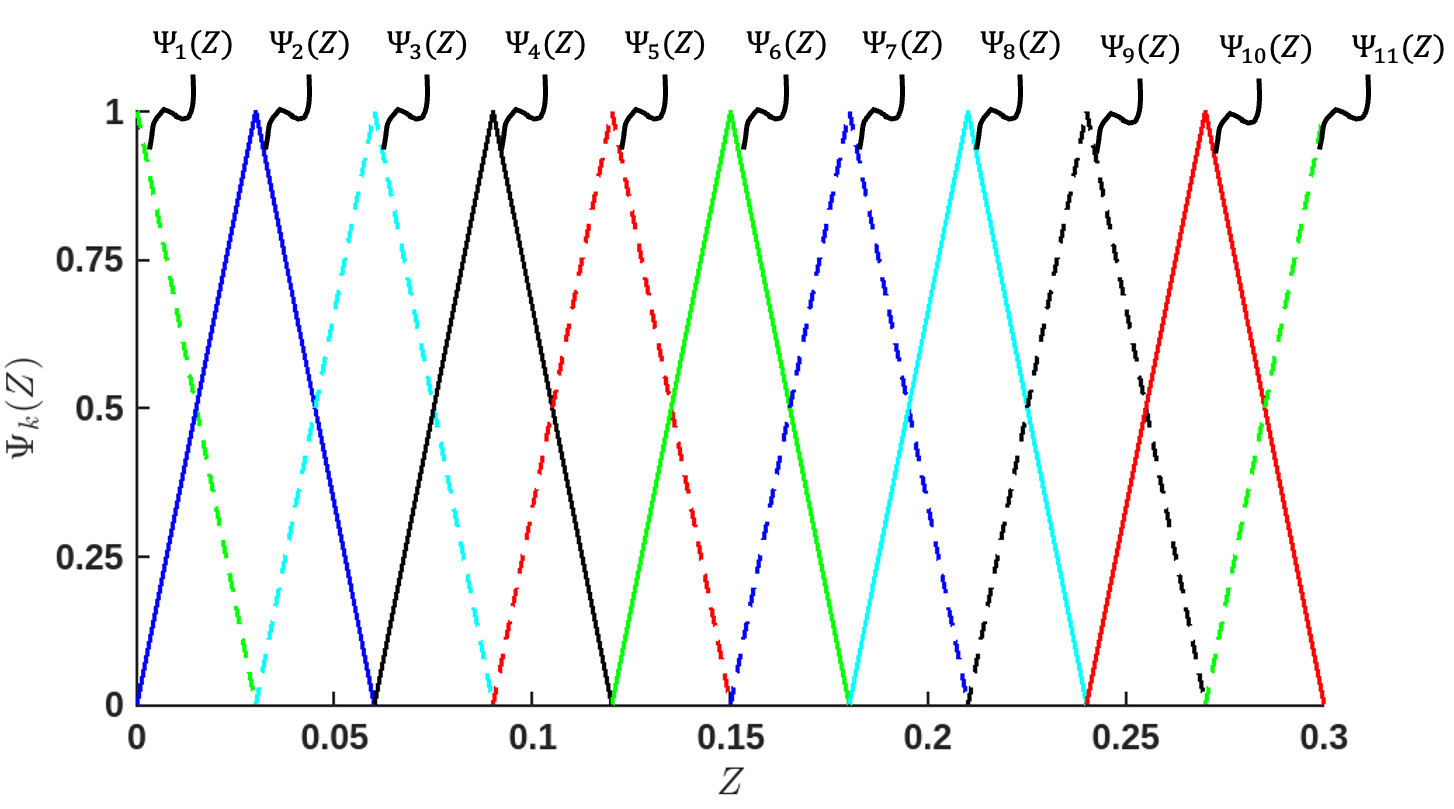}
	% \vspace*{-6mm}
	\caption{
	Experiments-3: Linear basis functions, $\{\Psi_1, ..., \Psi_{11}\}$ used to represent the unknown zooplankton mortality function, $F(Z; \omega)$, using a rich stochastic space (Eq.~\ref{eq: Expt 3 unknown function}).
 }
	\label{fig:Exp-3 basis functions}
\end{figure}

From prior knowledge \cite{newberger2003analysis}, the non-dimensional value of zooplankton is assumed non negative and its maximum value to be 0.3. 
Thus, $F(Z; \omega)$ is set to be composed of any continuous piece-wise linear segments in the interval $Z \in [0, 0.3]$. 
Dividing this interval $[0, 0.3]$ into 10 equal non-overlapping sections, such that, $0 = Z_L^1<Z_R^{1} = 0.03 = Z_L^1<...<Z_R^{9} = 0.27 = Z_L^{10}< Z_R^{10} = 0.3$, $F(Z; \omega)$ is thus represented as,
\begin{equation}
\label{eq: Expt 3 unknown function}
F(Z; \omega) = \sum_{k = 1}^{11} \gamma_k(\omega)\Psi_k(Z)
\end{equation}
where the linear basis functions, $\{\Psi_1, ..., \Psi_{11}\}$ defined by Eq.~\ref{eq: linear basis functions} are shown in Fig.~\ref{fig:Exp-3 basis functions}.

% \begin{equation}
% \begin{split}
% \Psi_1(Z) &= \begin{cases} 
% \frac{1}{0.03}(0.03 - Z) & \text{if} \; 0 \leq Z \leq 0.03 \;, \\
% 0 & \text{otherwise} 
% \end{cases} \\
% \Psi_k(Z) &= \begin{cases} \frac{1}{(Z_R^{k-1} - Z_L^{k-1})}(Z-Z_L^{k-1}) & \text{if} \; Z_L^{k-1} \leq Z \leq Z_R^{k-1} \;, \\
% \frac{1}{(Z_R^{k} - Z_L^{k})}(Z_R^{k} - Z) & \text{if} \; Z_L^{k} \leq Z \leq Z_R^{k} \;, \\
% 0 & \text{otherwise} 
% \end{cases} \quad \text{for} \; k \in \{2, ..., 10\} \;, \\
% \Psi_{11}(Z) &= \begin{cases} \frac{1}{0.03}(Z-0.27) & \text{if} \; 0.27 \leq Z \leq 0.3 \;, \\
% 0 & \text{otherwise} 
% \end{cases}
% \end{split}
% \end{equation}
%
% \begin{eqnarray}
% \phi_i(Z) = \begin{cases}
% \frac{1}{(Z^i - Z^{i-1})}(Z - Z^{i-1}), & \quad Z \in [Z^{i-1}, Z^{i}] \\
% \frac{1}{(Z^{i+1} - Z^{i})}(Z^{i+1}- Z), & \quad Z \in [Z^{i+1}, Z^{i}] \\
% 0 & \text{otherwise}
% \end{cases}
% \end{eqnarray}
%
%\PFJL{Need to check/fix usage of summation indices $i$, $j$, $k$, $\ell$, etc., for the different learning parameters, such that $i$ is for one type, $j$ another type, etc., if at all possible.}
%\AG{Done.}
%
Each set of realizations of $\gamma_k$'s, $k\in\{1, ..., 11\}$, are sampled so as to avoid a prior with unnatural highly fluctuating functions. 
The function range is set within 0 and 0.08; it is non-negative as mortality is negative in the zooplankton equation (Eq.~\ref{eq:Modified NPZ model functional learning}).
To initialize the tracer fields, we find equilibrium solutions corresponding to each realization of the zooplankton mortality function. 
%The stochastic NPZ reactions (Eq.~\ref{eq:Modified NPZ model functional learning}) are coupled with the RANS flow PDEs and used in the stochastic ADR PDEs that are solved with the DO methodology (Sects.\ \ref{sec: Stochastic Dynamically-Orthogonal PDEs in the bio-detailed paper}--\ref{sec: Numerical Schemes}). 
%
The stochastic ADR PDEs with the stochastic NPZ reactions (Eq.~\ref{eq:Modified NPZ model functional learning}) are coupled with the RANS flow PDEs, and solved with the DO methodology (Sects.\ \ref{sec: Stochastic Dynamically-Orthogonal PDEs in the bio-detailed paper}--\ref{sec: Numerical Schemes}).
Table \ref{table:DA Setup information} provides the values of other known model and hyper- parameters.
The learning objective of these experiments is to learn $F(Z;\omega)$ by estimating $\gamma_k's$ along with the biological tracer fields. 

\emph{True solution generation:}
The true solution contains quadratic zooplankton mortality, with values of the other parameters provided in Table \ref{table:DA Setup information}.

\emph{Observations and learning parameters:}
The simulated observations remain sparse in time and space, but here they consist of the nutrient field at 8 spatial locations, starting at $t=1$ and occurring every two non-dimensional times. 
In these experiments, we start the assimilation at the earlier $t=1$ time in order to limit the exploding growth of uncertainty in the system, because each function realization leads to very different biological dynamics, several of which would lead to nonphysical biological states. 
The non-dimensional $N$-data error standard deviation is 0.35.
Other hyper-parameters related to the GMM-DO filtering are provided in Table \ref{table:DA Setup information}.
%
%\AG{Numerical method:}
%As for the last two sets of experiments, the DO equations and the deterministic governing equations for the true solution are solved using the modular finite-volume framework described in Sect.\ \ref{sec: Numerical Schemes}.
%

\emph{Learning metrics:}
We compare the true fields and parameters to their DO estimates. %(mean and most probable values).
To quantify performance, we also examine the evolution of the normalized RMSEs and
pdf and realizations of the stochastic piece-wise linear functions. 
%
%\AG{The headings in this para need to be removed.}

\subsubsection{Learning results}

\begin{figure}[h]
	\centering
	\includegraphics[width=\textwidth]{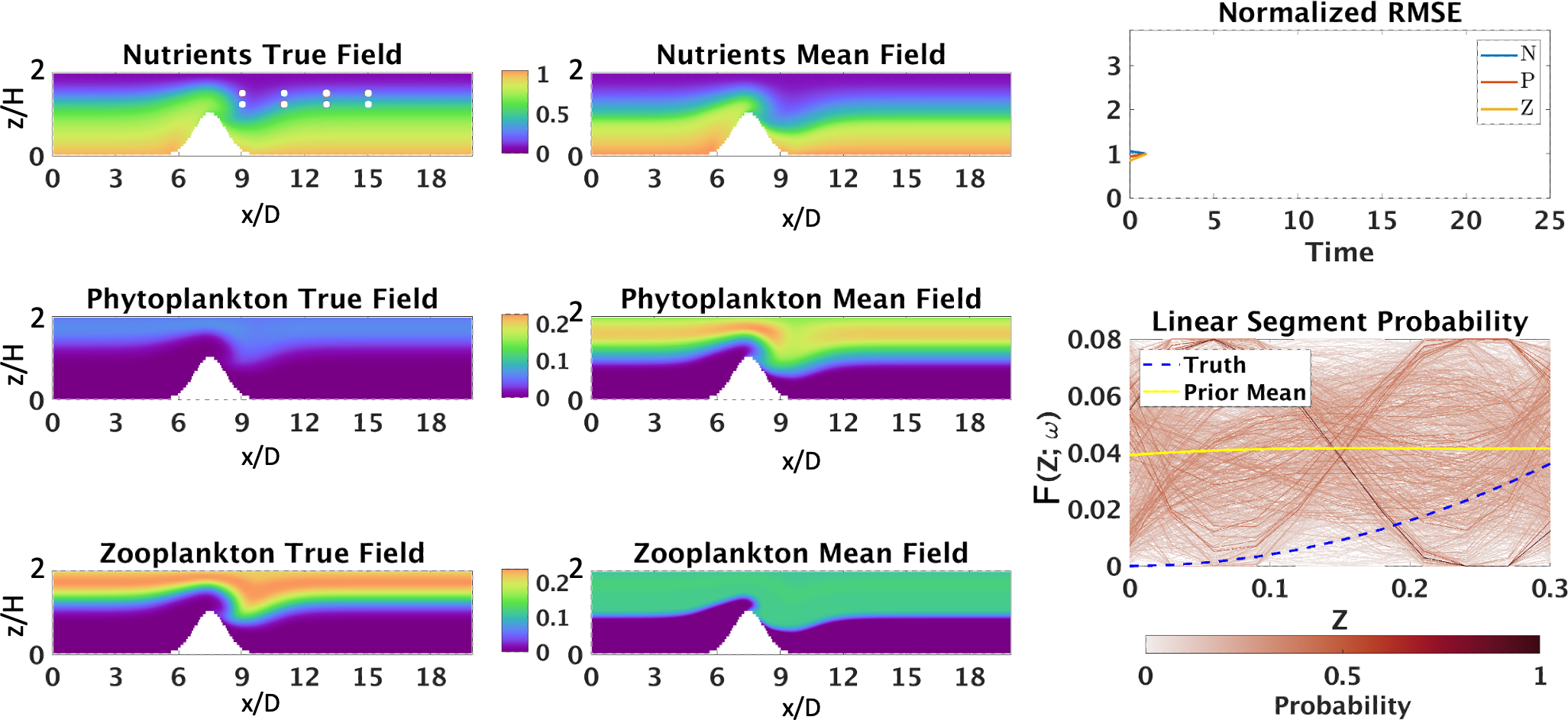}
	\vspace*{-6mm}
	\caption{
	Experiments-3: State of the true and prior estimate NPZ fields and parameters at $t = 1$ (i.e.\ just before the 1st assimilation).
    The first two columns consist of the non-dimensionalized true (left) and mean estimate (right) tracer fields of $N$, $P$, and $Z$. In the third column, the first panel shows the evolution of normalized RMSE for all the stochastic state variables. 
    The second panel contains all the realizations of the unknown functional form approximated by piece-wise linear segments. The function realizations are colored according to their respective pdf values (the 0-1 probability colorbar is under the panel). 
    The mean functional form estimate is marked with a solid yellow line, while the true functional form is marked with a dashed blue line. 
    The velocity field is deterministic with $Re=1$. The white circles on the nutrient true field mark the 8 observation locations.}
	\label{fig:NPZ Functional T=1 Prior}
\end{figure}

Figure \ref{fig:NPZ Functional T=1 Prior} illustrates the prior at $t=1$. 
Realizations in the space of the unknown function are provided (third column, bottom panel). Each of the function realizations is colored proportional to the joint probability density of the stochastic expansion parameters ($\gamma_k$'s in Eq.~\ref{eq: Expt 3 unknown function}). 
For the prior, $\gamma_k$'s are considered independent of each other and sampled uniformly, hence each piece-wise linear segment is equiprobable. 
However, the piece-wise linear segments that constitute unnatural highly fluctuating functions are eliminated. 
This leads to deviations from each function being equally likely and thus their pdf values are not equal, as seen in the third column, bottom panel, of 
Fig,\ \ref{fig:NPZ Functional T=1 Prior}.
In general, mortality being 0 for $Z=0$ is common knowledge. 
Otherwise, it could act as a sink for zooplankton and lead to negative tracer values. We let this be discovered by the learning algorithm.
The DO biogeochemical mean fields are quite far from the unknown true fields, and the prior function realizations are not similar to the true quadratic mortality. 

\begin{figure}[h]
	\centering
	\includegraphics[width=\textwidth]{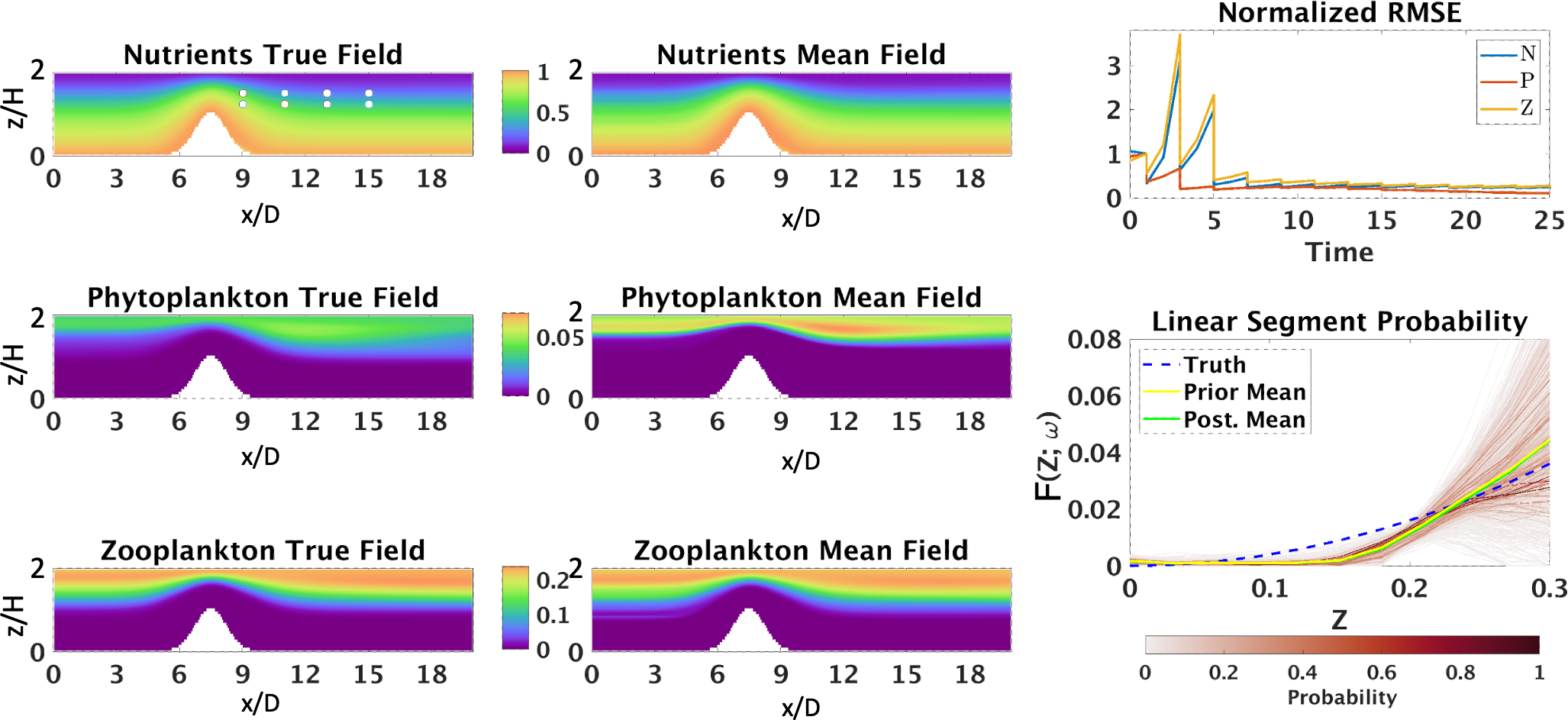}
	\vspace*{-6mm}
	\caption{
	Experiments-3: As Fig.\ \ref{fig:NPZ Functional T=1 Prior} but for posterior fields and function 
	at $t=25$ (i.e.\ just after the 13th assimilation). 
    In the second panel of the third column, the prior mean functional form estimate at $t=25$ is marked with a solid yellow line, while the posterior mean functional form estimate is marked with a solid green line.
    In the first two columns, the axis limits for the state variables have changed so as to follow the bloom, but in the third column, they remain as in Fig.\ \ref{fig:NPZ Functional T=1 Prior} so as to directly highlight the uncertainty evolution.}
	\label{fig:NPZ Functional T=25 Post}
\end{figure}

As the eight noisy $N$-observations are assimilated every two non-dimensional times, nearly all the piece-wise linear function realizations converge to the true quadratic mortality. 
Results after 13 GMM-DO assimilation in Fig.\ \ref{fig:NPZ Functional T=25 Post} show this.
The posterior function realizations (third column) are constructed by sampling the posterior joint probability density of the stochastic expansion parameters ($\gamma_k$'s in Eq.~\ref{eq: Expt 3 unknown function}) and colored proportional to the probability value.
We find however that the $N$ data are not as informative about mortality function for $Z$ beyond 0.25. 
This is in part because the maximum value reached in the true $Z$ field is $\sim 0.2$, which limits the uncertainty reduction in the larger $Z$ regime. 
The mean fields also converge to the true fields.
The normalized RMSEs of biogeochemical fields decrease at each assimilation. The learned phytoplankton mean field however remains a bit higher than true fields, in part because they were much higher initially. It is also because the observed data (here eight $N$ data) are not equally informative about all the learning objectives. As in \cite{lermusiaux_et_al_TheSea2017_modified,lermusiaux_et_al_Oceanog2017_modified,lin_PhDThesis2020}, this is confirmed by mutual information fields (not shown).

\emph{Sensitivity Studies.}
Other experiments included studying the effect of incorporating or excluding prior knowledge such as the function value being 0 for $Z=0$ and using smoothly varying function realizations. For the former,
sampling $\gamma_k$'s independent of each other led to highly fluctuating function realizations which completely impaired the learnability of the unknown function. 
For the latter, enforcing $\gamma_0 = 0$ sets $F(0; \omega) = 0$ for all realizations improved the convergence among the learned function realizations and the true function.
Finally, increasing the number of independent observations (more $N$ data, data for $Z$ or $P$ as well, etc.) also improved the sharpness of our GMM-DO inference: in all examples we show, we highlight cases with sparse observations as seen in real ocean applications.

%%%%%%%%%%%%%%%%%%%%%%%%%%%%%%%%%%%%%%%%%%%%%%%%%%%%%%%%%%%%%%%%%%%%%%%%%%%%%%%%%%%%%%%%%%%%%%%%%%%%%%%%%%%%%%%%%%%%%%%%%%%%%%%%%%%%%%%%%%%%%%

\subsection{Experiments 4: Learning in chaotic dynamics}

In the last set of experiments, in order to robustly test our algorithms, the aim is to learn a five-component NNPZD model with a flow of Reynolds number $Re = 500$. At such high $Re$, vortices start to shed in the wake of the ridge and the flow chaotic. The learning objectives include all 5 biogeochemical fields, the Ivlev grazing parameter ($\Lambda$), the phytoplankton-specific mortality rate ($\Xi$), the zooplankton maximum grazing rate ($R_m$), the zooplankton specific mortality ($\Gamma$), and the presence or absence of the quadratic zooplankton mortality term. 
Once again, the ambiguity in the quadratic zooplankton mortality function corresponds to the $\mathcal{\widehat{L}}$ term introduced in Eq.~\ref{eq:model uncertainty type 1}.
The stochastic NNPZD reactions, with all the uncertain parameters explicitly containing $\omega$ as an argument, are given by,
\begin{equation}
\label{eq:NNPZD Model modified}
\begin{split}
S^{NO_3} =& ~\Omega NH_4 - G\left[\frac{NO_3}{NO_3+K_u}\exp^{-\Psi_I NH_4}\right]P  \,, \\
S^{NH_4} =& ~-\Omega NH_4 + \Phi D +\Gamma(\omega) Z + \alpha(\omega)\underbrace{(\Gamma_2 Z^2) }_{\text{Quad.~Z~Mort.}}  - G\left[\frac{NH_4}{NH_4+K_u}\right]P  \,, \\
S^{P} =& ~G\left[\frac{NO_3}{NO_3+K_u}\exp^{-\Psi_I NH_4} + \frac{NH_4}{NH_4+K_u}\right]P - \Xi(\omega)  P  \,,\\ 
&- R_m(\omega)  Z(1-\exp^{-\Lambda(\omega)  P})  \,,\\
S^{Z} =&  ~ R_m(\omega)  (1-\gamma) Z(1-\exp^{-\Lambda(\omega)  P}) - \Gamma(\omega)  Z + \alpha(\omega)\underbrace{(\Gamma_2 Z^2) }_{\text{Quad.~Z~Mort.}} \,, \\
S^{D} =&~ R_m(\omega)  \gamma Z(1-\exp^{-\Lambda(\omega)  P}) + \Xi(\omega)  P - \Phi D \,.
\end{split}
\end{equation} 
Initially, we assume uniform and independent pdfs for the four uncertain regular parameters and equiprobability for the quadratic zooplankton mortality term to be present or absent.
%
%The NNPZD reactions (\ref{eq:NNPZD Model modified}) are coupled with the deterministic RANS flow PDEs and used in the stochastic ADR PDEs that are solved with the DO methodology (Sect.\ \ref{sec: Stochastic Dynamically-Orthogonal PDEs in the bio-detailed paper}).
%
The stochastic ADR PDEs with the stochastic NNPZD reactions (\ref{eq:NNPZD Model modified}) are coupled with the deterministic RANS flow PDEs, and solved with the DO methodology (Sects.\ \ref{sec: Stochastic Dynamically-Orthogonal PDEs in the bio-detailed paper}--\ref{sec: Numerical Schemes}).
The other known physical-biogeochemical model parameters as well as the hyper-parameters for the DO equations 
% and the GMM-DO filter 
are provided in Table \ref{table:DA Setup information}.

\emph{True solution generation:}
The true solution from which observations are extracted, corresponds to the non-dimensional values, 1.5 for $\Lambda$, 0.04 for $\Xi$, 0.6 for $R_m$, 0.14 for $\Gamma$, and 0 for $\alpha$, i.e.\ the quadratic mortality term absent. The state fields are initialized and evolved as described in Sect. \ref{sec:True Solution Generation}.

\emph{Observations and learning parameters:}
The noisy observations remain sparse and univariate, but due to the unstable and fast dynamics of the flow, there is a need for slightly more frequent data than in other experiments. The phytoplankton field is observed at nine locations starting at $t=2$ and subsequently every one non-dimensional time. In total, we assimilate 24 times, i.e.\ until $t=25$, with a non-dimensional $P$-data error standard deviation of 0.04.
Other hyper-parameters related to the GMM-DO filtering are given in Table \ref{table:DA Setup information}.
%\AG{Numerical method:}
%The DO equations and the deterministic governing equations for the true solution remain solved using the finite-volume framework (Sect.\ \ref{sec: Numerical Schemes}). 

\emph{Learning metrics:} 
%
% To both quantitatively and qualitatively analyse the performance of our learning framework, we make use of the normalized RMSE and posterior pdf of the stochastic parameters as described in Sect. \ref{sec:Learning Metrics}.
We compare the true fields and parameters
to their DO estimates.
To quantify performance, we compute the evolution of the normalized RMSEs for all five biological fields and five stochastic parameters. 
We also analyze the evolution of pdfs of the regular and formulation parameters, and the variances of DO coefficients.

\subsubsection{Learning results}

\begin{figure}[h]
	\centering
	\includegraphics[width=\textwidth]{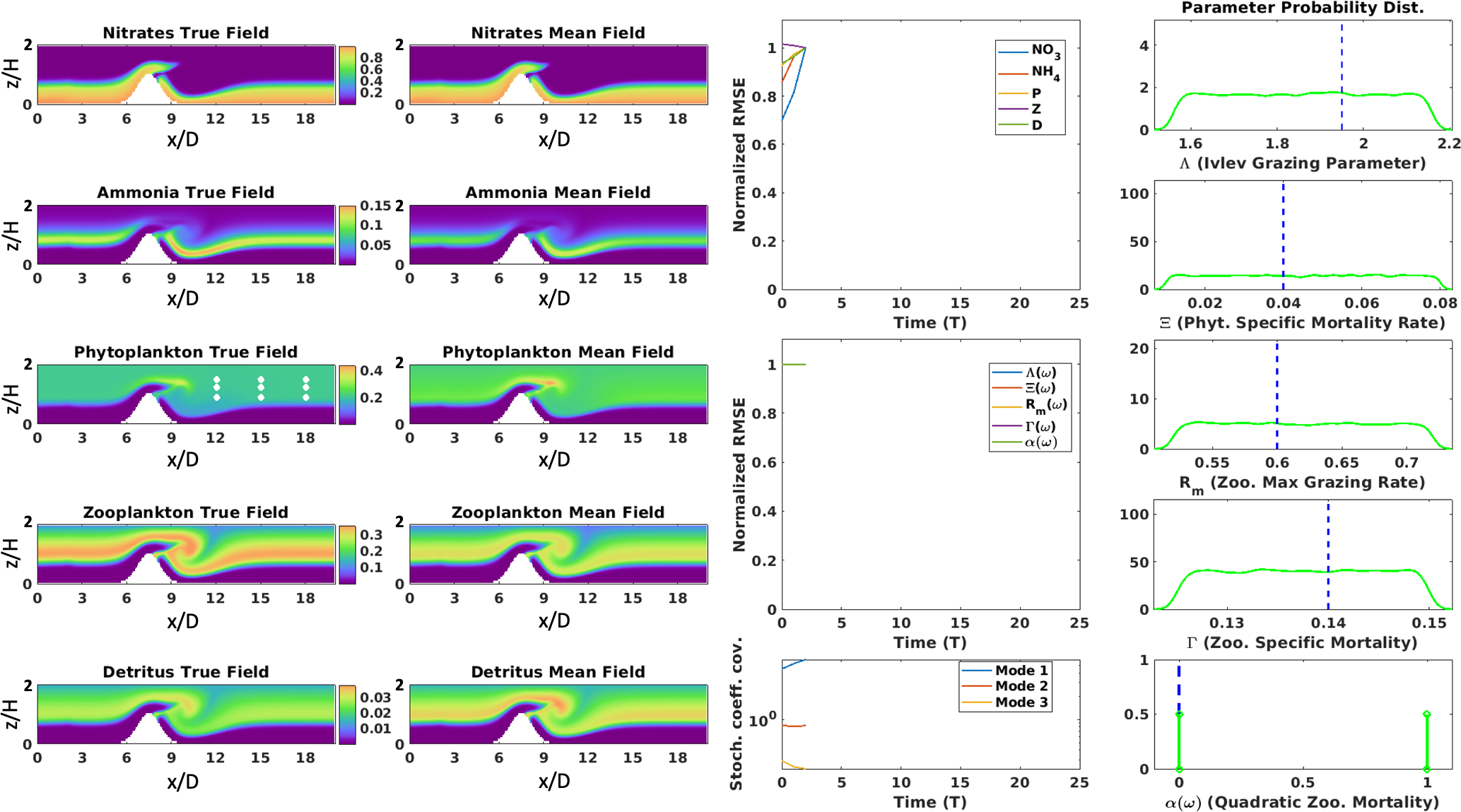}
	\vspace*{-7mm}
	\caption{
	Experiments-4: State of the true and prior estimate NNPZD fields and parameters at $t = 2$ (i.e.\ just before the 1st assimilation).
    The first two columns consist of the non-dimensionalized true (left) and mean estimate (right) fields of $NO_3$, $NH_4$, $P$, $Z$, and $D$. 
	In the third column, the first two panels show the evolution of the normalized RMSEs for the five state variables and five parameters.
    The third panel shows the evolution of variance of the top 3 DO modes. 
    In the fourth column, the pdfs of the non-dimensional $\Lambda(\omega)$, $\Xi(\omega)$, $R_m(\omega)$, $\Gamma(\omega)$, and $\alpha(\omega)$ (learns the presence or absence of quadratic zooplankton mortality) are marked with solid green lines, with the true unknown parameter values marked with blue dotted lines. The velocity field is deterministic with $Re=500$. Additionally, the white circles on the phytoplankton true field mark the 9 observation locations.}
	\label{fig:NNPZD T=2 Prior}
\end{figure}

Figure \ref{fig:NNPZD T=2 Prior} shows the prior estimates at $t=2$. 
The flow has just started to develop. There are  significant differences between the true and mean biogeochemical fields. The normalized RMSEs are equal to 1 by construction.
The pdfs of all parameters remained as they were initially since no data has been assimilated.
%We assume uniform and independent probability distributions for the 4 uncertain regular parameters, and equilikely probability for the quadratic zooplankton mortality term to be present or absent. 

\begin{figure}[h]
	\centering
	\includegraphics[width=\textwidth]{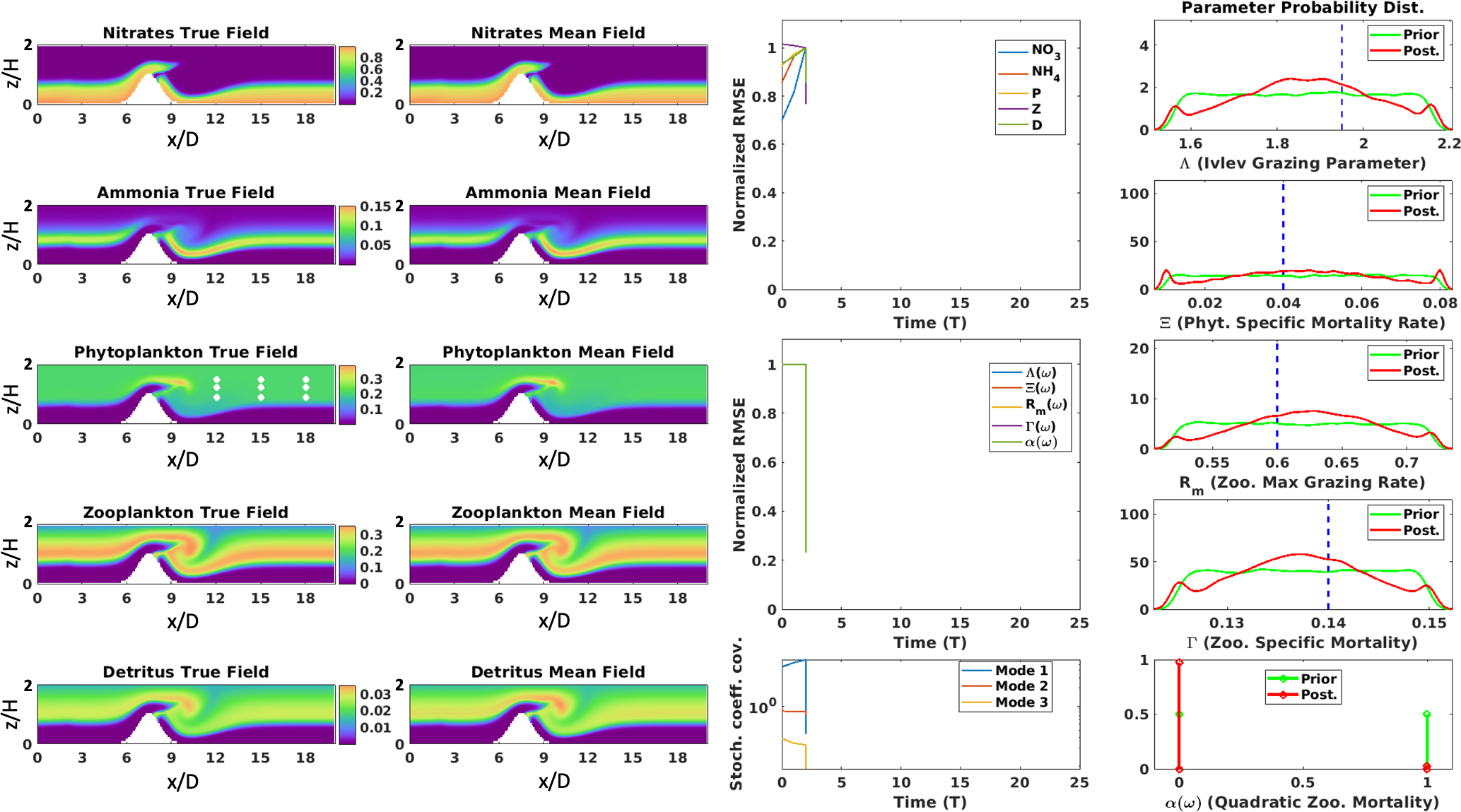}
	\vspace*{-7mm}
	\caption{
	Experiments-4: As Fig.\ \ref{fig:NNPZD T=2 Prior}, but for posterior fields and parameters at $t=2$ (i.e.\ just after the 1st assimilation). 
    In the fourth column, the prior pdfs of the non-dimensional $\Lambda(\omega)$, $\Xi(\omega)$, $R_m(\omega)$, $\Gamma(\omega)$ at $t=2$ are marked with solid green lines, while the posterior pdfs are marked with solid red lines.
    In the first two columns, the axis limits for the state variables have changed so as to follow the bloom and decay events, but in the last two columns, they remain as in Fig.\ \ref{fig:NNPZD T=2 Prior} so as to directly highlight the uncertainty evolution.}
	\label{fig:NNPZD T=2 Post}
\end{figure}

Figure \ref{fig:NNPZD T=2 Post} illustrates the posterior estimates at $t=2$, just after the first assimilation. 
Large corrections were made to the mean tracer fields (also visible in their RMSEs that decay by about 15 to 25\%), and the GMM-DO learning already predicts the absence of quadratic zooplankton term. 
These first 9 noisy $P$-observations are not as informative however about the other parameters (their RMSEs only decay by about 4\% to 8\%).

\begin{figure}
	\centering
	\includegraphics[width=\textwidth]{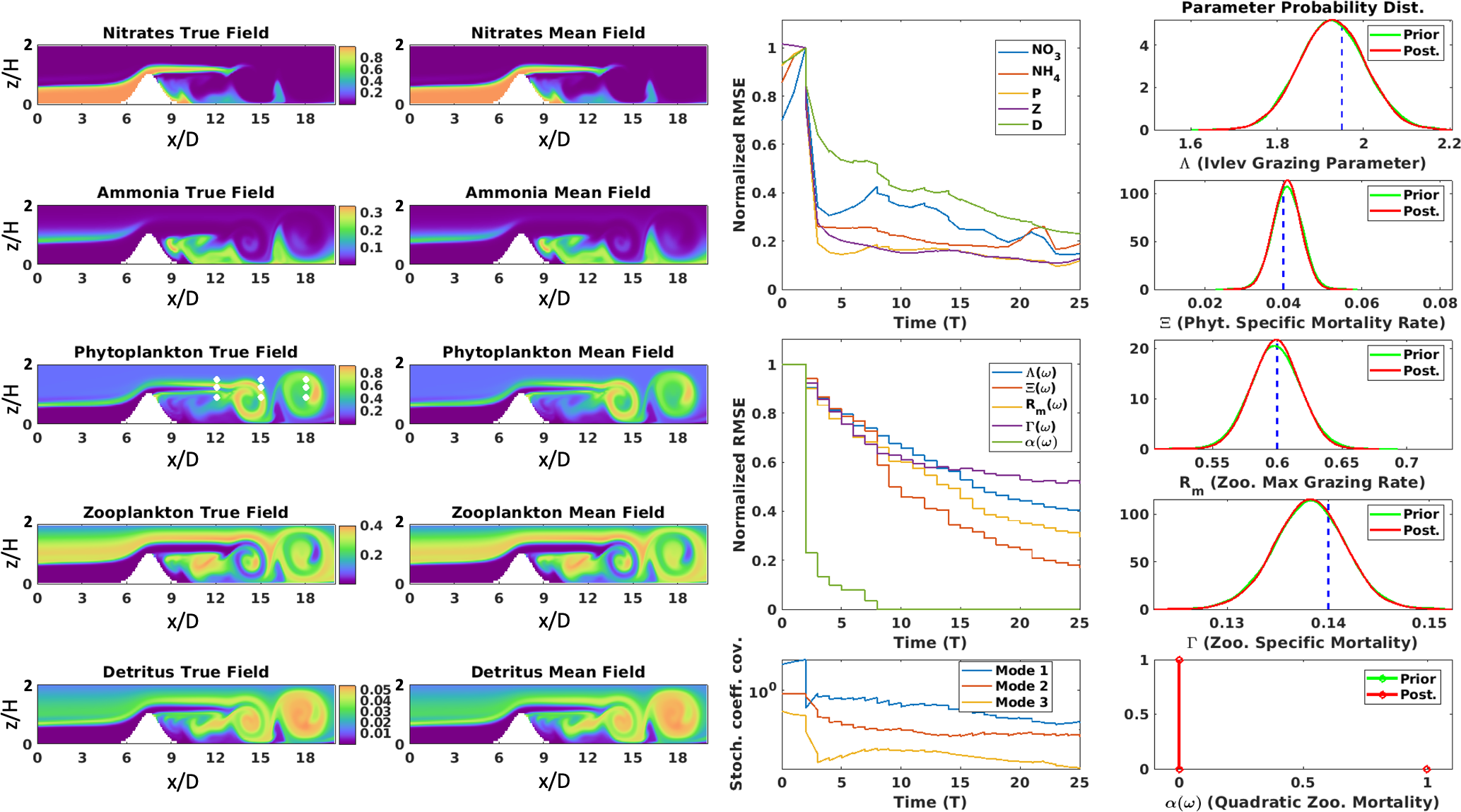}
	\vspace*{-7mm}
	\caption{
	Experiments-4: As Fig.\ \ref{fig:NNPZD T=2 Prior}~\&~ \ref{fig:NNPZD T=2 Post}, but for posterior fields and parameters at $t=25$ (i.e.\ just after the 24th assimilation).
    In the first two columns, the axis limits for the state variables have changed so as to follow bloom and decay events, but in the last two columns, they remain as in Fig.\ \ref{fig:NNPZD T=2 Prior} so as to directly highlight the uncertainty evolution.
    %In the third column, the prior pdf associated with the non-dimensional $\Lambda(\omega)$, $\Xi(\omega)$, $R_m(\omega)$, $\Gamma(\omega)$ at $t=25$ is marked with solid green lines, while the posterior pdf is marked with solid red lines.
    }
	\label{fig:NNPZD T=25 Post}
\end{figure}

Figure \ref{fig:NNPZD T=25 Post} shows the estimates at $t=25$, after 24 GMM-DO assimilation steps. In addition to the mean fields, our augmented filter has been learning the four regular parameters.
Their posterior pdfs have become Gaussian which has occurred in intermediate assimilation steps (not shown). 
We also show the evolution of variance of the top three modes. We find that the total variance on average either decreases or remains similar, while that of individual modes in general decreases but may also increase in accord with the stochastic dynamics.
The velocity field being chaotic renders the learning more challenging in this experiment but our framework can still meet all the learning objectives, even with sparse and univariate data. 

\begin{figure}
	\centering
        \includegraphics[width=0.64\textwidth]{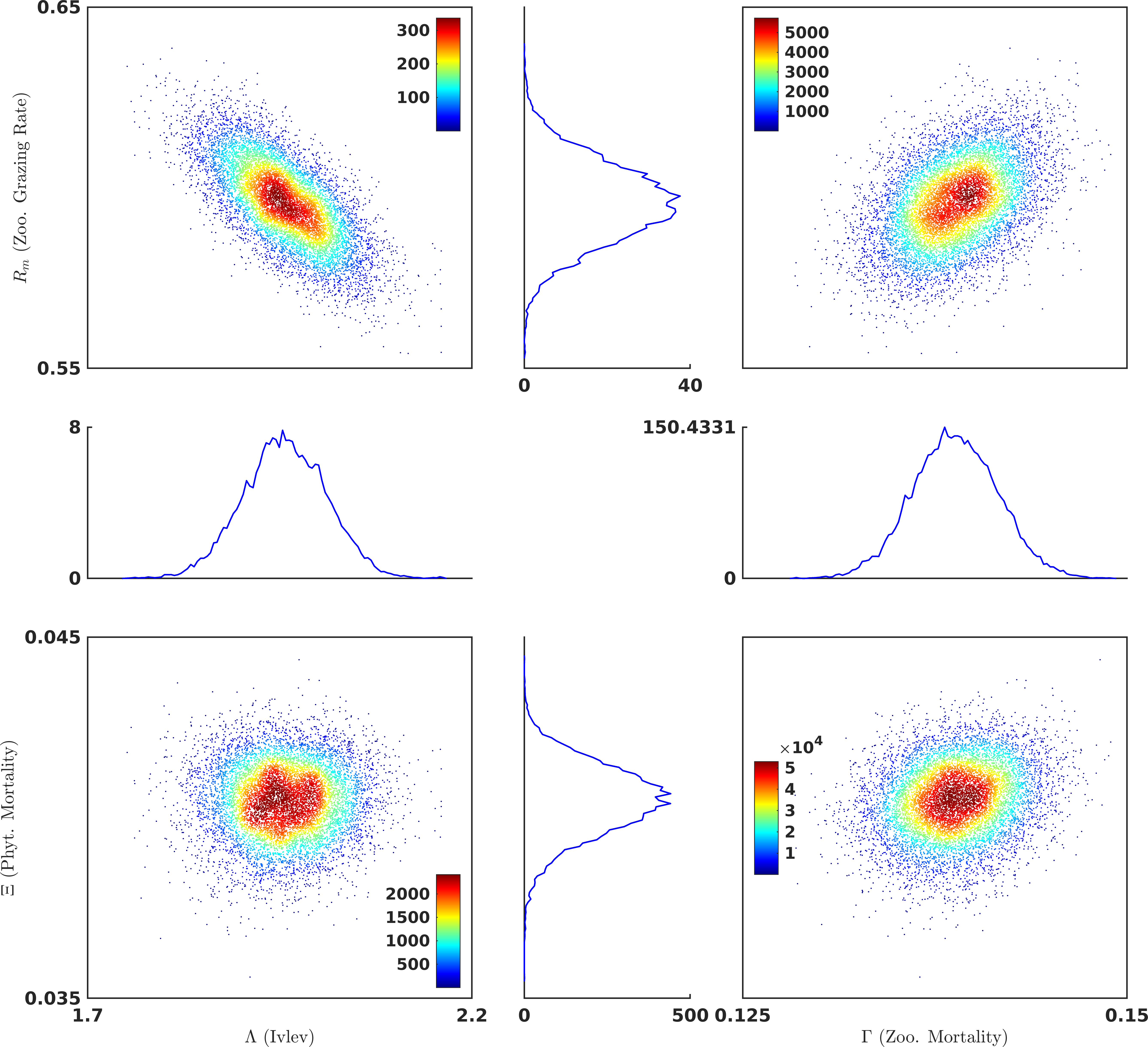}
	\vspace*{-3mm}
	\caption{
	Experiments-4: Posterior joint-double pdf (colored scatter plots)
    and single pdf (line plots) marginals of the four non-dimensional regular stochastic parameters $\Lambda(\omega)$, $\Xi(\omega)$, $R_m(\omega)$, $\Gamma(\omega)$ at $t=25$ (i.e.\ just after the 24th assimilation). 
    The joint-double pdfs highlight non-Gaussian details and dependencies among parameters.
    %, even if the single marginals are somewhat Gaussian.
    %
    %\AG{The opacity of the face-color for each of the blue-filled circles in the scatter plots is proportional to the probability density obtained from respective \textit{ksdensity} fits. Pierre, there is another version without edge-color for the scatter plot cirles, Joint\_para\_NNPZD\_t25\_post\_noedge.png. Please use that if you prefer it.}
 }
	\label{fig: NNPZD Joint Parameters T=25}
\end{figure}

As the stochastic states, parameters, and model formulations are estimated jointly, our augmented GMM-DO Bayesian learning can provide interesting insights into the co-dependence, biases, and equifinality of all the quantities being estimated. To showcase this capability, we provide joint distributions for combinations of the four uncertain regular parameters in Figure~\ref{fig: NNPZD Joint Parameters T=25} at $t=25$. We find that $\Lambda(\omega)$ and $R_m(\omega)$ are negatively correlated and that $\Gamma(\omega)$ and $R_m(\omega)$ are positively correlated, while $\Xi(\omega)$, $\Lambda(\omega)$, and $\Gamma(\omega)$ are nearly independent of each other. 
We note that such a negative correlation between $\Lambda(\omega)$ and $R_m(\omega)$ can also be inferred from the zooplankton grazing term in the NNPZD model (equation~\ref{eq:NNPZD Model modified}). Similarly, $\Gamma(\Omega)$ and $R_m(\omega)$ need to simultaneously increase or decrease to maintain phytoplankton concentration levels.
Thus the joint estimation of all the variables of interest can provide the researcher with an essential tool for additional analysis and discoveries. %especially in real applications.

% \vspace*{-5mm}

% \vspace*{-4mm}

\emph{Sensitivity Studies}.
Other experiments were performed. As expected, they demonstrated sensitivity to the schedule, type, noise, and quantity of observations.
With only nine noisy, sparse, and univariate data, starting them after the chaos sets in, or sampling even less frequently than every one non-dimensional time, led to posterior pdf of some stochastic parameters that were not concentrated around their respective true values. 
Similar results were found even when less than nine data were collected. Adding other observation types improved the learning. For other sensitivity studies, trends similar to other experiments were found. 
%and in turn the stochastic simulation completely diverging from the truth.
%\PFJL{for the above, this likely means the DO estimate is not correct and prior uncertainty has collapsed? Perhaps you could check if the prior std fields still match the prior errors. If this is the case, we should likely tone down the above.} \AG{I plotted variation of variance of first 5 modes vs. time which will also give us an idea of too much decrease in prior uncertainty. For the 2 additional runs where observations were started at $T=1$, there was a sharper decrease in variance values (and values remained lower in magnitude) as compared to the one presented in the paper. Also confirmed prior uncertainty collapse by plotting prior std. dev. and prior errors. However, there is a caveat that the initial uncertainty was slightly off (a little lower in the additional runs) as compared to the one presented in the paper. For the run where observations were started at $T=5$, I think the realization correspondence might have gotten too blurry due to DO errors, thus, observations do not help.}

%%%%%%%%%%%%%%%%%%%%%%%%%%%%%%%%%%%%%%%%%%%%%%%%%%%%%%%%%%%%%%%%%%%%%%%%%%%%%%%%%%%%%%%%%%%%%%%%%%%%%%%%%%%%%%%%%%%%%%%%%%%%%%%%%%%%%%%%%%%%%%

\subsection{Remarks and Discussions}

\paragraph{\textit{Computational Costs and Implementation}} Our Bayesian model learning framework can be broadly divided into 3 components, construction and initialization of a general model combining all compatible and compatible-embedded candidate models, probabilistic prediction using the DO differential equations, and multivariate Bayesian filtering to update the augmented state, all as 
summarized in Fig.\ \ref{fig: overview general Bayesian model learning methodology}. 
The first component ensures that a single stochastic model is evolved, theoretically encompassing infinitely many candidate models. Additionally, our overall framework is general enough to be employed in any existing data-assimilation systems capable of joint state and parameter estimation, without much computational overhead. 
If only state uncertainty exists in the data-assimilation system, then state augmentation (\ref{app:state_aug}) should be leveraged. 
Employing our framework with existing stochastic models will involve code modification to implement the derived general model. 
In the Bayesian filtering step,
the stochastic formulation, complexity, and expansion parameters can be treated similarly to other regular stochastic parameters. 
The number of these additional stochastic parameters is expected to be similar to the number of existing regular stochastic parameters. 
In the current work, the uncertainty evolution is achieved using the DO methodology, while the non-Gaussian Bayesian filtering uses the GMM-DO filter. 
The computational cost will depend on the problem and software implementation. In general, the highest cost is the probabilistic DO prediction as it commonly adds costs of the order of the cost of $N_s$ deterministic model simulation.
%The overall framework is implemented in MATLAB, and experiments were run on single nodes with configuration, 24 CPUs (Intel(R) Xeon(R) CPU E5-2680 v3 @ 2.50GHz), and 264Gb RAM. Each of the experiments took a total of $\mathcal{O}$(1-2) days to run depending on the number of biological state variables and DO modes, and a single GMM-DO filtering step took $\mathcal{O}$(1) minute depending on the number of DO modes and stochastic parameters. 
Both DO and GMM-DO have been well studied and rigorously compared to competing algorithms. We refer the reader to \cite{sapsis_lermusiaux_PD2009, sapsis_lermusiaux_PHYSD2012, feppon_lermusiaux_SIMAX2018a, sondergaard_lermusiaux_MWR2013_part1, sondergaard_lermusiaux_MWR2013_part2} for more details on efficiency and implementation.  

\paragraph{\textit{Extension to 3D in Space, High-dimensionality, and Smoothing}} As mentioned above, our framework for Bayesian state estimation and model discovery could be used with uncertainty forecasting and data assimilation schemes developed for 3D modeling systems of high numerical dimensions. On the one hand, the DO methodology has been extended to 3D ocean primitive equations with a nonlinear free-surface \cite{subramani_PhDThesis2018,subramani_lermusiaux_prep,gkirgkis_MSThesis2021,gkirgkis_lermusiaux_prep}. 
On the other hand, the GMM-DO filter performs the Bayesian update in the DO coefficient subspace with a dimension very small compared to that of the numerical state vector and it has remained efficient for 3D in-space modeling systems. 
The precursor to the GMM-DO filter, the error subspace statistical estimation (ESSE) scheme \cite{lermusiaux_robinson_MWR1999,lermusiaux_MWR1999} which also completes its assimilation update in a subspace (a Gaussian update), has been successfully used in 3D for multiple ocean regions, e.g., \cite{lermusiaux_DAO1999,
lermusiaux_et_al_ICTCA2002,cossarini_et_al_JGR2009,
lermusiaux_et_al_Oceanog2011,lermusiaux_et_al_BBN_Oceans2020}. 
Its non-Gaussian versions, the GMM-ensemble and GMM-ESSE filtering schemes, have also been very useful, for example to update subsurface fields based on surface observations (e.g., satellite sea surface temperature or color). 
We confirmed in our sensitivity studies that surface-only observations can update subsurface fields as long as they contain mutual information \cite{lermusiaux_et_al_TheSea2017}.
The 3D extension of our novel GMM-DO learning of states and models is thus very promising for use with real ocean data \cite{haley_et_al_Oceans2020,gupta_PhDThesis2022} as well as for the combination with deep learning \cite{gupta_lermusiaux_SR2023}. The extension to Bayesian smoothing for updating the state variable pdfs backward in time is another direct opportunity \cite{lolla_lermusiaux_MWR2017_partI,lolla_lermusiaux_MWR2017_partII,gupta_PhDThesis2022}.

\paragraph{\textit{Constraints, Biases, and Equifinality}} 
The GMM-DO Bayesian estimation utilizes scientific knowledge in the form of the dynamic joint prior distribution over the states, parameters, and models, and it can accommodate constraints \cite{sondergaard_lermusiaux_MWR2013_part1,lolla_lermusiaux_MWR2017_partI,lu_lermusiaux_PhysD2021}.
Indeed, Eq.~\ref{eq:post model dist} shows for example that posterior distributions will always be contained within the support of the priors. 
Thus, constraints incorporated in the priors such as always-positive parameters will be maintained in the posterior.
The joint posterior distribution also captures non-Gaussian and nonlinear dynamical relationships between the states, parameters, and models. 
For example, in Experiments-4, we provided such an analysis by examining the joint posterior of the regular stochastic parameters (Fig.~\ref{fig: NNPZD Joint Parameters T=25}).
The joint posterior pdfs can also be useful to find different combinations of parameters, model functions, and complexities that lead to the same solution, also known as equifinality \cite{duda2006pattern}. Biases can be discovered in a similar fashion \cite{lu_lermusiaux_MSEAS2014,lu_lermusiaux_PhysD2021}. For example,
the posterior multimodal distributions and thus the possible presence of biases or parameter combinations with equifinality were clearly visible in the pdfs of the Ivlev parameter in Experiments-1 and 2, see Figs.\ \ref{fig: NPZ_Parameter_Quad_Mort T=5 Post},~ \ref{fig: NPZ_Parameter_Quad_Mort T=15 Post},~ \ref{fig: NPZ_Parameter_Quad_Mort T=25 Post}, and \ref{fig: NPZD_Complexity T=25 Post}.

\paragraph{\textit{Most Informative Observations for Model Learning}} 
What, when, and where to sample for optimal information are critical questions for the efficient use of resource-constrained observation platforms \cite{lermusiaux_et_al_TheSea2017}. 
Our research group and collaborators have developed and employed ``adaptive sampling" methods for varied purposes and ocean regions \cite{evangelinos_et_al_ICCS2003,heaney_et_al_OD2016,heaney_et_al_JFR2007,lermusiaux_et_al_ST2007,lermusiaux_PhysD2007,ramp_et_al_DSR2009,wang_et_al_JMS2009,petillo_et_al_Oceans2015,Cococcioni_et_al_Oceans2015,rajan_et_al_MTSJ2021}. 
Using the DO methodology, GMM-DO filter, and reachability schemes, adaptive sampling based on mutual information (MI) has been derived and applied to fluid flows and ocean fields and parameters \cite{lolla_PhDThesis2016,lermusiaux_et_al_Oceanog2017}. 
The often intractable MI computations are made feasible within the DO subspace, while still accounting for both nonlinear dynamics and non-Gaussian prior pdfs. 
This MI-based scheme was recently extended to adaptive sampling
for optimal learning of dynamical models \cite{lin_PhDThesis2020}, so as to identify the data that best constrain the model formulation.  
First, the scheme predicts the MI field between the possible measurement types, locations, and times and the adequacy of model formulations.
The optimal sampling scheme then utilizes these predicted MI fields to determine where, when, and what to sample for collecting the most information about the adequacy of competing or unknown model formulations \cite{lin_PhDThesis2020,lermusiaux_et_al_TheSea2017}.
The present work does not employ these MI-based optimal sampling schemes, but this can be done, as highlighted in \cite{lermusiaux_et_al_TheSea2017}.

\section{Conclusions} \label{sec:conclusions}

Biogeochemical-physical models for the ocean are inherently uncertain due to the inability to capture all the complex marine interactions and processes with a single mathematical model. Uncertainty manifests itself in many different forms including the initial conditions, boundary conditions, parameters, parameterizations, state variables, and the model complexity and equations themselves. 
% In general, Bayesian approaches are advantageous for melding observations with the model, as they provide the ability to take into account all the existing prior knowledge in the learning process, accompanied by the associated uncertainty estimates. Thus, we built upon the approach developed in Lu and Lermusiaux \cite{lu_lermusiaux_MSEAS2014, lu_lermusiaux_PhysD2021} for the simultaneous estimation of states and parameters along with discrimination among candidate models in high-dimensional stochastic dynamical systems using sparse observations.
% However, often none of the candidate models is exactly equal to the true model, or the functional form is yet completely elusive to scientists.
% Also, the candidate models are mostly compatible with each other, for example, only certain functional terms in a model are unknown, or low complexity models are embedded in higher complexity models.
%     Especially in such cases, we introduce additional parameterizations to unify all the candidates into a general single model. 
% It might also be the case that no candidate models are available, which would require a search in a very large model space. 
%
In this work,
we develop a Bayesian model learning methodology that interpolates in the space of candidate dynamical models and discovers model formulations, all while estimating state variable fields and parameter values, as well as the joint probability distributions of all learned quantities.
It employs the GMM-DO filter and state augmentation to predict and update pdfs of
high-dimensional and multidisciplinary ecosystem dynamics governed by PDEs.
Using noisy, sparse, and indirect univariate observations and Bayes' law,
the complete joint probabilities of biogeochemical-physical fields and parameters, and of known, uncertain, and unknown model formulations are updated. 
Non-Gaussian statistics, ambiguity, and biases are captured. The parameter values, model functional forms, and model complexities that best explain the data are identified.
The first crucial innovations are the stochastic formulation and complexity parameters that unify compatible candidate models, possibly of different complexities, into a single general stochastic PDEs system. 
A second is the use of stochastic expansion parameters within piecewise function approximations that generate dense candidate model spaces. 
Our new methodology is generalizable and
interpretable, and provides marginal pdfs for all learned model quantities.
At the cost of a single stochastic DO model simulation with parameter estimation, it seamlessly and
rigorously discriminates among many existing models, possibly none of which are accurate, but also extrapolates out of the space of models to discover new ones.

%These situations were addressed in two novel ways: first, using special stochastic parameters to unify all the candidates into a single general model; second, parameterizing unknown functions using stochastic piece-wise linear functions, allowing us to search in an infinite candidate space.
%Our new methodology not only seamlessly and rigorously discriminated between existing models, but also extrapolated out of the
% space of models to discover newer ones.
%In all cases, the results were generalizable and interpretable, and our Bayesian estimations provided much more than maximum likelihood estimates: they predicted and updated
%the complete joint probability distribution of states, parameters, and models. All of this was achieved just at the cost of single stochastic model simulation with parameter estimation, enabling both discrimination and discovery of models. 
% Our rigorous PDE-based Bayesian learning framework combines the Dynamically Orthogonal (DO) equations for efficient reduced-dimension uncertainty evolution; and the Gaussian mixture model (GMM) DO filtering algorithm for the nonlinear, non-Gaussian inference of the states, parameters, and model equations simultaneously.

% We developed a rigorous PDE-based Bayesian learning framework by combining Dynamically Orthogonal (DO) equations for efficient reduced dimension uncertainty evolution; and the Gaussian mixture model (GMM) DO filtering algorithm for the simultaneous nonlinear inference of the states, parameters, and model equations.

The performance of our Bayesian learning framework was evaluated using a series of twin experiments
based on flows past a ridge
with compatible and embedded PDEs for the three-component NPZ model (nutrients ($N$), phytoplankton ($P$), and zooplankton ($Z$)), four-component NPZD model ($N$, $P$, $Z$ and Detritus ($D$)), and five-component NNPZD model (ammonia ($NH_4$), nitrate ($NO_3$), $P$, $Z$, and $D$). 
In the first set of experiments, we use the NPZ model with uncertain initial conditions, unknown Ivlev grazing parameter value, and ambiguity in the presence or absence of the quadratic zooplankton mortality term.  
Our new Bayesian learning simultaneously estimated the state variables, Ivlev parameter, and unknown functional form, using noisy sparse Z observations in space and time (only six data points every two non-dimensional times). The posterior pdf of the parameter contained secondary peaks,  indicating that alternative combinations of parameter values  could explain the observed data. This showcased the ability of our framework to capture non-Gaussian statistics including ambiguity and biases. 
In the second set of experiments, assimilating just eight noisy N-data every two non-dimensional times, we demonstrated the ability to learn the complexity of the model. We identified the true model within NPZ and NPZD, along with the uncertain fields and Ivlev grazing parameter. 
In the third set of experiments, we assumed no prior knowledge about the functional form of zooplankton mortality and generated a dense function space using stochastic piece-wise linear approximations.
Assimilating just eight noisy N-data every two non-dimensional times, our framework then searched in this rich functional space, estimated the fields and regular parameter values, and was shown capable of discovering the mortality function. 
The last set of experiments involved learning the complex NNPZD model in an unsteady chaotic deterministic flow with vortex shedding. The NNPZD model had uncertainty in all the tracer fields, four parameters, and in the presence or absence of the zooplankton mortality term. All of the learning objectives were achieved simultaneously, using only nine noisy P-data every non-dimensional time.
In all cases, 
we quantified the learning skill, and evaluated convergence and the sensitivity to hyper-parameters.

These four sets of experiments were complementary, allowing us to showcase the features of
our PDE Bayesian learning framework. It successfully discriminates among functional forms and model complexities and also learns in the absence of prior knowledge by searching in dense function spaces. 
The next steps include applying this framework to more complex ocean applications, especially to realistic ocean models and to real ocean data. 
Even though we demonstrate our learning framework using biogeochemical models, it is applicable to other domains with model uncertainty, for example, medicine, economics, energy, etc. 
Our framework can provide scientists not only the ability to choose between competing existing hypotheses but to also discover new ones in a fundamental Bayesian manner.

\section*{Acknowledgments}
We thank the members of our MSEAS group for insightful discussions, including A.\;Babu for proofreading as well as C.\ Mirabito and P.J.\ Haley Jr.\ for help with the figures. We are grateful to
the Office of Naval Research for partial support under grants N00014-19-1-2693 (IN-BDA) and 
N00014-20-1-2023 (MURI ML-SCOPE), and to Sea Grant and NOAA for support under grant  NA18OAR4170105 (BIOMAPS), all  to the
Massachusetts Institute of Technology. We also thank MathWorks and the Mechanical Eng.\ Dept.\ at MIT for awarding a competitive 2020-2021 MathWorks Mechanical Eng.\ Fellowship for A.G.

\appendix
\section{Dynamically Orthogonal (DO) Equations} \label{app:DO}

In this appendix, we derive the dynamically orthogonal (DO) equations \cite{sapsis_lermusiaux_PD2009, sapsis_lermusiaux_PHYSD2012, feppon_lermusiaux_SIREV2018, feppon_lermusiaux_SIMAX2018a} for optimal reduced-order probabilistic evolution of high-dimensional stochastic dynamical systems with regular and new parametric uncertainties for
known, uncertain, and unknown model formulations.

The general stochastic nonlinear dynamical system 
%which encompasses the different model uncertainty scenarios encountered in this paper 
governs the dynamics of $\mbs{\phi}(\mbs{x},t;\omega): \mathbb{R}^n\times [0, T]\rightarrow \mathbb{R}^{N_\phi}$, the stochastic state vector comprising $N_{\phi}$ physical-biogeochemical fields defined on a spatial domain $\mathcal{D}$, where $\omega$ is the realization index belonging to a measurable sample space $\Omega$. It is is given by,
\begin{equation}
\label{eq:stoch dyn model}
\begin{split}
\frac{\Par \mbs{\phi}(\mbs{x},t; \omega)}{\Par t}
&= \mathcal{L}[\mbs{\phi}(\mbs{x},t; \omega),\mbs{\theta}(\omega), \mbs{\beta}(\omega), \mbs{x}, t; \omega] + \widehat{\mathcal{L}}[\mbs{\phi}(\mbs{x},t; \omega),\mbs{\alpha}(\omega), \mbs{x}, t; \omega] + \widetilde{\mathcal{L}}[\mbs{\phi}(\mbs{x},t; \omega),\mbs{\gamma}(\omega), \mbs{x}, t; \omega], \\
& \hspace{0.5\textwidth} \mbs{x}\in \mathcal{D}, \: t\in [0, T], \: \omega\in\Omega \;, \\
\text{with} \quad & \mbs{\phi}(\mbs{x}, 0;\omega) = \mbs{\phi}_o(\mbs{x}; \omega)\,, \\
 \text{and} \quad & \mathcal{B}[\mbs{\phi}(\mbs{x}, t; \omega)] = \mbs{b}(\mbs{x}, t; \omega), \; \mbs{x} \in \partial\mathcal{D}, \: t\in [0, T], \: \omega\in\Omega \;,
\end{split}
\end{equation}
where $\mbs{\phi}_o(\mbs{x}; \omega)$, $\mathcal{B}$, and $\mbs{b}(\mbs{x}, t; \omega)$ are the stochastic initial conditions, boundary condition operators, and boundary values respectively.
The functional form of the first dynamics term
$\mathcal{L}[\bullet]$ is assumed to be known, however it contains $N_\theta$ uncertain regular parameters $\mbs{\theta}(\omega)$.
The second term $\mathcal{\widehat{L}}[\bullet]$ is uncertain: it belongs to a family of candidate functions, parameterized using $N_\alpha$ stochastic formulation parameters $\mbs{\alpha}(\omega)$. $\widehat{\mathcal{L}}[\bullet]$ can also contain uncertain regular parameters $\mbs{\theta}(\omega)$. 
The candidate models of different complexities are combined using $N_\beta$ stochastic complexity parameters $\mbs{\beta}(\omega)$. The $\beta_k(\omega)$'s multiplied with the original state variables (as described in Sect.\ \ref{Sec: Special Stochastic Parameters}) are absorbed into $\phi_i$'s and not explicitly shown; however, $\beta_k(\omega)$'s can still appear on the right-hand-side (RHS) in $\mathcal{L}[\bullet]$ and $\widehat{\mathcal{L}}[\bullet]$. 
The third term $\widetilde{\mathcal{L}}[\bullet]$ has a functional form completely unknown, and is parameterized using $N_\gamma$ stochastic expansion parameters $\mbs{\gamma}(\omega)$. 

%For efficient reduced-dimension probabilistic evolution of high-dimensional systems, 
The DO methodology %\cite{sapsis_lermusiaux_PD2009, sapsis_lermusiaux_PHYSD2012, feppon_lermusiaux_SIREV2018, feppon_lermusiaux_SIMAX2018a} 
employs a generalized, time-dependent Karhunen-Lo\'{e}ve decomposition of $\mbs{\phi}(\mbs{x},t;\omega)$ into a mean, $\bar{\mbs{\phi}}(\mbs{x},t)\in \mathbb{R}^{N_\phi}$, $N_s$ deterministic modes, $\mbs{\tilde{\phi}}_i(\mbs{x},t)\in \mathbb{R}^{N_\phi}$, and stochastic coefficients, $Y_i(t;\omega)\in \mathbb{R}$,
%up to arbitrary precision,
\begin{equation}
\label{eq:KL decomp}
	\mbs{\phi}(\mbs{x},t;\omega) =
	\bar{\mbs{\phi}}(\mbs{x},t) + \sum_{i=1}^{N_s} Y_i(t;\omega) \mbs{\tilde{\phi}}_i(\mbs{x},t) \;.
\end{equation}
We define the stochastic subspace $\mbs{V_{S}} = span\{\mbs{\tilde{\phi}}_i(\mbs{x},t)\}_{i=1}^{N_s}$ as the linear space spanned by the $N_s$ deterministic modes that evolve to capture the dominant uncertainty in $\mbs{V_{S}}$. In general, the number of modes $N_s$
is orders of magnitude smaller than the dimension of the discretized state variables or of the domain grid $N_x$, i.e.\ $N_s\ll N_\phi N_x$.
Similarly, uncertain regular and new parameters are split into means and deviations, $\mbs{\theta}(\omega) = \bar{\mbs{\theta}} + \mbs{\mathfrak{D}}^{\theta}(\omega)$, $\mbs{\alpha}(\omega) = \bar{\mbs{\alpha}} + \mbs{\mathfrak{D}}^{\alpha}(\omega)$, and $\mbs{\beta}(\omega) = \bar{\mbs{\beta}} + \mbs{\mathfrak{D}}^{\beta}(\omega)$.
% , and $\mbs{\gamma}(\omega) = \bar{\mbs{\gamma}} + \mbs{E}^{\gamma}(\omega)$.

Nonlinear terms on the RHS are handled using local Taylor series expansion around the statistical means of states and parameters. We use first order Taylor series expansion for the $\mathcal{L}[\bullet]$ and $\mathcal{\widehat{L}}[\bullet]$ terms,
\begin{equation}
    \begin{split}
        \mathcal{L}[\mbs{\phi}(\mbs{x},t; \omega),\mbs{\theta}(\omega), \mbs{\beta}(\omega), \mbs{x}, t; \omega] & \approx \mathcal{L}|_{\begin{smallmatrix}\mbs{\phi} = \mbs{\bar{\phi}}, \\\mbs{\theta} = \mbs{\bar{\theta}}, \\\mbs{\beta} = \mbs{\bar{\beta}}\end{smallmatrix}} +
        \frac{\partial\mathcal{L}}{\partial \mbs{\phi}}\bigg|_{\begin{smallmatrix}\mbs{\phi} = \mbs{\bar{\phi}}, \\\mbs{\theta} = \mbs{\bar{\theta}}, \\\mbs{\beta} = \mbs{\bar{\beta}}\end{smallmatrix}} \sum_{i=1}^{N_s}\mbs{\tilde{\phi}}_i Y_i + 
        \sum_{i=1}^{N_{\theta}}\frac{\partial\mathcal{L}}{\partial \theta_i}\bigg|_{\begin{smallmatrix}\mbs{\phi} = \mbs{\bar{\phi}}, \\\mbs{\theta} = \mbs{\bar{\theta}}, \\\mbs{\beta} = \mbs{\bar{\beta}}\end{smallmatrix}}  \mathfrak{D}_i^{\theta} +
        \sum_{i=1}^{N_{\beta}}\frac{\partial\mathcal{L}}{\partial \beta}\bigg|_{\begin{smallmatrix}\mbs{\phi} = \mbs{\bar{\phi}}, \\\mbs{\theta} = \mbs{\bar{\theta}}, \\\mbs{\beta} = \mbs{\bar{\beta}}\end{smallmatrix}}  \mathfrak{D}_i^{\beta}
        \;, \\
        \mathcal{\widehat{L}}[\mbs{\phi}(\mbs{x},t; \omega),\mbs{\alpha}(\omega), \mbs{x}, t; \omega] & \approx \mathcal{\widehat{L}}|_{\begin{smallmatrix}\mbs{\phi} = \mbs{\bar{\phi}}, \\\mbs{\alpha} = \mbs{\bar{\alpha}}\end{smallmatrix}} +
        \frac{\partial\mathcal{\widehat{L}}}{\partial \mbs{\phi}}\bigg|_{\begin{smallmatrix}\mbs{\phi} = \mbs{\bar{\phi}}, \\\mbs{\alpha} = \mbs{\bar{\alpha}}\end{smallmatrix}} \sum_{i=1}^{N_s}\mbs{\tilde{\phi}}_i Y_i + 
        \sum_{i=1}^{N_{\alpha}}\frac{\partial\mathcal{\widehat{L}}}{\partial \alpha_i}\bigg|_{\begin{smallmatrix}\mbs{\phi} = \mbs{\bar{\phi}}, \\\mbs{\alpha} = \mbs{\bar{\alpha}}\end{smallmatrix}}  \mathfrak{D}_i^{\alpha} \;.
    \end{split}
\end{equation}
Using a higher-order polynomial approximation leads to higher accuracy for the DO evolution, but also increases computational costs. 
For analyses of the scaling of computational costs with the order of polynomial approximation, we refer to \citep{gupta_MSThesis2016,gupta_ali_lermusiaux_2016prep}.
Handling the $\mathcal{\widetilde{L}}[\bullet]$ term is less straightforward because of the need to evaluate the interval in which each state realization value lies at all discrete times and spatial locations in the domain (see Sect.\ \ref{sec: Piece-wise Linear Function Approximations}). 
Thus, for maximum accuracy, we directly evaluate the $\mathcal{\widetilde{L}}[\bullet]$ terms for every state realization in a Monte-Carlo fashion. 
%However, this could potentially be circumvented and made more efficient in the future by using techniques such as clustering \cite{charous_clustering}.
%
To increase efficiency without much loss of accuracy, recent techniques such as dynamic clustering \cite{humara_MSThesis2020,humara_et_al_Oceans2022,charous_clustering} %\AG{Please check if the arron-clustering citation is the correct one.} 
could also be used.

To derive the DO equations, we substitute the KL decomposition (Eq.~\ref{eq:KL decomp}) into the stochastic system (Eq.~\ref{eq:stoch dyn model}). To obtain an efficient closed-form dynamical system, without loss of generality, we impose 
%additional constraints on the modes. As shown in Sapsis and Lermusiaux, 2009 (\cite{sapsis_lermusiaux_PD2009}), an appropriate constraint is 
the DO condition \cite{sapsis_lermusiaux_PD2009}: the rate of change of the stochastic subspace is orthogonal to itself,
 \begin{equation}
 \label{eq:DO cond}
 \frac{d\mbs{V_{S}}}{dt} \perp \mbs{V_{S}} \Leftrightarrow \left<\frac{\Par \mbs{\tilde{\phi}}_i(\mbs{x},t)}{\Par t}  , \mbs{\tilde{\phi}}_j(\mbs{x},t) \right> = 0  \quad \forall i,j \in \{1,...,N_s\} \,,
 \end{equation}
 where $\langle\mbs{a}, \mbs{b}\rangle = \int_\mathcal{D}\sum_i(a^ib^i)d\mathcal{D}$ denotes the spatial inner-product of vectors $\mbs{a} = [a^1, a^2,...]^T$ and $\mbs{b} = [b^1, b^2, ...]^T$.
 %defined by $\langle\mbs{a}, \mbs{b}\rangle = \int_\mathcal{D}\sum_i(a^ib^i)d\mathcal{D}$. 
 Note that the DO condition (\ref{eq:DO cond}) also implies the preservation of orthogonality for the basis $\{\mbs{\tilde{\phi}}_i(\mbs{x},t)\}_{i=1}^{N_s}$ themselves \cite{ueckermann_et_al_JCP2013}.
 %since,
 %\begin{eqnarray}
 %\frac{\Par }{\Par t} \langle\mbs{\tilde{\phi}}_i(\mbs{x},t), \mbs{\tilde{\phi}}_j(\mbs{x},t)\rangle= \left\langle\frac{\Par \mbs{\tilde{\phi}}_i(\mbs{x},t)}{\Par t}, \mbs{\tilde{\phi}}_j(\mbs{x},t)\right\rangle + \left\langle\mbs{\tilde{\phi}}_i(\mbs{x},t), \frac{\Par \mbs{\tilde{\phi}}_j(\mbs{x},t)}{\Par t}\right\rangle = 0, \nonumber\\ \forall i, j \in \{1, ..., N_s\} \,.
 %\end{eqnarray}
 %Substituting the expansion (Eq.~\ref{eq:KL decomp}) into the stochastic dynamical model (Eq.~\ref{eq:stoch dyn model}) with the help of DO condition (Eq. \ref{eq:DO cond}), a unique set of independent evolution equations can be derived for mean, modes, and stochastic coefficients. These are the DO evolution equations (omitting function arguments for brevity),
 %
 Substituting Eq.\ (\ref{eq:KL decomp}) into  Eq.\ (\ref{eq:stoch dyn model}), and using Eq.\ (\ref{eq:DO cond}) and the above schemes for nonlinear terms, we derive independent evolution equations for the DO mean, modes, and stochastic coefficients. These are the DO evolution equations (omitting function arguments for brevity),
 \begin{equation}
    \begin{split}
        \frac{\partial \bar{\mbs{\phi}}}{\partial t} &= \mathcal{L}|_{\begin{smallmatrix}\mbs{\phi} = \mbs{\bar{\phi}}, \\\mbs{\theta} = \mbs{\bar{\theta}}, \\\mbs{\beta} = \mbs{\bar{\beta}}\end{smallmatrix}} 
        +
        \mathcal{\widehat{L}}|_{\begin{smallmatrix}\mbs{\phi} = \mbs{\bar{\phi}}, \\
        \mbs{\theta} = \mbs{\bar{\theta}}, \\
        \mbs{\alpha} = \mbs{\bar{\alpha}}, \\
        \mbs{\beta} = \mbs{\bar{\beta}} \end{smallmatrix}}
        +
        \mathbb{E}[\mathcal{\widetilde{L}}]\,, \\
        \frac{\partial \tilde{\mbs{\phi}}_i}{\partial t} &= \mbs{Q}_i - \sum_{j=1}^{N_s}\langle \mbs{Q}_i, \tilde{\mbs{\phi}}_j \rangle \tilde{{\mbs{\phi}}}_j \,, \\
        \frac{d Y_i}{dt} &= \sum_{m=1}^{N_s} \langle \mbs{F}_m, \tilde{\mbs{\phi}}_i \rangle Y_m 
        + \sum_{m=1}^{N_{\theta}} \left\langle \frac{\partial\mathcal{L}}{\partial \theta_i}\bigg|_{\begin{smallmatrix}\mbs{\phi} = \mbs{\bar{\phi}}, \\\mbs{\theta} = \mbs{\bar{\theta}}, \\\mbs{\beta} = \mbs{\bar{\beta}}\end{smallmatrix}} , \tilde{\mbs{\phi}}_i \right\rangle \mathfrak{D}_m^{\theta} 
        + \sum_{m=1}^{N_{\beta}} \left\langle \frac{\partial\mathcal{L}}{\partial \beta}\bigg|_{\begin{smallmatrix}\mbs{\phi} = \mbs{\bar{\phi}}, \\\mbs{\theta} = \mbs{\bar{\theta}}, \\\mbs{\beta} = \mbs{\bar{\beta}}\end{smallmatrix}}  , \tilde{\mbs{\phi}}_i \right\rangle \mathfrak{D}_m^{\beta}
        \\
        & + \sum_{m=1}^{N_{\theta}} \left\langle \frac{\partial\mathcal{\widehat{L}}}{\partial \theta_i}\bigg|_{\begin{smallmatrix}\mbs{\phi} = \mbs{\bar{\phi}}, \\
        \mbs{\theta} = \mbs{\bar{\theta}}, \\
        \mbs{\alpha} = \mbs{\bar{\alpha}}, \\
        \mbs{\beta} = \mbs{\bar{\beta}} \end{smallmatrix}} , \tilde{\mbs{\phi}}_i \right\rangle \mathfrak{D}_m^{\theta}
        + \sum_{m=1}^{N_{\alpha}} \left\langle \frac{\partial\mathcal{\widehat{L}}}{\partial \alpha_i}\bigg|_{\begin{smallmatrix}\mbs{\phi} = \mbs{\bar{\phi}}, \\
        \mbs{\theta} = \mbs{\bar{\theta}}, \\
        \mbs{\alpha} = \mbs{\bar{\alpha}}, \\
        \mbs{\beta} = \mbs{\bar{\beta}} \end{smallmatrix}} , \tilde{\mbs{\phi}}_i \right\rangle \mathfrak{D}_m^{\alpha}
        + \sum_{m=1}^{N_{\beta}} \left\langle \frac{\partial\mathcal{\widehat{L}}}{\partial \beta_i}\bigg|_{\begin{smallmatrix}\mbs{\phi} = \mbs{\bar{\phi}}, \\
        \mbs{\theta} = \mbs{\bar{\theta}}, \\
        \mbs{\alpha} = \mbs{\bar{\alpha}}, \\
        \mbs{\beta} = \mbs{\bar{\beta}} \end{smallmatrix}} , \tilde{\mbs{\phi}}_i \right\rangle \mathfrak{D}_m^{\beta}
        \\
        & +
        \left\langle \mathcal{\widetilde{L}} - \mathbb{E}[\mathcal{\widetilde{L}}], \tilde{\mbs{\phi}}_i \right\rangle
        \,,
    \end{split}
    \label{eq:DO_Eq1 appendix}
\end{equation}
where,
\begin{equation}
    \begin{split}
       \mbs{Q}_i &=  \frac{\partial\mathcal{L}}{\partial \mbs{\phi}}\bigg|_{\begin{smallmatrix}\mbs{\phi} = \mbs{\bar{\phi}}, \\\mbs{\theta} = \mbs{\bar{\theta}}, \\\mbs{\beta} = \mbs{\bar{\beta}}\end{smallmatrix}} \mbs{\tilde{\phi}}_i 
       + \sum_{j=1}^{N_s}\sum_{n=1}^{N_{\theta}}C_{Y_iY_j}^{-1}C_{\mathfrak{D}_n^{\theta}Y_j}\frac{\partial \mathcal{L}}{\partial \theta_n}\bigg|_{\begin{smallmatrix}\mbs{\phi} = \mbs{\bar{\phi}}, \\\mbs{\theta} = \mbs{\bar{\theta}}, \\\mbs{\beta} = \mbs{\bar{\beta}}\end{smallmatrix}} 
       + \sum_{j=1}^{N_s}\sum_{n=1}^{N_{\beta}}C_{Y_iY_j}^{-1}C_{\mathfrak{D}_n^{\beta}Y_j}\frac{\partial \mathcal{L}}{\partial \beta_n}\bigg|_{\begin{smallmatrix}\mbs{\phi} = \mbs{\bar{\phi}}, \\\mbs{\theta} = \mbs{\bar{\theta}}, \\\mbs{\beta} = \mbs{\bar{\beta}}\end{smallmatrix}} \\
       & +
       \frac{\partial\mathcal{\widehat{L}}}{\partial \mbs{\phi}}\bigg|_{\begin{smallmatrix}\mbs{\phi} = \mbs{\bar{\phi}}, \\
        \mbs{\theta} = \mbs{\bar{\theta}}, \\
        \mbs{\alpha} = \mbs{\bar{\alpha}}, \\
        \mbs{\beta} = \mbs{\bar{\beta}} \end{smallmatrix}} \mbs{\tilde{\phi}}_i 
       +
       \sum_{j=1}^{N_s}\sum_{n=1}^{N_{\theta}}C_{Y_iY_j}^{-1}C_{\mathfrak{D}_n^{\theta}Y_j}\frac{\partial \mathcal{\widehat{L}}}{\partial \theta_n}\bigg|_{\begin{smallmatrix}\mbs{\phi} = \mbs{\bar{\phi}}, \\
        \mbs{\theta} = \mbs{\bar{\theta}}, \\
        \mbs{\alpha} = \mbs{\bar{\alpha}}, \\
        \mbs{\beta} = \mbs{\bar{\beta}} \end{smallmatrix}}
       +
       \sum_{j=1}^{N_s}\sum_{n=1}^{N_{\alpha}}C_{Y_iY_j}^{-1}C_{\mathfrak{D}_n^{\alpha}Y_j}\frac{\partial \mathcal{\widehat{L}}}{\partial \alpha_n}\bigg|_{\begin{smallmatrix}\mbs{\phi} = \mbs{\bar{\phi}}, \\
        \mbs{\theta} = \mbs{\bar{\theta}}, \\
        \mbs{\alpha} = \mbs{\bar{\alpha}}, \\
        \mbs{\beta} = \mbs{\bar{\beta}} \end{smallmatrix}} 
       \\
       & + \sum_{j=1}^{N_s}\sum_{n=1}^{N_{\beta}}C_{Y_iY_j}^{-1}C_{\mathfrak{D}_n^{\beta}Y_j}\frac{\partial \mathcal{\widehat{L}}}{\partial \beta_n}\bigg|_{\begin{smallmatrix}\mbs{\phi} = \mbs{\bar{\phi}}, \\
        \mbs{\theta} = \mbs{\bar{\theta}}, \\
        \mbs{\alpha} = \mbs{\bar{\alpha}}, \\
        \mbs{\beta} = \mbs{\bar{\beta}} \end{smallmatrix}}
       +
       \sum_{j=1}^{N_s} C_{Y_iY_j}^{-1} \mathbb{E}[Y_j\mathcal{\widetilde{L}}]
       \,, \\
       \mbs{F}_m &= \frac{\partial\mathcal{L}}{\partial \mbs{\phi}}\bigg|_{\begin{smallmatrix}\mbs{\phi} = \mbs{\bar{\phi}}, \\\mbs{\theta} = \mbs{\bar{\theta}}, \\\mbs{\beta} = \mbs{\bar{\beta}}\end{smallmatrix}} \mbs{\tilde{\phi}}_m  + \frac{\partial\mathcal{\widehat{L}}}{\partial \mbs{\phi}}\bigg|_{\begin{smallmatrix}\mbs{\phi} = \mbs{\bar{\phi}}, \\
        \mbs{\theta} = \mbs{\bar{\theta}}, \\
        \mbs{\alpha} = \mbs{\bar{\alpha}}, \\
        \mbs{\beta} = \mbs{\bar{\beta}} \end{smallmatrix}} \mbs{\tilde{\phi}}_m \,,
    \end{split}
    \label{eq:DO_Eq2 appendix}
\end{equation}
and $\mathbb{E}[\bullet]$ represents the expectation operator,  $C^{-1}_{Y_iY_j}$ the inverse of the cross-covariance between the $i^{th}$ and $j^{th}$ stochastic coefficients, and $C_{Y_iY_j}$ is given by, 
 \begin{eqnarray}
 C_{Y_i,Y_j} = \mathbb{E}[Y_i(t;\omega)Y_j(t;\omega)] \;.
 \end{eqnarray} 
As discussed in (\citep{gupta_MSThesis2016,gupta_ali_lermusiaux_2016prep,gupta_PhDThesis2022}), 
the boundary conditions are also obtained by inserting DO decompositions in Eq.\ (\ref{eq:stoch dyn model}). This yields for the mean fields, 
 \begin{eqnarray}
\mathcal{B}[\mbs{\bar{\phi}}(\mbs{x},t)]|_{\mbs{x}\in \Par \mathcal{D}} = \mathbb{E}[\mbs{b}(\mbs{x},t; \omega)] \;,
 \end{eqnarray}
and for the modes fields,
\begin{eqnarray}
  \mathcal{B}[\mbs{\tilde{\phi}}_i(\mbs{x},t)]|_{\mbs{x}\in \Par \mathcal{D}} = \sum_{j=1}^{N_s}\mathbb{E}[Y_j(t;\omega)\mbs{b}(\mbs{x},t; \omega)]C^{-1}_{Y_iY_j} \;.
\end{eqnarray}
Similarly, the initial conditions in Eq.\ (\ref{eq:KL decomp}) are approximated by
using the DO decomposition of
the initial stochastic fields $\mbs{\phi}_o(\mbs{x};\omega)$. 
%Complete derivation of the DO equations, along with discussion on computational cost saving can be found in several of the existing papers on DO methodology \cite{sapsis_lermusiaux_PD2009, sapsis_lermusiaux_PHYSD2012, feppon_lermusiaux_SIREV2018, feppon_lermusiaux_SIMAX2018a, feppon_lermusiaux_PhysD2021}. 
%For a more detailed discussion on handling stochastic boundary condition, please refer to Gupta, 2016~\&~2022 (\citep{gupta_MSThesis2016, gupta_PhDThesis2022}).

Finally, the stochastic dynamical system (Eq.\ \ref{eq:stoch dyn model}) is multivariate and we normalize the spatial inner-product operator using appropriate scaling, so as to account for the different uncertainty magnitudes of state variables \cite{lermusiaux_MWR1999,lermusiaux_et_al_QJRMS2000,
lermusiaux_JAOT2002,subramani_PhDThesis2018,subramani_lermusiaux_prep}. 
For the present DO modes $\mbs{\tilde{\phi}}_i(\mbs{x},t) = [\tilde{\phi}_i^1(\mbs{x},t), ..., \tilde{\phi}_i^{N_\phi}(\mbs{x},t)]$,  
the normalized spatial inner-product is,
\begin{equation}
\begin{split}
\langle\mbs{\tilde{\phi}}_i(\mbs{x},t), \mbs{\tilde{\phi}}_j(\mbs{x},t)\rangle = \frac{1}{|\mathcal{D}|}\int_\mathcal{D}\sum_{k=1}^{N_\phi}\left(\frac{1}{\sigma_{nd, k}^2}\tilde{\phi}_i^k \tilde{\phi}_j^k\right)d\mathcal{D} \;,
\end{split}
\end{equation}
where $|\mathcal{D}|$ is the volume (area) of the domain and $\sigma_{nd, k}$ is the expected volume-averaged standard deviations of state variable $k$. These $\sigma_{nd, k}$'s normalize the relative weights given to state variables in the inner-product.

\section{Gaussian Mixture Model (GMM)-DO Filter} \label{app:GMM-DO}
 %The DO methodology introduced in the appendix \ref{app:DO} helps to effectively evolve uncertainty between assimilation steps, i.e.\ provide prior probability distributions for the state variables. For the assimilation step, we employ a framework based on Gaussian Mixture Models (in order to preserve non-Gaussian statistics of state variables) and Bayes law, called the GMM-DO filter. 

The GMM-DO filter \cite{sondergaard_lermusiaux_MWR2013_part1,sondergaard_lermusiaux_MWR2013_part2}
consists of a recursive succession in time of two steps: a forecast DO step (\ref{app:DO}) and a Bayesian update step. 
Using the affine transformation between stochastic coefficients and state variables (Eq.~\ref{eq:KL decomp}), the GMM-DO filter obtains the Bayesian update of the state variable distribution through an equivalent update of the stochastic coefficient distribution. 
The result is an efficient reduced-dimension Bayesian state variable inference \cite{sondergaard_lermusiaux_MWR2013_part1}.
Next, we assume the DO coefficients of the discrete state variables are augmented with the regular and new parameters (see \ref{app:state_aug}).

For the Bayesian update, the GMM-DO filter first represents the prior probability distribution of the stochastic coefficients in the DO subspace using a GMM,
\begin{eqnarray}
p_{\mbs{Y}^f} (Y^f) \approx \sum_{j=1}^{N_{\text{GMM}}} \pi^f_{\mbs{Y},j} \times \mathcal{N}(Y^f;\boldsymbol{\mu}^f_{\mbs{Y},j}, \mbs{\Sigma}^f_{\mbs{Y},j}) \qquad \forall Y^f\in \mathbb{R}^{N_s} \,,
\label{eq:DO-coef-GMM-prior}
\end{eqnarray}
where $N_{\text{GMM}}$ is the to-be-determined number of GMM components, $\pi^f_{\mbs{Y},j} \in [0,1]$ the $j^{th}$ component weight (also $\sum_{j=1}^{N_{\text{GMM}}}\pi^f_{\mbs{Y},j}=1$), $\boldsymbol{\mu}^f_{\mbs{Y},j}$ the $j^{th}$ component mean vector, and $\boldsymbol{\Sigma}^f_{\mbs{Y},j}$ the $j^{th}$ component covariance matrix. This approximation is found by performing a semiparametric fit to the Monte-Carlo samples used to numerically evolve the stochastic coefficients. Specifically, the expectation-maximization (EM) algorithm for GMMs \cite{bilmes1998gentle} is used to find maximum likelihood estimate for the GMM parameters $\pi^f_{\mbs{Y},j}$, $\boldsymbol{\mu}^f_{\mbs{Y},j}$ and $\boldsymbol{\Sigma}^f_{\mbs{Y},j}$, while the selection of the number of GMM components ($N_{\text{GMM}}$) is determined by the Bayesian Information Criterion (BIC) \cite{stoica2004model} by successively fitting GMMs of varying complexity (e.g.\ GMM = 1, 2, 3, ...) until a minimum of the BIC is obtained. 

Using the Gaussian observation model (Eq.~\ref{eq:observation model}), the GMM for the prior stochastic coefficients is updated by Bayesian update, using conjugacy \cite{sondergaard_lermusiaux_MWR2013_part1}. The resulting GMM of the posterior stochastic coefficients is,
\begin{eqnarray}
\label{eq: post GMM dist stoch coeff}
p_{\mbs{Y}^a} (Y^a) \approx \sum_{j=1}^{N_{\text{GMM}}} \pi^a_{\mbs{Y},j} \times \mathcal{N}(Y^a;\boldsymbol{\mu}^a_{\mbs{Y},j}, \mbs{\Sigma}^a_{\mbs{Y},j}) \;, \qquad \forall Y^a\in \mathbb{R}^{N_s} \;,
\end{eqnarray}
where,
\begin{eqnarray}
\begin{split}
\pi_{\mbs{Y},j}^a &= \frac{\pi_{\mbs{Y},j}^f \times \mathcal{N}(\mbs{\tilde{y}};\mbs{\tilde{H}}\boldsymbol{\mu}^f_{\mbs{Y},j}, \mbs{\tilde{H}}\mbs{\Sigma}^f_{\mbs{Y},j}\mbs{\tilde{H}}^T + \mbs{R})}{\sum_{m=1}^{N_{\text{GMM}}}\pi_{\mbs{Y},m}^f \times \mathcal{N}(\mbs{\tilde{y}};\mbs{\tilde{H}}\boldsymbol{\mu}^f_{\mbs{Y},m}, \mbs{\tilde{H}}\mbs{\Sigma}^f_{\mbs{Y},m}\mbs{\tilde{H}}^T + \mbs{R})} \;, \qquad \forall j\in \{1,...,N_{\text{GMM}}\} \;, \\
\boldsymbol{\mu}^a_{\mbs{Y},j} &= \boldsymbol{\hat{\mu}}^a_{\mbs{Y},j} - \sum_{m=1}^{N_{\text{GMM}}} \pi_{\mbs{Y},m}^a \times \boldsymbol{\hat{\mu}}^a_{\mbs{Y},m} \;, \qquad \forall j\in \{1,...,N_{\text{GMM}}\} \;, \\
\mbs{\Sigma}^a_{\mbs{Y},j} &= (\mbs{I} - \mbs{\tilde{K}}_j\mbs{\tilde{H}})\mbs{\Sigma}^f_{\mbs{Y},j} \;, \qquad \forall j\in \{1,...,N_{\text{GMM}}\} \;,
\end{split}
\end{eqnarray}
with the following definitions,
\begin{eqnarray}
\label{innovation vector def chap2}
\begin{split}
\mbs{\tilde{H}} &= \mbs{H} \boldsymbol{\tilde{\Phi}} \;, \\
\mbs{\tilde{y}} &= \mbs{y} - \mbs{H}\mbs{\bar{\Phi}}^f \;, \\
\boldsymbol{\hat{\mu}}^a_{\mbs{Y},j} &= \boldsymbol{\mu}^f_{\mbs{Y},j} + \mbs{\tilde{K}}_j(\mbs{\tilde{y}} - \mbs{\tilde{H}}\boldsymbol{\mu}^f_{\mbs{Y},j}) \;, \qquad \forall j\in \{1,...,N_{\text{GMM}}\}  \;, \\
\mbs{\tilde{K}}_j &= \mbs{\Sigma}^f_{\mbs{Y},j}\mbs{\tilde{H}}^T(\mbs{\tilde{H}}\mbs{\Sigma}^f_{\mbs{Y},j}\mbs{\tilde{H}}^T + \mbs{R})^{-1} \equiv \boldsymbol{\tilde{{\Phi}}}^T \mbs{K}_j \;, \qquad \forall j\in \{1,...,N_{\text{GMM}}\} \;.
\end{split}
\end{eqnarray}
%
%Using an affine transformation, we can show that the posterior GMM stochastic coefficient distribution (Eq.~\ref{eq: post GMM dist stoch coeff}) is equivalent to the posterior GMM state space distribution, if the state vector mean is updated according to,
%
The posterior GMM state space distribution is obtained from Eq.\ (\ref{eq: post GMM dist stoch coeff}) by updating the state vector mean,
\begin{eqnarray}
\mbs{\bar{\Phi}}^a = \mbs{\bar{\Phi}}^f + \boldsymbol{\tilde{{\Phi}}}\sum_{j=1}^{N_{\text{GMM}}}\pi_{\mbs{Y},j}^a\times \boldsymbol{\hat{\mu}}^a_{\mbs{Y},j} \;.
\end{eqnarray}
In the GMM-DO update step, no matrices of size larger than $N_{\phi}N_x\times S\ll (N_{\phi} N_x)^2$ are manipulated. The GMM-DO filter is thus computationally feasible for high-dimensional multivariate PDE systems (Eq.\ \ref{eq:stoch dyn model}).

At last, new Monte-Carlo samples are drawn from the posterior GMM  (Eq.~\ref{eq: post GMM dist stoch coeff}) and  dynamically evolved using the DO evolution Eqs.\ (\ref{eq:DO_Eq1 appendix}) until new observations come in and the filtering process is repeated.
%Hence, the GMM-DO filter provides an efficient and computationally feasible Bayesian inference methodology for high-dimensional, non-linear stochastic dynamical systems. 

\section{State Augmentation} \label{app:state_aug}
%\subsection{State Augmentation}
%\label{sec:State Augmentation}

To simultaneously estimate the uncertain parameters and states, we employ state augmentation \cite{gelb1974applied}. 
% This technique premises itself on the fact that observations are correlated to states, and states are correlated to parameters. Thus, there exists indirect correlation between observations and parameters, and hence, we can treat all the uncertain parameters as additional state variables. 
We start by decomposing the stochastic regular parameters ($\mbs{\theta}(\omega) \in \mathbb{Re}^{N_\theta}$), formulation and complexity parameters ($\mbs{\alpha}(\omega) \in \mathbb{Re}^{N_\alpha}$ and $\mbs{\beta}(\omega) \in \mathbb{Re}^{N_\beta}$), and expansion parameters ($\mbs{\gamma}(\omega) \in \mathbb{Re}^{N_\gamma}$) into their means and uncertain parts, 
%\PFJL{The symbol $\mbs{E^{\alpha}}$ likely needs to be replaced, see same comment in main text.} 
%\AG{Done. Also in the notation table}
\begin{equation}
\label{eq: parameter vector decomposition}
\begin{split}
\boldsymbol{\theta}(\omega) &= \boldsymbol{\bar{\theta}} + \mbs{\mathfrak{D}^{\theta}}(\omega) \;, \\
\boldsymbol{\alpha}(\omega) &= \boldsymbol{\bar{\alpha}} + \mbs{\mathfrak{D}^{\alpha}}(\omega) \;,\\
\boldsymbol{\beta}(\omega) &= \boldsymbol{\bar{\beta}} + \mbs{\mathfrak{D}^{\beta}}(\omega) \;,\\
\boldsymbol{\gamma}(\omega) &= \boldsymbol{\bar{\gamma}} + \mbs{\mathfrak{D}^{\gamma}}(\omega) \;.
\end{split}
\end{equation}
The augmented state vector can be written as,
\begin{equation}
\mbs{\Phi}_{\text{aug}}(t;\omega) = \begin{bmatrix} \boldsymbol{\theta}(\omega) \\
\boldsymbol{\alpha}(\omega) \\
\boldsymbol{\beta}(\omega) \\
\boldsymbol{\gamma}(\omega) \\
\mbs{\Phi}(t;\omega) 
\end{bmatrix} \in \mathbb{R}^{N_{\phi}N_x + N_\theta + N_\alpha + N_\beta + N_\gamma} \;.
\end{equation}
Now, let us write the DO decomposition for this new augmented system. We define a new coefficient matrix in which each parameter uncertainty amounts to an additional scalar stochastic coefficient,
\begin{equation}
\label{eq: augmented stoch coeff}
\mbs{DY} (t;\omega) = \begin{bmatrix}
\mbs{\mathfrak{D}^{\theta}} (\omega) | \mbs{\mathfrak{D}^{\alpha}} (\omega) | \mbs{\mathfrak{D}^{\beta}} (\omega) | \mbs{\mathfrak{D}^{\gamma}} (\omega) | \mbs{Y}(t;\omega)
\end{bmatrix} \in \mathbb{R}^{N_s + N_\theta + N_\alpha + N_\beta + N_\gamma} \;,
\end{equation}
a new modes matrix with parameters having unit modes,
\begin{equation}
\label{eq: augmented mode}
\boldsymbol{\tilde{\Phi}}_{\text{aug}} (t) = \begin{bmatrix}	\mbs{I} & \mbs{0} \\ \mbs{0} & \mbs{\tilde{{\Phi}}}(t)	\end{bmatrix} \in \mathbb{R}^{(N_{\phi} N_x + N_\theta + N_\alpha + N_\beta + N_\gamma) \times (N_s + N_\theta + N_\alpha + N_\beta + N_\gamma)} \;,
\end{equation}
and a new augmented mean vector,
\begin{equation}
\label{eq: augmented mean}
\mbs{\bar{\Phi}}_{\text{aug}}(t) = \begin{bmatrix} \boldsymbol{\bar{\theta}} \\
\boldsymbol{\bar{\alpha}} \\
\boldsymbol{\bar{\beta}} \\
\boldsymbol{\bar{\gamma}} \\
\mbs{\bar{\Phi}}(t) 
\end{bmatrix} \in \mathbb{R}^{N_{\phi}N_x + N_\theta + N_\alpha + N_\beta + N_\gamma} \;.
\end{equation}
Thus, the DO decomposition of the augmented state is given by,
\begin{equation}
\begin{split}
\mbs{\Phi}_{\text{aug}} (t;\omega) &= \mbs{\bar{\Phi}}_{\text{aug}}(t) + \sum_{i=1}^{\substack{N_s + N_\theta + N_\alpha \\ + N_\beta + N_\gamma}}\mbs{\tilde{\Phi}}_{\text{aug},i}(t) DY_i(t;\omega) \\
 &= \mbs{\bar{\Phi}}_{\text{aug}}(t) + \boldsymbol{\tilde{{\Phi}}}_{\text{aug}} (t) \mbs{DY} (t;\omega) \;.
 \end{split}
\end{equation}

We can also define the augmented observation model as,
\begin{equation}
 \begin{split}
  \label{observation model augmented}
  \mbs{\mathcal{Y}} &=  \begin{bmatrix}
  \mbs{0} &\mbs{H}
  \end{bmatrix}\mbs{\Phi}_{\text{aug}} + \mbs{V}, \qquad \mbs{V} \sim \mathcal{N}(0,\mbs{R}) \\
  &=\mbs{H}_{\text{aug}} \mbs{\Phi}_{\text{aug}} + \mbs{V} \;,
   \end{split}
\end{equation}
where $\mbs{H}$ is the original observation matrix, and $\mbs{H}_{\text{aug}}\in \mathbb{R}^{N_y\times (N_{\phi}N_x + N_\theta + N_\alpha + N_\beta + N_\gamma)}$ the augmented observation matrix, while $\mbs{\Phi}_{\text{aug}}$ is the augmented state ensemble. 

We can consider the above augmented state vector as forecast for time $t_k$, and employ the GMM-DO filter (\ref{app:GMM-DO}) to obtain joint posterior distributions of all parameters and state variables. The GMM fit is completed jointly for the (normalized) parameter realizations and DO stochastic coefficients realizations of the discrete state variables. If the observations, commonly of state variables, are informative about some parameter values, the pdf of these parameters will be updated by the Bayesian update, jointly with the pdf of the DO stochastic coefficients of the discrete state variables. 
%\PFJL{Is a separate GMM fit or double-GMM fit used for parameters? I don't remember.}
%\AG{The parameters and the state. coefficients share the same GMM fit}
%A main advantage of the above joint state-parameter methodology is that each parameter has its own sample realizations. The stochastic coefficients for the states and parameters are not shared.
%Sometimes there can be orders of magnitude differences between the uncertainty in state variables and in parameter values, while the number of parameters is often much less than the number of discrete state variables. 
% As a result, accurate normalization (\ref{app:DO}) can be challenging, and separate coefficients avoid this.
% In such cases, sharing of stochastic coefficients can lead to inefficient DO representation and evolution as they might require additional modes to effectively capture and evolve the joint uncertainty between states and parameters, thus increasing our computational cost. 
% It might also be the case, that the presence of large numbers of uncertain parameters requires additional GMM components while fitting the augmented coefficient matrix. 
%Because assimilation happens only at sparse times, this would not cause any significant increase in the overall computational costs. 

\section{Notations}

We define the notation used throughout the paper in Table~\ref{table: Notation compendium}, without repeating definitions already given in Table~\ref{table:DA Setup information}.

%\PFJL{I worked on the layout, also adding hfill and break. However, I think the dimensions in the exponents of $ \mathbb{R}$ that have linebreaks may still be a bit hard for readers. I removed several, but there are a few more we could perhaps do.} 
%\AG{Hopefully now they are more legible.}

\begin{footnotesize}
\begin{longtable}{|p{0.18\textwidth}|p{0.32\textwidth}|p{0.5\textwidth}|}
	\caption{Notation Compendium}
	\label{table: Notation compendium}
% 	\centering
% 	\begin{tabular}
\\
		\hline
		\multicolumn{3}{|c|}{\textbf{General}} \\\hline
		$\mbs{x}$     &  $\in \mathbb{R}^n$   &   Spatial coordinate vector \\\hline
		$t$     &  $\in \mathbb{R}$   &   Temporal coordinate \\\hline
		$T$     &  $\in \mathbb{R}$   &   Total simulation time \\\hline
		$\mathcal{D}$     &     &   Simulation domain \\\hline
		$\partial\mathcal{D}$     &     &   Simulation domain boundary \\\hline
		$\omega$     &    &   Realization index \\\hline
		$\Omega$     &     &   Measurable sample space \\\hline
		$N_x$     &  $\in \mathbb{N}$   &   Size of the discretized domain \\\hline
		$\mbs{\phi}$     &  $\in \mathbb{R}^{N_v}$ or $\mathbb{R}^{N_{\phi}}$   &   General state vector or biological tracer fields \\\hline
		$\mbs{\phi}_o$     &  $\in \mathbb{R}^{N_v}$ or $\mathbb{R}^{N_{\phi}}$   &   Initial condition of $\mbs{\phi}$ \\\hline
		$N_v$     &  $\in \mathbb{N}$   &   Number of state variables \\\hline
		$N_v(i)$     &  $\in \mathbb{N}$   &   Number of state variables in the $i^{th}$ candidate model of different complexity \\\hline
		$\{\phi_1^i, ..., \phi_{N_v(i)}^i\}$     &  $\in \mathbb{R}^{N_v(i)}$   &   State variables for the $i^{th}$ candidate model of different complexity \\\hline
		$\mbs{\Phi}$     &  $\in \mathbb{R}^{N_vN_x}$   &   Discretized state vector of $\mbs{\phi}$ \\\hline
		$\mbs{b}$     &   &   Boundary values \\\hline
		$\mathcal{M}$     &    &   Candidate model \\\hline
		$N_{m}$     &  $\in \mathbb{N}$   &   Number of candidate models \\\hline
		$\mbs{\theta}$     &  $\in \mathbb{R}^{N_{\theta}}$   &   Uncertain regular parameters \\\hline
		$N_\theta$     &  $\in \mathbb{N}$   &   Number of uncertain regular parameters \\\hline
		$\mbs{u}$     &  $\in \mathbb{R}^{N_{2}}$   &   Velocity field \\\hline
		%
		% $\mbs{\phi}_o$     &  $\in \mathbb{R}^{N_{\phi}}$   &   Initial condition of $\mbs{\phi}$ \\\hline
		%
		$N_\phi$     &  $\in \mathbb{N}$   &   Number of biological tracers \\\hline
		$p$     &  $\in \mathbb{R}^2$   &   Pressure field \\\hline
		%
% 		$\nu_E$     &  $\in \mathbb{R}$   &   Turbulent eddy viscosity \\\hline
% 		%
% 		$\mathcal{K}_E$     &  $\in \mathbb{R}$   &   Eddy diffusivity \\\hline
		%
		$u$ and $v$     &  $\in \mathbb{R}$   &   Horizontal and vertical velocity \\\hline
		$x$ and $z$     &  $\in \mathbb{R}$   &   Horizontal and vertical direction \\\hline
		$i$, $j$, $m$, and $n$     &  $\in \mathbb{N}$   &   Miscellaneous index \\\hline
		%%%%%%%%%%%%%%%%%%%%%%%%%%%%%%%%%%%
		\multicolumn{3}{|c|}{\textbf{Bayesian model learning}} \\\hline
		$\mbs{\alpha}$     &  $\in \mathbb{R}^{N_\alpha}$   &   Stochastic formulation parameters for combining candidate models with different functional forms \\\hline
		$N_\alpha$     &  $\in \mathbb{N}$   &   Number of stochastic formulation parameters $\alpha_k$'s \\\hline
		$\mbs{\beta}$     &  $\in \mathbb{R}^{N_\beta}$   &   Stochastic complexity parameters for combining candidate models of different complexities \\\hline
		$N_\beta$     &  $\in \mathbb{N}$   &   Number of stochastic complexity parameters $\beta_k$'s \\\hline
		$\mathcal{H}$     &     &   Range of values taken by the state variable \\\hline
		$I_i$     &     &   Interval with non-zero measure \\\hline
		$N_I$     &   $\mathbb{N}$  &   Number of intervals \\\hline
		$\mbs{\gamma}$     &  $\in \mathbb{R}^{N_\gamma}$   &   Stochastic expansion parameters \\\hline
		$N_\gamma$     &  $\in \mathbb{N}$   &   Number of stochastic expansion parameters $\gamma_k$'s \\\hline
		$k$     &  $\in \mathbb{N}$   &   Index for uncertain regular parameters and stochastic formulation, complexity, and expansion parameters \\\hline
		%%%%%%%%%%%%%%%%%%%%%%%%%%%%%%%%%%
		\multicolumn{3}{|c|}{\textbf{DO evolution equations}} \\\hline
		$N_s$     &  $\in \mathbb{N}$   &   Number of DO modes % $\gamma_k$'s
		\\\hline
		$N_r$     &  $\in \mathbb{N}$   &   Number of Monte-Carlo samples \\\hline
		$\mbs{\bar{\phi}}$     &  $\in \mathbb{R}^{N_{\phi}}$   &   Biological tracer DO mean \\\hline
		$\mbs{\bar{\Phi}}$     &  $\in \mathbb{R}^{N_{\phi}N_x}$   &   Discretized biological tracer DO mean $\mbs{\bar{\phi}}$ \\\hline
		$\mbs{\tilde{\phi}}_i$     &  $\in \mathbb{R}^{N_{\phi}}$   &   $i^{th}$ biological tracer DO mode \\\hline
		$\mbs{\tilde{\Phi}}_i$     &  $\in \mathbb{R}^{N_{\phi}N_x \times N_s}$   &   Discretized biological tracer DO modes matrix \\\hline
		${Y}_i$     &  $\in \mathbb{R}^{N_{s}}$   &   $i^{th}$ DO stochastic coefficient \\\hline
		$\mbs{Y}$     &  $\in \mathbb{R}^{N_r \times N_{s}}$   &   DO stochastic coefficient matrix \\\hline
		$\bar{\mbs{\theta}}$, $\bar{\mbs{\alpha}}$, $\bar{\mbs{\beta}}$, and $\bar{\mbs{\gamma}}$     &  $\in \mathbb{R}^{N_{\theta}}$, $\mathbb{R}^{N_{\alpha}}$, $\mathbb{R}^{N_{\beta}}$, and $\mathbb{R}^{N_{\gamma}}$ resp.  &   Mean vectors of uncertain parameters  \\\hline
		${\mbs{\mathfrak{D}^\theta}}$, ${\mbs{\mathfrak{D}^\alpha}}$, ${\mbs{\mathfrak{D}^\beta}}$, and ${\mbs{\mathfrak{D}^\gamma}}$     &  $\in \mathbb{R}^{N_{\theta}}$, $\mathbb{R}^{N_{\alpha}}$, $\mathbb{R}^{N_{\beta}}$, and $\mathbb{R}^{N_{\gamma}}$ resp.  &   Mean removed (deviation) part of uncertain parameters  \\\hline
		$\sigma_{nd, \bullet}$     &  $\in \mathbb{R}$   &   Weight of different state variables in inner-product computation \\\hline
		%%%%%%%%%%%%%%%%%%%%%%%%%%%%%%%%%%%
		\multicolumn{3}{|c|}{\textbf{GMM-DO filter}} \\\hline
		$\mbs{\mathcal{Y}}$     &  $\in \mathbb{R}^{N_y}$   &   Observation vector \\\hline
		$\mbs{{y}}$     &  $\in \mathbb{R}^{N_y}$   &   Observation vector realization \\\hline
		$N_y$     &  $\in \mathbb{N}$   &   Number of observations \\\hline
		$\mbs{H}$     &  $\in \mathbb{R}^{N_y \times N_vN_x}$ or $\mathbb{R}^{N_y \times N_{\phi}N_x}$   &   Linear observation matrix \\\hline
		$\mbs{V}$     &  $\in \mathbb{R}^{N_y}$   &   Measurement noise vector \\\hline
		$\mbs{R}$     &  $\in \mathbb{R}^{N_y \times N_y}$   &   Covariance matrix of measurement noise \\\hline
		$N_{\text{GMM}}$     &  $\in \mathbb{N}$   &   Number of Gaussian mixture model (GMM) components \\\hline
		$\pi^{\bullet}_{\bullet, j}$     &  $\in [0, 1]$   &   $j^{th}$ GMM component weight \\\hline
		$\mbs{\mu}^{\bullet}_{\bullet, j}$     &  $\in \mathbb{R}^{N_s}$ or \vspace{0.1cm} \hfill \break 
        $~~ \mathbb{R}^{N_{\phi}N_x + N_{\theta} + N_{\alpha} + N_{\beta} + N_{\gamma}}$   &   $j^{th}$ GMM component mean vector \\\hline
		$\mbs{\Sigma}^{\bullet}_{\bullet, j}$     &  $\in \mathbb{R}^{N_s \times N_s}$ or \vspace{0.1cm} \hfill \break
        $\mathbb{R}^{\left(\substack{ N_{\phi}N_x + N_{\theta} + N_{\alpha} \\ + N_{\beta} + N_{\gamma}}\right) \times \left(\substack{ N_{\phi}N_x + N_{\theta} + N_{\alpha} \\ + N_{\beta} + N_{\gamma}}\right)}$  &   $j^{th}$ GMM component covariance matrix \\\hline
		$\mbs{K}_j$     &  $\in \mathbb{R}^{N_v N_x \times N_y}$ or $\mathbb{R}^{N_{\phi} N_x \times N_y}$ or \vspace{0.1cm} \hfill \break $\mathbb{R}^{\smallskip \left( N_{\phi}N_x + N_{\theta} + N_{\alpha}  + N_{\beta} + N_{\gamma} \right) \times N_y}$   &   $j^{th}$ Kalman gain matrix \\\hline
		$p_{\bullet}(\bullet)$     &  $\in \mathbb{R}$   &   Probability distribution value \\\hline
		$\mbs{\tilde{y}}$     &  $\in \mathbb{R}^{N_y}$   &   Transformed observation vector realization \\\hline
		$\mbs{\tilde{H}}$     &  $\in \mathbb{R}^{N_y \times N_s}$   &   Transformed linear observation matrix \\\hline
		$\mbs{\tilde{K}}_j$     &  $\in \mathbb{R}^{N_s \times N_y}$  &   Transformed $j^{th}$ Kalman gain matrix \\\hline
		$\mbs{\hat{\mu}}^{\bullet}_{\bullet, j}$     &  $\in \mathbb{R}^{N_s}$   &  Intermediate $j^{th}$ GMM component mean vector \\\hline
		%
		%%%%%%%%%%%%%%%%%%%%%%%%%%%%%%%%%%%
		\multicolumn{3}{|c|}{\textbf{State augmentation}} \\\hline
		$\mbs{\Phi}_{\text{aug}}$     &  $\in \mathbb{R}^{\substack{N_{\phi}N_x + N_{\theta} + N_{\alpha}  + N_{\beta} + N_{\gamma}}}$   &  Augmented discretized state vector \\\hline
		$\mbs{\bar{\Phi}}_{\text{aug}}$     &  $\in \mathbb{R}^{\substack{N_{\phi}N_x + N_{\theta} + N_{\alpha} + N_{\beta} + N_{\gamma}}}$   &  Augmented discretized DO mean vector \\\hline
		$\mbs{\tilde{\Phi}}_{\text{aug}}$     &  $\in \mathbb{R}^{\substack{N_{\phi}N_x + N_{\theta} + N_{\alpha}  + N_{\beta} + N_{\gamma}}}$   &  Augmented discretized DO modes matrix \\\hline
		$\mbs{DY}$     &  $\in \mathbb{R}^{\substack{N_s + N_{\theta} + N_{\alpha} + N_{\beta} + N_{\gamma}}}$   &  Augmented DO stochastic coefficient matrix \\\hline
		$\mbs{H}_{\text{aug}}$     &  $\in \mathbb{R}^{\substack{N_y \times  \left(N_{\phi}N_x + N_{\theta} + N_{\alpha} + N_{\beta} + N_{\gamma}\right)}}$   &  Augmented linear observation matrix  \\\hline
		$\mbs{I}$     &  $\in \mathbb{R}^{\substack{\left(N_{\theta} + N_{\alpha} + N_{\beta} + N_{\gamma}\right) \times  \left((N_{\theta} + N_{\alpha} + N_{\beta} + N_{\gamma}\right)}}$   &  Identity matrix  \\\hline
		$\mbs{0}$     &    &  Matrix of zeroes of appropriate size  \\\hline
		%%%%%%%%%%%%%%%%%%%%%%%%%%%%%%%%%%%
		\multicolumn{3}{|c|}{\textbf{Operators, functions, and indicators}} \\\hline
		$\mathcal{L}[\bullet]$     &    &   Functional form of known dynamics  \\\hline
		$\widehat{\mathcal{L}}[\bullet]$     &    &   Unknown dynamics belonging to a set of candidate functional forms \\\hline
		$\{\widehat{\mathcal{L}}_1[\bullet], ..., \widehat{\mathcal{L}}_{N_m}[\bullet]\}$     &    &   Set of candidate functional forms \\\hline
		$\widetilde{\mathcal{L}}[\bullet]$     &    &   Functional form of completely unknown dynamics \\\hline
		$\{\mathcal{L}^i_1[\bullet], ..., \mathcal{L}^i_{N_v(i)}[\bullet]\}$     &    &   Right-hand-sides of $i^{th}$ candidate model of different complexity  \\\hline
		$\mathcal{N}(\bullet, \bullet)$     &     &   Multivariate Gaussian distribution \\\hline
		$\Psi_k(\bullet)$     &     &  $k^{th}$  Linear function \\\hline
		$\{S^{\phi^1}(\bullet), ..., S^{\phi^{N_{\phi}}}(\bullet)\}$     &    &   Biological reaction terms \\\hline
		$\mbs{\widehat{S}}^{\mbs{\phi}}(\bullet)$     &    &   Unknown biological reaction terms belonging to a set of candidate functional forms \\\hline
		$\mbs{\widetilde{S}}^{\mbs{\phi}}(\bullet)$     &    &   Completely unknown biological reaction terms \\\hline
		$\nabla . (\bullet)$     &    &   Gradient operator \\\hline
		$\nabla^2 (\bullet)$     &    &   Diffusion operator \\\hline
		$\langle \bullet, \bullet \rangle$     &    &   $i^{th}$ Spatial inner-product \\\hline
		$\mathbb{E} [\bullet]$     &    &   Expectation \\\hline
		$C_{\bullet, \bullet}$     &    &   Cross-covariance \\\hline
		$\mathcal{B}[\bullet]$     &    &   Boundary condition operator \\\hline
		$\mbs{V_S}$     &    &   DO Subspace \\\hline
		$(\bullet)^{f}$     &    &  Prior \\\hline
		$(\bullet)^{a}$     &    &  Posterior \\\hline
% 	\end{tabular}
\end{longtable}
\end{footnotesize}

%\section*{References}
\bibliographystyle{elsarticle-num}
\bibliography{Bibliography/mseas_new,Bibliography/biology, Bibliography/ml, Bibliography/ocean, Bibliography/refs_pl,Bibliography/misc}

\begin{thebibliography}{100}
\expandafter\ifx\csname url\endcsname\relax
  \def\url#1{\texttt{#1}}\fi
\expandafter\ifx\csname urlprefix\endcsname\relax\def\urlprefix{URL }\fi
\expandafter\ifx\csname href\endcsname\relax
  \def\href#1#2{#2} \def\path#1{#1}\fi

\bibitem{lalli1997biological}
C.~Lalli, T.~R. Parsons, Biological {O}ceanography: {A}n {I}ntroduction,
  Elsevier Butterworth-Heinemann, 1997.

\bibitem{fennel2014introduction}
W.~Fennel, T.~Neumann, Introduction to the {M}odelling of {M}arine
  {E}cosystems:(with {MATLAB} programs on accompanying {CD-ROM}), Vol.~72 of
  Oceanography, Elsevier, 2014.

\bibitem{hofmann2002predictive}
E.~Hofmann, M.~A. Friedrichs, Predictive modeling for marine ecosystems, The
  Sea 12 (2002) 537--565.

\bibitem{fennel2022ocean}
K.~Fennel, J.~P. Mattern, S.~C. Doney, L.~Bopp, A.~M. Moore, B.~Wang, L.~Yu,
  Ocean biogeochemical modelling, Nature Reviews Methods Primers 2~(1) (2022)
  76.

\bibitem{franks2002npz}
P.~J.~S. Franks, {NPZ} models of plankton dynamics: their construction,
  coupling to physics, and application, Journal of Oceanography 58~(2) (2002)
  379--387.

\bibitem{ward2010parameter}
B.~A. Ward, M.~A.~M. Friedrichs, T.~R. Anderson, A.~Oschlies, Parameter
  optimisation techniques and the problem of underdetermination in marine
  biogeochemical models, Journal of Marine Systems 81~(1) (2010) 34--43.

\bibitem{denman2003modelling}
K.~L. Denman, Modelling planktonic ecosystems: parameterizing complexity,
  Progress in Oceanography 57~(3-4) (2003) 429--452.

\bibitem{franks1986behavior}
P.~J.~S. Franks, J.~S. Wroblewski, G.~R. Flierl, Behavior of a simple plankton
  model with food-level acclimation by herbivores, Marine Biology 91~(1) (1986)
  121--129.

\bibitem{flierl2002mesoscale}
G.~Flierl, D.~J. McGillicuddy, Mesoscale and submesoscale physical-biological
  interactions, The sea 12 (2002) 113--185.

\bibitem{davis1994biological}
C.~S. Davis, J.~H. Steele, Biological/physical modeling of upper ocean
  processes, Tech. rep., Woods Hole Oc. Inst. (1994).

\bibitem{fasham1990nitrogen}
M.~J.~R. Fasham, H.~W. Ducklow, S.~M. McKelvie, A nitrogen-based model of
  plankton dynamics in the oceanic mixed layer, Journal of Marine Research
  48~(3) (1990) 591--639.

\bibitem{besiktepe_et_al_JMS2003}
{\c S}.~T. {Be{\c s}iktepe}, P.~F.~J. {Lermusiaux}, A.~R. {Robinson}, {Coupled
  physical and biogeochemical data-driven simulations of Massachusetts Bay in
  late summer: Real-time and post-cruise data assimilation}, Journal of Marine
  Systems 40--41 (2003) 171--212.
\newblock \href {https://doi.org/10.1016/S0924-7963(03)00018-6}
  {\path{doi:10.1016/S0924-7963(03)00018-6}}.

\bibitem{baretta1995european}
J.~W. Baretta, W.~Ebenh{\"o}h, P.~Ruardij, The {European Regional Seas
  Ecosystem Model}, a complex marine ecosystem model, Netherlands Journal of
  Sea Research 33~(3) (1995) 233--246.

\bibitem{baretta1997preface}
J.~W. Baretta, Preface to the {European Regional Seas Ecosystem Model} {II},
  Journal of Sea Research 38~(3) (1997) 169--171.

\bibitem{blackford2004ecosystem}
J.~C. Blackford, J.~I. Allen, F.~J. Gilbert, Ecosystem dynamics at six
  contrasting sites: a generic modelling study, Journal of Marine Systems
  52~(1) (2004) 191--215.

\bibitem{lermusiaux_et_al_CS2004}
P.~F.~J. {Lermusiaux}, C.~{Evangelinos}, R.~{Tian}, P.~J. {Haley}, Jr, J.~J.
  {McCarthy}, N.~M. {Patrikalakis}, A.~R. {Robinson}, H.~{Schmidt}, Adaptive
  coupled physical and biogeochemical ocean predictions: A conceptual basis,
  in: Computational Science - ICCS 2004, Vol. 3038 of Lecture Notes in Computer
  Science, Springer Berlin Heidelberg, 2004, pp. 685--692.
\newblock \href {https://doi.org/10.1007/978-3-540-24688-6\_89}
  {\path{doi:10.1007/978-3-540-24688-6\_89}}.

\bibitem{lermusiaux_PhysD2007}
P.~F.~J. {Lermusiaux}, Adaptive modeling, adaptive data assimilation and
  adaptive sampling, Physica D: Nonlinear Phenomena 230~(1) (2007) 172--196.
\newblock \href {https://doi.org/10.1016/j.physd.2007.02.014}
  {\path{doi:10.1016/j.physd.2007.02.014}}.

\bibitem{robinson_lermusiaux_Sea2002}
A.~R. Robinson, P.~F.~J. Lermusiaux, Data assimilation for modeling and
  predicting coupled physical--biological interactions in the sea, in: A.~R.
  Robinson, J.~J. McCarthy, B.~J. Rothschild (Eds.), Biological-Physical
  Interactions in the Sea, Vol.~12 of The Sea, John Wiley and Sons, New York,
  2002, chapter~12, pp. 475--536.

\bibitem{dowd2014statistical}
M.~Dowd, E.~Jones, J.~Parslow, A statistical overview and perspectives on data
  assimilation for marine biogeochemical models, Environmetrics 25~(4) (2014)
  203--213.

\bibitem{friedrichs2007assessment}
M.~A.~M. Friedrichs, et~al., Assessment of skill and portability in regional
  marine biogeochemical models: {R}ole of multiple planktonic groups, Journal
  of Geophysical Research: Oceans 112~(C8) (2007).

\bibitem{losa2004weak}
S.~N. Losa, G.~A. Kivman, V.~A. Ryabchenko, Weak constraint parameter
  estimation for a simple ocean ecosystem model: what can we learn about the
  model and data?, Journal of Marine Systems 45~(1) (2004) 1--20.

\bibitem{mattern2012estimating}
J.~P. Mattern, K.~Fennel, M.~Dowd, Estimating time-dependent parameters for a
  biological ocean model using an emulator approach, Journal of Marine Systems
  96 (2012) 32--47.

\bibitem{toyoda2013improved}
T.~Toyoda, et~al., Improved state estimations of lower trophic ecosystems in
  the global ocean based on a green's function approach, Progress in
  Oceanography 119 (2013) 90--107.

\bibitem{mattern2010sequential}
J.~P. Mattern, M.~Dowd, K.~Fennel, Sequential data assimilation applied to a
  physical--biological model for the bermuda atlantic time series station,
  Journal of Marine Systems 79~(1) (2010) 144--156.

\bibitem{allen2003ensemble}
J.~I. Allen, M.~Eknes, G.~Evensen, An {E}nsemble {K}alman {F}ilter with a
  complex marine ecosystem model: hindcasting phytoplankton in the {C}retan
  {S}ea, in: Annales Geophysicae, Vol.~21, 2003, pp. 399--411.

\bibitem{natvik2003assimilation}
L.~J. Natvik, G.~Evensen, {Assimilation of ocean colour data into a biochemical
  model of the North Atlantic: Part 1. Data assimilation experiments}, Journal
  of Marine Systems 40 (2003) 127--153.

\bibitem{hu2012data}
J.~Hu, K.~Fennel, J.~P. Mattern, J.~Wilkin, Data assimilation with a local
  ensemble kalman filter applied to a three-dimensional biological model of the
  middle atlantic bight, Journal of Marine Systems 94 (2012) 145--156.

\bibitem{doron2011stochastic}
M.~Doron, P.~Brasseur, J.-M. Brankart, Stochastic estimation of biogeochemical
  parameters of a {3D} ocean coupled physical--biogeochemical model: {T}win
  experiments, Journal of Marine Systems 87~(3) (2011) 194--207.

\bibitem{jones2010bayesian}
E.~Jones, J.~Parslow, L.~Murray, A bayesian approach to state and parameter
  estimation in a phytoplankton-zooplankton model, Australian Meteorological
  and Oceanographic Journal 59~(SP) (2010) 7--16.

\bibitem{mattern2013particle}
J.~P. Mattern, M.~Dowd, K.~Fennel, Particle filter-based data assimilation for
  a three-dimensional biological ocean model and satellite observations,
  Journal of Geophysical Research: Oceans 118~(5) (2013) 2746--2760.

\bibitem{giricheva2015aggregation}
E.~Giricheva, Aggregation in ecosystem models and model stability, Progress in
  Oceanography 134 (2015) 190--196.

\bibitem{ward2013biogeochemical}
B.~A. Ward, et~al., When is a biogeochemical model too complex? objective model
  reduction and selection for north atlantic time-series sites, Progress in
  Oceanography 116 (2013) 49--65.

\bibitem{evangelinos_et_al_ICCS2003}
C.~Evangelinos, R.~Chang, P.~F.~J. Lermusiaux, N.~M. Patrikalakis, Rapid
  real-time interdisciplinary ocean forecasting using adaptive sampling and
  adaptive modeling and legacy codes: Component encapsulation using {XML}, in:
  Computational Science--ICCS 2003, Springer, 2003, pp. 375--384.
\newblock \href {https://doi.org/10.1007/3-540-44864-0_39}
  {\path{doi:10.1007/3-540-44864-0_39}}.

\bibitem{tian_etal_2004}
R.~C. Tian, P.~F.~J. Lermusiaux, J.~J. McCarthy, A.~R. Robinson, A generalized
  prognostic model of marine biogeochemical-ecosystem dynamics: {S}tructure,
  parameterization and adaptive modeling, Harvard Reports in
  Physical/Interdisciplinary Ocean Science~67, Department of Earth and
  Planetary Sciences, Harvard University, Cambridge, MA (May 2004).

\bibitem{lermusiaux_et_al_Oceanog2011}
P.~F.~J. Lermusiaux, P.~J. Haley, W.~G. Leslie, A.~Agarwal, O.~Logutov, L.~J.
  Burton, Multiscale physical and biological dynamics in the {P}hilippine
  {A}rchipelago: Predictions and processes, Oceanography 24~(1) (2011) 70--89,
  {S}pecial Issue on the Philippine Straits Dynamics Experiment.
\newblock \href {https://doi.org/10.5670/oceanog.2011.05}
  {\path{doi:10.5670/oceanog.2011.05}}.

\bibitem{brunton2016discovering}
S.~L. Brunton, J.~L. Proctor, J.~N. Kutz, Discovering governing equations from
  data by sparse identification of nonlinear dynamical systems, PNAS 113~(15)
  (2016) 3932--3937.

\bibitem{rudy2019data}
S.~Rudy, A.~Alla, S.~L. Brunton, J.~N. Kutz, Data-driven identification of
  parametric partial differential equations, SIAM Journal on Applied Dynamical
  Systems 18~(2) (2019) 643--660.

\bibitem{messenger2021weak}
D.~A. Messenger, D.~M. Bortz, Weak sindy for partial differential equations,
  Journal of Computational Physics (2021) 110525.

\bibitem{kulkarni_et_al_DDDAS2020}
C.~S. Kulkarni, A.~Gupta, P.~F.~J. Lermusiaux, Sparse regression and adaptive
  feature generation for the discovery of dynamical systems, in: F.~Darema,
  E.~Blasch, S.~Ravela, A.~Aved (Eds.), Dynamic Data Driven Application
  Systems. {DDDAS} 2020., Vol. 12312 of Lecture Notes in Computer Science,
  Springer, Cham, 2020, pp. 208--216.
\newblock \href {https://doi.org/10.1007/978-3-030-61725-7_25}
  {\path{doi:10.1007/978-3-030-61725-7_25}}.

\bibitem{niven2020bayesian}
R.~K. Niven, A.~Mohammad-Djafari, L.~Cordier, M.~Abel, M.~Quade, Bayesian
  identification of dynamical systems, in: Multidisciplinary Digital Publishing
  Institute Proceedings, Vol.~33, 2020, p.~33.

\bibitem{gupta_lermusiaux_PRSA2021}
A.~Gupta, P.~F.~J. Lermusiaux, Neural closure models for dynamical systems,
  Proceedings of The Royal Society A 477~(2252) (2021) 1--29.
\newblock \href {https://doi.org/10.1098/rspa.2020.1004}
  {\path{doi:10.1098/rspa.2020.1004}}.

\bibitem{gupta_lermusiaux_SR2023}
A.~Gupta, P.~F.~J. Lermusiaux, Generalized neural closure models with
  interpretability, Scientific ReportsSub-judice (2023).
\newblock \href {https://doi.org/10.48550/arXiv.2301.06198}
  {\path{doi:10.48550/arXiv.2301.06198}}.

\bibitem{maslyaev1903data}
M.~Maslyaev, A.~Hvatov, A.~Kalyuzhnaya, Data-driven pde discovery with
  evolutionary approach.(2019), arXiv preprint arXiv:1903.08011.

\bibitem{bassenne2019computational}
M.~Bassenne, A.~Lozano-Dur{\'a}n, Computational model discovery with
  reinforcement learning, arXiv preprint arXiv:2001.00008 (2019).

\bibitem{novati2021automating}
G.~Novati, H.~L. de~Laroussilhe, P.~Koumoutsakos, Automating turbulence
  modelling by multi-agent reinforcement learning, Nature Machine Intelligence
  3~(1) (2021) 87--96.

\bibitem{wang2019learning}
Y.~Wang, Z.~Shen, Z.~Long, B.~Dong, Learning to discretize: solving 1d scalar
  conservation laws via deep reinforcement learning, arXiv preprint
  arXiv:1905.11079 (2019).

\bibitem{raissi2018hidden}
M.~Raissi, G.~E. Karniadakis, Hidden physics models: Machine learning of
  nonlinear partial differential equations, Journal of Computational Physics
  357 (2018) 125--141.

\bibitem{lu_lermusiaux_MSEAS2014}
P.~G.~Y. Lu, P.~F.~J. Lermusiaux, Pde-based bayesian inference of
  high-dimensional dynamical models, MSEAS Report~19, Department of Mechanical
  Engineering, Massachusetts Institute of Technology, Cambridge, MA, USA
  (2014).

\bibitem{lu_lermusiaux_PhysD2021}
P.~Lu, P.~F.~J. Lermusiaux, Bayesian learning of stochastic dynamical models,
  Physica D: Nonlinear Phenomena 427 (2021) 133003.
\newblock \href {https://doi.org/10.1016/j.physd.2021.133003}
  {\path{doi:10.1016/j.physd.2021.133003}}.

\bibitem{sondergaard_lermusiaux_MWR2013_part1}
T.~Sondergaard, P.~F.~J. Lermusiaux, Data assimilation with {Gaussian Mixture
  Models using the Dynamically Orthogonal} field equations. {Part I}: Theory
  and scheme., Monthly Weather Review 141~(6) (2013) 1737--1760.
\newblock \href {https://doi.org/10.1175/MWR-D-11-00295.1}
  {\path{doi:10.1175/MWR-D-11-00295.1}}.

\bibitem{sondergaard_lermusiaux_MWR2013_part2}
T.~Sondergaard, P.~F.~J. Lermusiaux, Data assimilation with {Gaussian Mixture
  Models using the Dynamically Orthogonal} field equations. {Part II}:
  Applications., Monthly Weather Review 141~(6) (2013) 1761--1785.
\newblock \href {https://doi.org/10.1175/MWR-D-11-00296.1}
  {\path{doi:10.1175/MWR-D-11-00296.1}}.

\bibitem{janjic2018representation}
T.~Janji{\'c}, N.~Bormann, M.~Bocquet, J.~Carton, S.~Cohn, S.~L. Dance,
  S.~Losa, N.~K. Nichols, R.~Potthast, J.~A. Waller, et~al., On the
  representation error in data assimilation, Quarterly Journal of the Royal
  Meteorological Society 144~(713) (2018) 1257--1278.

\bibitem{lermusiaux_et_al_QJRMS2000}
P.~F.~J. Lermusiaux, D.~G.~M. Anderson, C.~J. Lozano, On the mapping of
  multivariate geophysical fields: Error and variability subspace estimates,
  Quarterly Journal of the Royal Meteorological Society 126~(565) (2000)
  1387--1429.
\newblock \href {https://doi.org/10.1256/smsqj.56509}
  {\path{doi:10.1256/smsqj.56509}}.

\bibitem{lermusiaux_JAOT2002}
P.~F.~J. Lermusiaux, On the mapping of multivariate geophysical fields:
  Sensitivities to size, scales, and dynamics, Journal of Atmospheric and
  Oceanic Technology 19~(10) (2002) 1602--1637.
\newblock \href
  {https://doi.org/10.1175/1520-0426(2002)019<1602:OTMOMG>2.0.CO;2}
  {\path{doi:10.1175/1520-0426(2002)019<1602:OTMOMG>2.0.CO;2}}.

\bibitem{mr1763essay}
M.~Bayes, M.~Price, An {Essay towards Solving a Problem in the Doctrine of
  Chances}. {B}y the {L}ate {R}ev. {M}r. {B}ayes, {F. R. S.} {C}ommunicated by
  {M}r. {P}rice, in a {L}etter to {J}ohn {C}anton, {A. M. F. R. S.},
  Philosophical Transactions (1683-1775) 53 (1763) 370--418.
\newblock \href {https://doi.org/10.1098/rstl.1763.0053}
  {\path{doi:10.1098/rstl.1763.0053}}.

\bibitem{bertsekas2008introduction}
D.~Bertsekas, J.~N. Tsitsiklis, Introduction to probability, Vol.~1, Athena
  Scientific, 2008.

\bibitem{trefethen2019approximation}
L.~N. Trefethen, Approximation Theory and Approximation Practice, Extended
  Edition, SIAM, 2019.

\bibitem{jim2023mat}
J.~Lambers, {MAT} 772 fall semester 2010-11 lecture 17 notes, Available at
  \url{https://www.math.usm.edu/lambers/mat772/fall10/lecture17.pdf}
  (2023/01/20).

\bibitem{sapsis_lermusiaux_PD2009}
T.~P. Sapsis, P.~F.~J. Lermusiaux, Dynamically orthogonal field equations for
  continuous stochastic dynamical systems, Physica D: Nonlinear Phenomena
  238~(23--24) (2009) 2347--2360.
\newblock \href {https://doi.org/10.1016/j.physd.2009.09.017}
  {\path{doi:10.1016/j.physd.2009.09.017}}.

\bibitem{sapsis_lermusiaux_PHYSD2012}
T.~P. Sapsis, P.~F.~J. Lermusiaux, Dynamical criteria for the evolution of the
  stochastic dimensionality in flows with uncertainty, Physica D: Nonlinear
  Phenomena 241~(1) (2012) 60--76.
\newblock \href {https://doi.org/10.1016/j.physd.2011.10.001}
  {\path{doi:10.1016/j.physd.2011.10.001}}.

\bibitem{feppon_lermusiaux_SIMAX2018a}
F.~Feppon, P.~F.~J. Lermusiaux, A geometric approach to dynamical model-order
  reduction, SIAM Journal on Matrix Analysis and Applications 39~(1) (2018)
  510--538.
\newblock \href {https://doi.org/10.1137/16M1095202}
  {\path{doi:10.1137/16M1095202}}.

\bibitem{gelb1974applied}
A.~Gelb, Applied optimal estimation, MIT Press, 1974.

\bibitem{casella2021statistical}
G.~Casella, R.~L. Berger, Statistical inference, Cengage Learning, 2021.

\bibitem{hart2000pattern}
P.~E. Hart, D.~G. Stork, R.~O. Duda, Pattern classification, Wiley Hoboken,
  2000.

\bibitem{newberger2003analysis}
P.~A. Newberger, J.~S. Allen, Y.~H. Spitz, Analysis and comparison of three
  ecosystem models, Journal of Geophysical Research: Oceans (1978--2012)
  108~(C3) (2003).

\bibitem{haley_lermusiaux_OD2010}
P.~J. Haley, Jr., P.~F.~J. Lermusiaux, Multiscale two-way embedding schemes for
  free-surface primitive equations in the ``{M}ultidisciplinary {S}imulation,
  {E}stimation and {A}ssimilation {S}ystem'', Ocean Dynamics 60~(6) (2010)
  1497--1537.
\newblock \href {https://doi.org/10.1007/s10236-010-0349-4}
  {\path{doi:10.1007/s10236-010-0349-4}}.

\bibitem{haley_et_al_OM2015}
P.~J. Haley, Jr., A.~Agarwal, P.~F.~J. Lermusiaux, Optimizing velocities and
  transports for complex coastal regions and archipelagos, Ocean Modeling 89
  (2015) 1--28.
\newblock \href {https://doi.org/10.1016/j.ocemod.2015.02.005}
  {\path{doi:10.1016/j.ocemod.2015.02.005}}.

\bibitem{mcwilliams2008nature}
J.~C. McWilliams, The nature and consequences of oceanic eddies, Ocean modeling
  in an eddying regime 177 (2008) 5--15.

\bibitem{hecht2013ocean}
M.~W. Hecht, H.~Hasumi, Ocean modeling in an eddying regime, Vol. 177, John
  Wiley \& Sons, 2013.

\bibitem{ferziger2002computational}
J.~H. Ferziger, M.~Peri{\'c}, R.~L. Street, Computational methods for fluid
  dynamics, Vol.~3, Springer, 2002.

\bibitem{kulkarni_lermusiaux_JCP2019}
C.~S. Kulkarni, P.~F.~J. Lermusiaux, Advection without compounding errors
  through flow map composition, Journal of Computational Physics 398 (2019)
  108859.
\newblock \href {https://doi.org/10.1016/j.jcp.2019.108859}
  {\path{doi:10.1016/j.jcp.2019.108859}}.

\bibitem{subramani_lermusiaux_OM2016}
D.~N. Subramani, P.~F.~J. Lermusiaux, Energy-optimal path planning by
  stochastic dynamically orthogonal level-set optimization, Ocean Modeling 100
  (2016) 57--77.
\newblock \href {https://doi.org/10.1016/j.ocemod.2016.01.006}
  {\path{doi:10.1016/j.ocemod.2016.01.006}}.

\bibitem{subramani_et_al_CMAME2018}
D.~N. Subramani, Q.~J. Wei, P.~F.~J. Lermusiaux, Stochastic time-optimal
  path-planning in uncertain, strong, and dynamic flows, Computer Methods in
  Applied Mechanics and Engineering 333 (2018) 218--237.
\newblock \href {https://doi.org/10.1016/j.cma.2018.01.004}
  {\path{doi:10.1016/j.cma.2018.01.004}}.

\bibitem{branicki_majda_limit_PCE_CMS2013}
M.~Branicki, A.~J. Majda, Fundamental limitations of polynomial chaos for
  uncertainty quantification in systems with intermittent instabilities,
  Communications in mathematical sciences 11~(1) (2013) 55--103.

\bibitem{mcgillicuddy1998adjoint}
D.~McGillicuddy, D.~Lynch, A.~Moore, W.~Gentleman, C.~Davis, C.~Meise, An
  adjoint data assimilation approach to diagnosis of physical and biological
  controls on pseudocalanus spp. in the gulf of maine--georges bank region,
  Fisheries Oceanography 7~(3-4) (1998) 205--218.

\bibitem{lermusiaux_JMS2001}
P.~F.~J. Lermusiaux, Evolving the subspace of the three-dimensional multiscale
  ocean variability: {M}assachusetts {B}ay, Journal of Marine Systems 29~(1)
  (2001) 385--422.
\newblock \href {https://doi.org/10.1016/S0924-7963(01)00025-2}
  {\path{doi:10.1016/S0924-7963(01)00025-2}}.

\bibitem{pineda2015whales}
J.~Pineda, V.~Starczak, J.~C. da~Silva, K.~Helfrich, M.~Thompson, D.~Wiley,
  Whales and waves: Humpback whale foraging response and the shoaling of
  internal waves at stellwagen bank, J. of Geophysical Res.: Oceans 120~(4)
  (2015) 2555--2570.

\bibitem{tian2015model}
R.~Tian, C.~Chen, J.~Qi, R.~Ji, R.~C. Beardsley, C.~Davis, Model study of
  nutrient and phytoplankton dynamics in the gulf of maine: patterns and
  drivers for seasonal and interannual variability, ICES J. of Marine Science
  72~(2) (2015) 388--402.

\bibitem{pershing_GoM_temp2050_2019}
A.~Pershing, et~al., Temperature and circulation conditions in the gulf of
  maine in 2050 and their expected impacts, Scientific scenario paper, Gulf of
  Maine 2050 International Symposium, 2019.

\bibitem{silva_SB_Sanct_2021}
T.~L. Silva, State of the science report: An addendum to the stellwagen bank
  national marine sanctuary 2020 condition report (2021).

\bibitem{ueckermann_and_lermusiaux_MSEAS2012}
M.~P. Ueckermann, P.~F.~J. Lermusiaux,
  \href{http://mseas.mit.edu/?p=2567}{{2.29 Finite Volume MATLAB Framework
  Documentation}}, MSEAS Report~14, Department of Mechanical Engineering,
  Massachusetts Institute of Technology, Cambridge, MA (2012).
\newline\urlprefix\url{http://mseas.mit.edu/?p=2567}

\bibitem{van1977towards}
B.~Van~Leer, Towards the ultimate conservative difference scheme. iv. a new
  approach to numerical convection, Journal of computational physics 23~(3)
  (1977) 276--299.

\bibitem{gupta_PhDThesis2022}
A.~Gupta, Scientific machine learning for dynamical systems: Theory and
  applications to fluid flow and ocean ecosystem modeling, Ph.D. thesis,
  Massachusetts Institute of Technology, Department of Mechanical Engineering,
  Cambridge, Massachusetts (Sep. 2022).

\bibitem{ueckermann_et_al_JCP2013}
M.~P. Ueckermann, P.~F.~J. Lermusiaux, T.~P. Sapsis, Numerical schemes for
  dynamically orthogonal equations of stochastic fluid and ocean flows, Journal
  of Computational Physics 233 (2013) 272--294.
\newblock \href {https://doi.org/10.1016/j.jcp.2012.08.041}
  {\path{doi:10.1016/j.jcp.2012.08.041}}.

\bibitem{feppon_lermusiaux_SIREV2018}
F.~Feppon, P.~F.~J. Lermusiaux, Dynamically orthogonal numerical schemes for
  efficient stochastic advection and {L}agrangian transport, {SIAM} Review
  60~(3) (2018) 595--625.
\newblock \href {https://doi.org/10.1137/16M1109394}
  {\path{doi:10.1137/16M1109394}}.

\bibitem{fsolve}
\href{https://www.mathworks.com/help/optim/ug/fsolve.html}{Solve system of
  nonlinear equations - {MATLAB}}, {A}ccessed on January 21, 2023.
\newline\urlprefix\url{https://www.mathworks.com/help/optim/ug/fsolve.html}

\bibitem{bengtsson1981dynamic}
L.~Bengtsson, M.~Ghil, E.~K{\"a}ll{\'e}n, Dynamic meteorology: data
  assimilation methods, Springer, 1981.

\bibitem{ide1998extended1}
K.~Ide, M.~Ghil, Extended kalman filtering for vortex systems. part 1:
  Methodology and point vortices, Dynamics of Atmospheres and Oceans 27~(1-4)
  (1998) 301--332.

\bibitem{ide1998extended2}
K.~Ide, M.~Ghil, Extended kalman filtering for vortex systems. part ii: Rankine
  vortices and observing-system design, Dynamics of Atmospheres and Oceans
  27~(1-4) (1998) 333--350.

\bibitem{lermusiaux_MWR1999}
P.~F.~J. Lermusiaux, Data assimilation via {E}rror {S}ubspace {S}tatistical
  {E}stimation, part {II}: Mid-{A}tlantic {B}ight shelfbreak front simulations,
  and {ESSE} validation, Monthly Weather Review 127~(7) (1999) 1408--1432.
\newblock \href
  {https://doi.org/10.1175/1520-0493(1999)127<1408:DAVESS>2.0.CO;2}
  {\path{doi:10.1175/1520-0493(1999)127<1408:DAVESS>2.0.CO;2}}.

\bibitem{lolla_et_al_OD2014_part2}
T.~Lolla, P.~J. Haley, Jr., P.~F.~J. Lermusiaux, Time-optimal path planning in
  dynamic flows using level set equations: Realistic applications, Ocean
  Dynamics 64~(10) (2014) 1399--1417.
\newblock \href {https://doi.org/10.1007/s10236-014-0760-3}
  {\path{doi:10.1007/s10236-014-0760-3}}.

\bibitem{gupta_et_al_Oceans2019}
A.~Gupta, P.~J. Haley, D.~N. Subramani, P.~F.~J. Lermusiaux, Fish modeling and
  {B}ayesian learning for the {L}akshadweep {I}slands, in: OCEANS 2019 MTS/IEEE
  SEATTLE, IEEE, Seattle, 2019, pp. 1--10.
\newblock \href {https://doi.org/10.23919/OCEANS40490.2019.8962892}
  {\path{doi:10.23919/OCEANS40490.2019.8962892}}.

\bibitem{bilmes1998gentle}
J.~A. Bilmes, et~al., A gentle tutorial of the {EM} algorithm and its
  application to parameter estimation for {G}aussian mixture and hidden
  {M}arkov models, International Computer Science Institute 4~(510) (1998) 126.

\bibitem{stoica2004model}
P.~Stoica, Y.~Selen, Model-order selection: a review of information criterion
  rules, Signal Processing Magazine, IEEE 21~(4) (2004) 36--47.

\bibitem{wornell2013}
G.~Wornell, Inference and information. lecture notes for mit course 6.437 in
  spring 2013 (May 2016).

\bibitem{lin_PhDThesis2020}
J.~Lin, Bayesian learning for high-dimensional nonlinear systems:
  {M}ethodologies, numerics and applications to fluid flows, Ph.D. thesis,
  Massachusetts Institute of Technology, Department of Mechanical Engineering,
  Cambridge, Massachusetts (Sep. 2020).

\bibitem{duda2006pattern}
R.~O. Duda, P.~E. Hart, et~al., Pattern classification, John Wiley \& Sons,
  2006.

\bibitem{gupta_MSThesis2016}
A.~Gupta, Bayesian inference of obstacle systems and coupled
  biogeochemical-physical models, Master's thesis, Indian Institute of
  Technology Kanpur, Kanpur, India (2016).

\bibitem{lermusiaux_et_al_TheSea2017}
P.~F.~J. Lermusiaux, D.~N. Subramani, J.~Lin, C.~S. Kulkarni, A.~Gupta,
  A.~Dutt, T.~Lolla, P.~J. Haley, Jr., W.~H. Ali, C.~Mirabito, S.~Jana, A
  future for intelligent autonomous ocean observing systems, Journal of Marine
  Research 75~(6) (2017) 765--813, the Sea. Volume 17, The Science of Ocean
  Prediction, Part 2.
\newblock \href {https://doi.org/10.1357/002224017823524035}
  {\path{doi:10.1357/002224017823524035}}.

\bibitem{lermusiaux_et_al_TheSea2017_modified}
P.~F.~J. Lermusiaux, et~al., A future for intelligent autonomous ocean
  observing systems, Journal of Marine Research 75~(6) (2017) 765--813, the
  Sea. Volume 17, The Science of Ocean Prediction, Part 2.
\newblock \href {https://doi.org/10.1357/002224017823524035}
  {\path{doi:10.1357/002224017823524035}}.

\bibitem{lermusiaux_et_al_Oceanog2017_modified}
P.~F.~J. Lermusiaux, et~al., Optimal planning and sampling predictions for
  autonomous and {L}agrangian platforms and sensors in the northern {A}rabian
  {S}ea, Oceanography 30~(2) (2017) 172--185, special issue on Autonomous and
  Lagrangian Platforms and Sensors ({ALPS}).
\newblock \href {https://doi.org/10.5670/oceanog.2017.242}
  {\path{doi:10.5670/oceanog.2017.242}}.

\bibitem{subramani_PhDThesis2018}
D.~N. Subramani, Probabilistic regional ocean predictions: Stochastic fields
  and optimal planning, Ph.D. thesis, Massachusetts Institute of Technology,
  Department of Mechanical Engineering, Cambridge, Massachusetts (Feb. 2018).

\bibitem{subramani_lermusiaux_prep}
D.~Subramani, P.~F.~J. Lermusiaux, Probabilistic ocean predictions with
  dynamically-orthogonal primitive equations, in preparation (\the\year{}).

\bibitem{gkirgkis_MSThesis2021}
K.~A. Gkirgkis, Stochastic ocean forecasting with the dynamically orthogonal
  primitive equations, Master's thesis, Massachusetts Institute of Technology,
  Department of Mechanical Engineering, Cambridge, Massachusetts (Jun. 2021).

\bibitem{gkirgkis_lermusiaux_prep}
K.~A. Gkirgkis, P.~F.~J. Lermusiaux, Massive probabilistic forecasts for the
  {G}ulf of {M}exico: Dynamically-orthogonal primitive equations, in
  preparation (\the\year{}).

\bibitem{lermusiaux_robinson_MWR1999}
P.~F.~J. Lermusiaux, A.~R. Robinson, Data assimilation via {E}rror {S}ubspace
  {S}tatistical {E}stimation, part {I}: Theory and schemes, Monthly Weather
  Review 127~(7) (1999) 1385--1407.
\newblock \href
  {https://doi.org/10.1175/1520-0493(1999)127<1385:DAVESS>2.0.CO;2}
  {\path{doi:10.1175/1520-0493(1999)127<1385:DAVESS>2.0.CO;2}}.

\bibitem{lermusiaux_DAO1999}
P.~F.~J. Lermusiaux, Estimation and study of mesoscale variability in the
  {S}trait of {S}icily, Dynamics of Atmospheres and Oceans 29~(2) (1999)
  255--303.
\newblock \href {https://doi.org/10.1016/S0377-0265(99)00008-1}
  {\path{doi:10.1016/S0377-0265(99)00008-1}}.

\bibitem{lermusiaux_et_al_ICTCA2002}
P.~F.~J. Lermusiaux, C.-S. Chiu, A.~R. Robinson, Modeling uncertainties in the
  prediction of the acoustic wavefield in a shelfbreak environment, in: E.-C.
  Shang, Q.~Li, T.~F. Gao (Eds.), Proceedings of the 5th International
  Conference on Theoretical and Computational Acoustics, World Scientific
  Publishing Co., 2002, pp. 191--200, refereed invited manuscript.
\newblock \href {https://doi.org/10.1142/9789812777362_0020}
  {\path{doi:10.1142/9789812777362_0020}}.

\bibitem{cossarini_et_al_JGR2009}
G.~Cossarini, P.~F.~J. Lermusiaux, C.~Solidoro, Lagoon of {V}enice ecosystem:
  Seasonal dynamics and environmental guidance with uncertainty analyses and
  error subspace data assimilation, Journal of Geophysical Research: Oceans
  114~(C6) (Jun. 2009).
\newblock \href {https://doi.org/10.1029/2008JC005080}
  {\path{doi:10.1029/2008JC005080}}.

\bibitem{lermusiaux_et_al_BBN_Oceans2020}
P.~F.~J. Lermusiaux, C.~Mirabito, P.~J. Haley, Jr., W.~H. Ali, A.~Gupta,
  S.~Jana, E.~Dorfman, A.~Laferriere, A.~Kofford, G.~Shepard, M.~Goldsmith,
  K.~Heaney, E.~Coelho, J.~Boyle, J.~Murray, L.~Freitag, A.~Morozov, Real-time
  probabilistic coupled ocean physics-acoustics forecasting and data
  assimilation for underwater {GPS}, in: OCEANS 2020 IEEE/MTS, IEEE, 2020, pp.
  1--9.
\newblock \href {https://doi.org/10.1109/IEEECONF38699.2020.9389003}
  {\path{doi:10.1109/IEEECONF38699.2020.9389003}}.

\bibitem{haley_et_al_Oceans2020}
P.~J. Haley, Jr., A.~Gupta, C.~Mirabito, P.~F.~J. Lermusiaux, Towards
  {B}ayesian ocean physical-biogeochemical-acidification prediction and
  learning systems for {M}assachusetts {B}ay, in: OCEANS 2020 IEEE/MTS, IEEE,
  2020, pp. 1--9.
\newblock \href {https://doi.org/10.1109/IEEECONF38699.2020.9389210}
  {\path{doi:10.1109/IEEECONF38699.2020.9389210}}.

\bibitem{lolla_lermusiaux_MWR2017_partI}
T.~Lolla, P.~F.~J. Lermusiaux, A {G}aussian mixture model smoother for
  continuous nonlinear stochastic dynamical systems: Theory and scheme, Monthly
  Weather Review 145 (2017) 2743--2761.
\newblock \href {https://doi.org/10.1175/MWR-D-16-0064.1}
  {\path{doi:10.1175/MWR-D-16-0064.1}}.

\bibitem{lolla_lermusiaux_MWR2017_partII}
T.~Lolla, P.~F.~J. Lermusiaux, A {G}aussian mixture model smoother for
  continuous nonlinear stochastic dynamical systems: Applications, Monthly
  Weather Review 145 (2017) 2763--2790.
\newblock \href {https://doi.org/10.1175/MWR-D-16-0065.1}
  {\path{doi:10.1175/MWR-D-16-0065.1}}.

\bibitem{heaney_et_al_OD2016}
K.~D. Heaney, P.~F.~J. Lermusiaux, T.~F. Duda, P.~J. Haley, Jr., Validation of
  genetic algorithm based optimal sampling for ocean data assimilation, Ocean
  Dynamics 66 (2016) 1209--1229.
\newblock \href {https://doi.org/10.1007/s10236-016-0976-5}
  {\path{doi:10.1007/s10236-016-0976-5}}.

\bibitem{heaney_et_al_JFR2007}
K.~D. Heaney, G.~Gawarkiewicz, T.~F. Duda, P.~F.~J. Lermusiaux, Nonlinear
  optimization of autonomous undersea vehicle sampling strategies for
  oceanographic data-assimilation, Journal of Field Robotics 24~(6) (2007)
  437--448.
\newblock \href {https://doi.org/10.1002/rob.20183}
  {\path{doi:10.1002/rob.20183}}.

\bibitem{lermusiaux_et_al_ST2007}
P.~F.~J. {Lermusiaux}, P.~J. {Haley}, Jr, N.~K. {Yilmaz}, Environmental
  prediction, path planning and adaptive sampling:~sensing and modeling for
  efficient ocean monitoring, management and pollution control, Sea Technology
  48~(9) (2007) 35--38.

\bibitem{ramp_et_al_DSR2009}
S.~R. Ramp, R.~E. Davis, N.~E. Leonard, I.~Shulman, Y.~Chao, A.~R. Robinson,
  J.~Marsden, P.~F.~J. Lermusiaux, D.~M. Fratantoni, J.~D. Paduan, F.~P.
  Chavez, F.~L. Bahr, S.~Liang, W.~Leslie, Z.~Li, Preparing to predict: The
  second {A}utonomous {O}cean {S}ampling {N}etwork ({AOSN-II}) experiment in
  the {M}onterey {B}ay, Deep Sea Research Part II: Topical Studies in
  Oceanography 56~(3--5) (2009) 68--86.
\newblock \href {https://doi.org/10.1016/j.dsr2.2008.08.013}
  {\path{doi:10.1016/j.dsr2.2008.08.013}}.

\bibitem{wang_et_al_JMS2009}
D.~Wang, P.~F.~J. Lermusiaux, P.~J. Haley, Jr., D.~Eickstedt, W.~G. Leslie,
  H.~Schmidt, Acoustically focused adaptive sampling and on-board routing for
  marine rapid environmental assessment, Journal of Marine Systems
  78~(Supplement) (2009) S393--S407.
\newblock \href {https://doi.org/10.1016/j.jmarsys.2009.01.037}
  {\path{doi:10.1016/j.jmarsys.2009.01.037}}.

\bibitem{petillo_et_al_Oceans2015}
S.~Petillo, D.~Y. H.~Schmidt, P.F.J.~Lermusiaux, A.~Balasuriya, Autonomous \&
  adaptive oceanographic front tracking on board autonomous underwater
  vehicles, in: Proceedings of IEEE OCEANS'15 Conference, IEEE, Genoa, 2015.
\newblock \href {https://doi.org/10.1109/oceans-genova.2015.7271616}
  {\path{doi:10.1109/oceans-genova.2015.7271616}}.

\bibitem{Cococcioni_et_al_Oceans2015}
M.~Cococcioni, B.~Lazzerini, P.~Lermusiaux, Adaptive sampling using fleets of
  underwater gliders in the presence of fixed buoys using a constrained
  clustering algorithm, in: Proceedings of IEEE OCEANS'15 Conference, IEEE,
  Genoa, 2015.
\newblock \href {https://doi.org/10.1109/oceans-genova.2015.7271446}
  {\path{doi:10.1109/oceans-genova.2015.7271446}}.

\bibitem{rajan_et_al_MTSJ2021}
K.~Rajan, F.~Aguado, P.~Lermusiaux, J.~B. de~Sousa, A.~Subramaniam, J.~Tintore,
  {METEOR}: A {M}obile (portable) oc{E}an robo{T}ic obs{E}rvat{OR}y, Marine
  Technology Society Journal 55~(3) (2021) 74--75.
\newblock \href {https://doi.org/10.4031/MTSJ.55.3.42}
  {\path{doi:10.4031/MTSJ.55.3.42}}.

\bibitem{lolla_PhDThesis2016}
S.~V.~T. Lolla, Path planning and adaptive sampling in the coastal ocean, Ph.D.
  thesis, Massachusetts Institute of Technology, Department of Mechanical
  Engineering, Cambridge, Massachusetts (Feb. 2016).

\bibitem{lermusiaux_et_al_Oceanog2017}
P.~F.~J. Lermusiaux, P.~J. Haley, Jr., S.~Jana, A.~Gupta, C.~S. Kulkarni,
  C.~Mirabito, W.~H. Ali, D.~N. Subramani, A.~Dutt, J.~Lin, A.~Shcherbina,
  C.~Lee, A.~Gangopadhyay, Optimal planning and sampling predictions for
  autonomous and {L}agrangian platforms and sensors in the northern {A}rabian
  {S}ea, Oceanography 30~(2) (2017) 172--185, special issue on Autonomous and
  Lagrangian Platforms and Sensors ({ALPS}).
\newblock \href {https://doi.org/10.5670/oceanog.2017.242}
  {\path{doi:10.5670/oceanog.2017.242}}.

\bibitem{gupta_ali_lermusiaux_2016prep}
A.~Gupta, W.~H. Ali, P.~F.~J. Lermusiaux, Boundary conditions for stochastic
  {DO} equations, {MSEAS Report}, Department of Mechanical Engineering,
  Massachusetts Institute of Technology, Cambridge, MA (2016).

\bibitem{humara_MSThesis2020}
M.~J. Humara, Stochastic acoustic ray tracing with dynamically orthogonal
  equations, Master's thesis, Massachusetts Institute of Technology, Joint
  Program in Applied Ocean Science and Engineering, Cambridge, Massachusetts
  (May 2020).

\bibitem{humara_et_al_Oceans2022}
M.~J. Humara, W.~H. Ali, A.~Charous, M.~Bhabra, P.~F.~J. Lermusiaux, Stochastic
  acoustic ray tracing with dynamically orthogonal differential equations, in:
  OCEANS 2022 IEEE/MTS, IEEE, Hampton Roads, VA, 2022, pp. 1--10.
\newblock \href {https://doi.org/10.1109/OCEANS47191.2022.9977252}
  {\path{doi:10.1109/OCEANS47191.2022.9977252}}.

\bibitem{charous_clustering}
A.~Charous, M.~J. Humara, W.~H. Ali, M.~S. Bhabra, A.~Gupta, P.~F. Lermusiaux,
  Dynamically orthogonal ray equations with adaptive reclustering, The Journal
  of the Acoustical Society of America 150~(4) (2021) A209--A209.
\newblock \href {https://doi.org/10.1121/10.0008139}
  {\path{doi:10.1121/10.0008139}}.

\end{thebibliography}

\end{document}